\documentclass[usenatbib]{mn2e}
\usepackage{color}
\usepackage{graphicx}
\usepackage{amsmath,amssymb,amsfonts}
\usepackage{bm}
\usepackage{natbib}
\usepackage{subfigure}
\usepackage{dcolumn}

\begin{document}

\title[$M_{\rm bh}$--$L,n$ -relations]{The near-IR $M_{\rm bh}$--$L$ and $M_{\rm bh}$--$n$  relations}

\author[Vika et al.]
{Marina Vika,$^{1,2}$ Simon P.~Driver,$^{1,2}$ Ewan Cameron ,$^{3}$ Lee Kelvin,$^{1,2}$ Aaron Robotham,$^{1,2}$\\
$^1$Scottish Universities Physics Alliance (SUPA)\\
$^2$School of Physics \& Astronomy, University of St Andrews, North Haugh, St Andrews, Fife, KY16 9SS\\
$^3$Department of Physics, Swiss Federal Institute of Technology(ETH-Z$\ddot{u}$rich), CH-8093 Z$\ddot{u}$rich, Switzerland\\
}

\date{Received XXXX; Accepted XXXX}
\pubyear{2006} \volume{000}
\pagerange{\pageref{firstpage}--\pageref{lastpage}}

\maketitle
\label{firstpage}

\begin{abstract}
We present near-IR surface photometry (2D-profiling) for a sample of 29 nearby galaxies for which super-massive black hole (SMBH) masses are constrained. The data is derived from the UKIDSS-LASS survey representing a significant improvement in image quality and depth over previous studies based on 2MASS data. We derive the spheroid luminosity and spheroid S{\'e}rsic index for each galaxy with GALFIT3 and use these data to construct SMBH mass -bulge luminosity ($M_{\rm bh}$--$L$) and SMBH - S{\'e}rsic index ($M_{\rm bh}$--$n$) relations. The best fit K-band relation for elliptical and disk galaxies is  $\log(M_{\rm bh}/M_{\odot})=~-0.36(\pm~0.03) ( M_{\rm K} + 18)~ + ~ 6.17(\pm~0.16)$ with an intrinsic scatter of 0.4$^{+0.09}_{-0.06}$dex whilst for elliptical galaxies we find $\log(M_{\rm bh}/M_{\odot})=~-0.42(\pm~0.06) (M_{\rm K} + 22)~ + ~  7.5(\pm~0.15)$ with an intrinsic scatter of 0.31$^{+0.087}_{-0.047}$dex. Our revised $M_{\rm bh}$--$L$ relation agrees closely with the previous near-IR constraint by \citet{tex:G07}.  The lack of improvement in the intrinsic scatter in moving to higher quality near-IR data suggests that the SMBH relations are not currently limited by the quality of the imaging data but is either intrinsic or a result of uncertainty in the precise number of required components required in the profiling process. Contrary to expectation (see \citealt{tex:GD07a}) a relation between SMBH mass and the S{\'e}rsic index was not found at near-IR wavelengths. This latter outcome is believed to be explained by the generic inconsistencies between 1D and 2D galaxy profiling which are currently under further investigation.
\end{abstract}


\section{Introduction}
\label{sec:1}

The number of SMBH mass measurements ($M_{\rm bh}$)  from inactive galaxies in the local Universe  have  rapidly increased over the last 15 years (see \citealt[hereafter GO10]{tex:GO10} for sample compilation).  \citet{tex:KR95} reviewed $M_{\rm bh}$ determinations  for a sample of 8 local galaxies and introduced the SMBH mass ($M_{\rm bh}$) - galaxy luminosity ($L$, or bulge luminosity in the case of disk galaxies) relation\footnote{Some other properties known to correlate with SMBH mass are the mean velocity dispersion (\citealt{tex:TG02}) and the stellar mass (\citealt{tex:HR04})}. Since this time, the $M_{\rm bh}$--$L$ correlation has been investigated by a number of groups from optical to near-IR passbands (\citealt{tex:MT98,tex:KG01,tex:MH03,tex:G07,tex:GR09b}).

Establishing an accurate $M_{\rm bh}$--$L$ relation is beneficial for two reasons: firstly to understand the physical basis as to why the relation exists  and secondly to provide a means for predicting SMBH masses for large samples of galaxies (e.g. \citealt{tex:GD07,tex:VD09}). Furthermore Active Galactic Nuclei (AGN) are the antecedent of most local inactive galaxies, studying the  $M_{\rm bh}$--$L$ relationship for both active and inactive galaxies can therefore provide information about the parallel evolution of black holes and their host galaxies.  By exploring the origin of the scaling relation  (\citealt{tex:WT06,tex:KH08}) at high redshift  (\citealt{tex:PI06a}), and within active galaxies (\citealt{tex:MD02,tex:BF03,tex:GK09b}), one can study the evolution of  SMBHs with time in comparison with the spheroid evolution. \citet{tex:BT10} found no change for evolution of the $M_{\rm bh}$--$L_{\rm tot}$ relation up to z$=$1,  in contradiction with the  $M_{\rm bh}$--$L_{\rm sph}$ evolution. Further research of SMBH evolution shows that up to z$\sim$3 host galaxies at fixed  $M_{\rm bh}$ are less massive as compared to local galaxies (\citealt{tex:TW07,tex:JB09}). This indicates SMBHs in early-type galaxies have reached their final mass during the very earliest  phases of  galaxy evolution (\citealt{tex:IH03}). As a result the $M_{\rm bh}$/ $M_{\rm gal}$ should be larger compared with the local ratio (\citealt{tex:GP10a}), an outcome confirmed by QSO observations up to z$\sim$6 (\citealt{tex:WC04,tex:SM06,tex:MB09}). However, other studies argue that this result may arise due to a selection bias (\citealt{tex:BS05,tex:AB08}).

Another aspect of the $M_{\rm bh}$--$L$ relation is the behaviour at the low luminosity end. Do SMBHs exist in low-mass galaxies (\citealt{tex:MFc01}) or is there a lower galaxy mass limit at which we can detect a SMBH.  \citet[hereafter H09]{tex:H09} argue that bulge-less galaxies (or pseudobulges) follow a distinct relation while \citet{tex:GH08} showed that a classical bulge is not necessary for a SMBH to exist. Pseudobulges are central components of late type galaxies with disk features and it is believed that they follow a separate formation path to the classical bulges with which they can coexist. For a review of the properties of pseudobulges see \citet{tex:KK04}.

The first estimation of the $M_{\rm bh}$--$L$ relation in the near-IR was established by \citet{tex:MH03} using three band images from the Two Micron All Sky Survey (2MASS)\footnote{ \citet{tex:JC00}} for a sample of 37 early- and late- type galaxies.  The intrinsic scatter of their correlation ranges from $\sim$0.5dex to $\sim$0.3dex depending on the subsample selection.  \citet{tex:G07} refined the \citet{tex:MH03} $M_{\rm bh}$--$L$ relation by using updated $M_{\rm bh}$ measurements and modifying the photometry of the data set and finding an intrinsic scatter of $\sim$0.30dex.

More recently and in addition to the $M_{\rm bh}$--$L$ relation \citet[hereafter GD07 following on from \citealt{tex:GE01}]{tex:GD07a} found a relation between the galaxy light concentration $n$ (S{\'e}rsic index)  and $M_{\rm bh}$ with a comparable intrinsic scatter of 0.31dex. In the review by \citet{tex:NF06} the $M_{\rm bh}$--$n$ relation was shown to be as accurate on predicting $M_{\rm bh}$ as the $M_{\rm bh}$--$\sigma$ relations.

\citet{tex:GD07} and \citet{tex:VD09} applied both the $M_{\rm bh}$--$n$ and $M_{\rm bh}$--$L$ relations  respectively to derive the nearby SMBH mass functions for the Millennium Galaxy Catalogue (\citealt{tex:LL03,tex:DL05}) and derived individual SMBH mass measurements for a sample of 1743 galaxies. While the mass functions agreed well within the cited errors,  the comparison of the derived SMBH masses on a galaxy by galaxy basis showed a low consistency between the two predictors. They explained that the lack of correlation is in part due to the scatter introduced by combining elliptical and disk galaxies, the uncertainty of separating the bulge component  from the disk component, and due to the intrinsic scatter of the $M_{\rm bh}$--$L,n$ correlations at optical wavelengths. 

In this paper we aim to reconstruct the $M_{\rm bh}$--$L,n$ relations in the near-IR by using high resolution  UKIRT Infrared Deep Sky Survey (UKIDSS, \citealt{tex:LW07}) images which extend significant deeper ($\sim2$mag/arcsec$^2$) than the previous studies based on 2MASS. We specifically choose near-IR photometry  because galaxy profiles should be less perturbed by young star populations and by dust attenuation relative to optical passbands (\citealt{tex:DP08}). Thereby yielding a lower intrinsic scatter and enabling more accurate SMBH mass function determinations from the application of these relations to large surveys (e.g. GAMA see \citealt{tex:DN09}). 

Section~\ref{sec:2} describes the data selection and the data reduction. Section~\ref{sec:3} describes the methodology for measuring \textbf{2D} radial surface-brightness profiles for our 29 galaxies using \texttt{GALFIT3} (\citealt{tex:PH10}) and presents information from the literature for our sample. In Section~\ref{sec:4} we explore the $M_{\rm bh}$--$L,n$ correlations and compare with previous studies. Finally, Section ~\ref{sec:5} summarises our conclusions and suggests possible direction for further study.

\begin{figure}
\centering
\includegraphics[height=6.0cm,width=6.0cm]{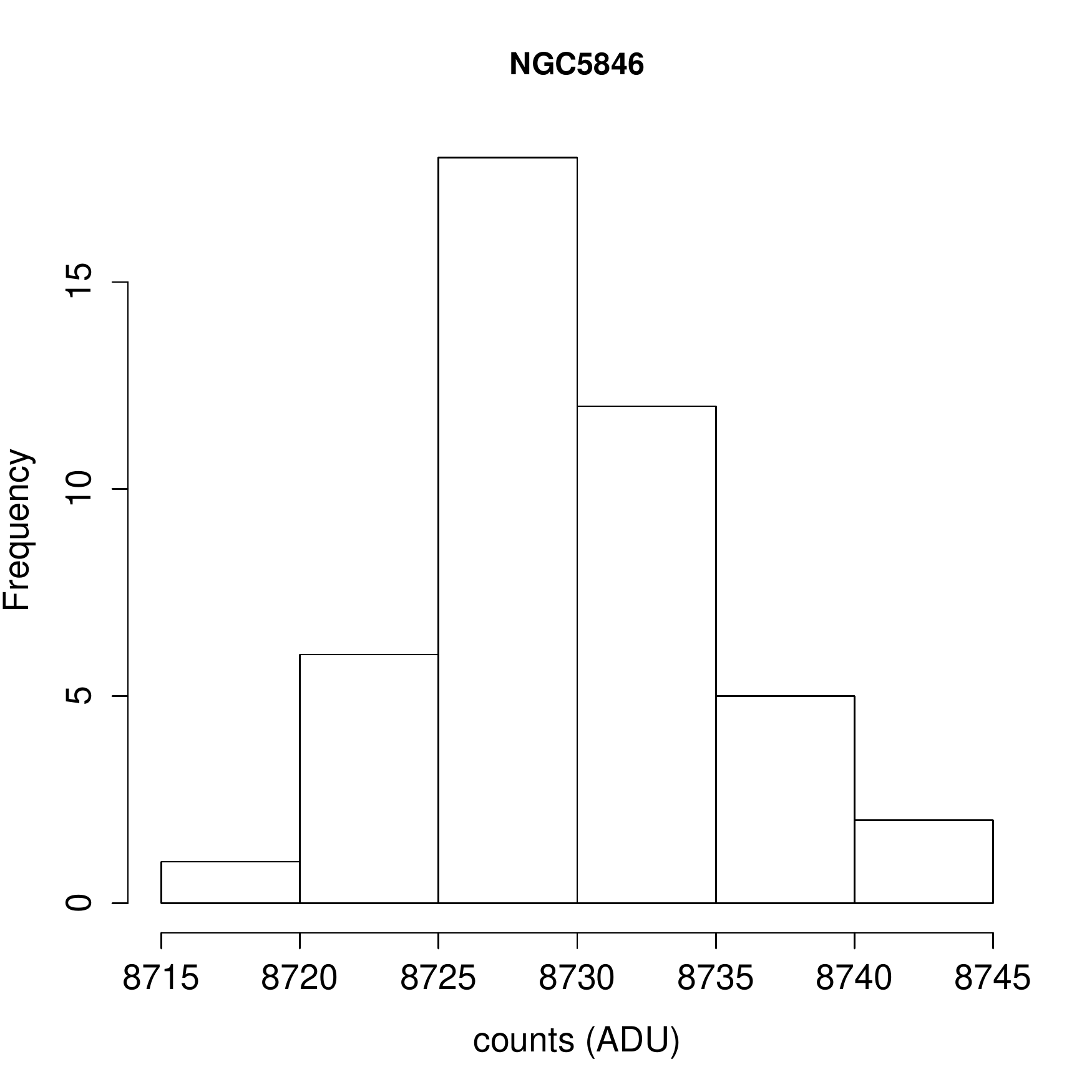}
\caption{The distribution of the sky background for the NGC5846. The mean 
sky value is 8730 ADU  with $\sigma=6$).}
\label{fig:sky}
\end{figure} 

 \section{Data Reduction}
\label{sec:2}
We extract calibrated K-band images from UKIDSS for 29 galaxies for which SMBH masses have been measured. These galaxies are a subsample of the host galaxy population, where the mass of the SMBH has been measured using a direct method. The full sample consists of 86 galaxies with SMBHs and 9 galaxies with intermediate massive black holes (IMBH), as presented by GO10  and references therein.
 
The wide-field images were obtained using the Wide Field Infrared Camera (WFCAM) (\citealt{tex:CA07}) on the 3.8-m United Kingdom Infra-red Telescope (UKIRT) as part of the Large Area Survey UKIDSS-LAS (\citealt{tex:LW07}). The pixel size of each detector is 0.4$~$arcsec with a gain of 4.5 e$^{-}$/ADU and a read noise of 25 ADU. 

The properties of our galaxy sample  are listed in Table~\ref{table:first}. We include galaxies with SMBH masses measured with stellar kinematics, gas kinematics, water masers, stellar proper motion and reverberation mapping (see Table~\ref{table:masses}). The masses for NGC2778, NGC4473, NGC4564, NGC4697 and NGC5845 have been modified from their initial published values due to an update of their  distances (see GO10). Our sample includes 15 elliptical galaxies and 14 disk galaxies (see Table 1).

We also include for reference only the Milky Way parameters derived from other studies. In particular, the SMBH mass is $(4.3 \pm 0.3)10^{6} \rm M_{\odot}$  for a distance of 0.0083Mpc (\citealt{tex:GE09}), the bulge luminosity is $(4.0 \pm 1.2)10^{8} \rm L_{\odot}$ at 2.2$\mu$m (\citealt{tex:DA95}) and the S{\'e}rsic index 1.32$^{+0.26 }_{- 0.22 }$ (\citealt{tex:KD91,tex:GD07a}).

 \begin{table*}
\caption{Galaxy Sample; Column(1): Galaxy name; Column(2) and (3): equatorial  coordinates (J2000); 
Column(4) and (5): K-Band and B-Band magnitudes from the SIMBAD Astronomical Database; Column(6): 
reddening estimate E(B-V) from \citet{tex:SF98}; Column(7): Hubble type from NED (and in bracket 
from GO10) ; Column(8): redshift from the NED; Column(9): Distance from \citet{tex:G08b}; 
Column(11): Activity (Sy:Seyfert, NLRG:Narrow Line Radio Galaxy, HII:Nuclear HII regions, 
L:Low Ionisation Nuclear Emission-Line Regions,LLAGN:low-luminosity AGN.}
  \smallskip
  \centering
  \begin{tabular}{lcccccccc}
  \hline
  \noalign{\smallskip}
Galaxy   &     RA    &  DEC     &  K	&    B   & A$_{\rm k}$ & Type     & Redshift & Activity \\
Name     &   (J2000) & (J2000)  &(mag)  &  (mag) &  (mag)      &	  &	     &  	\\
 (1)     &   (2)     & (3)      &  (4)  &   (5)  & (6)         & (7)      &  (8)     &   (9)	\\   
\noalign{\smallskip}
\hline
\hline
\noalign{\smallskip}       
NGC221   & 00 42 41.8  &  +40 51 57.2  & 5.09  & 9.2  &  0.057  & cE2	(S0)     & -0.0006   & -      \\
NGC863   & 02 14 33.56 &  -00 46 00.0  & 9.54  & 14.0 &  0.013 & SA(s)a          & 0.02638   &  Sy1-2 \\
NGC1068  & 02 42 40.83 &  -00 00 48.4  & 5.79  & 9.7  &  0.012 & SAb	         & 0.00379   &  Sy1-2 \\
NGC2778  & 09 12 24.35 &  +35 01 39.4  & 9.514 & 13.1 &  0.008 & E	(SB0)    & 0.00683   & -      \\
NGC2960  & 09 40 36.46 &  +03 34 36.6  & 9.783 & 13.6 &  0.016 & Sa	         & 0.01645   &  Sy3   \\
NGC3245  & 10 27 18.52 &  +28 30 24.8  & 7.862 & 11.6 &  0.009 & SA(r)  (S0)     & 0.00438   &  HII L \\
NGC4258  & 12 18 57.54 &  +47 18 14.3  & 5.464 & 9.6  &  0.006 & SABbc           & 0.00149   &  Sy1   \\
NGC4261  & 12 19 23.21 &  +05 49 29.7  & 7.26  & 12.0 &  0.006 & E2	         & 0.00746   &  Sy3-L \\
NGC4303  & 12 21 55.03 &  +04 28 28.7  & 6.843 & 10.9 &  0.008 & SAB(rs)bc       & 0.00522   & HII Sy2\\
NGC4342  & 12 23 39.12 &  +07 03 12.9  & 9.023 & 13.0 &  0.008 & SO	         & 0.0025    & -      \\
NGC4374  & 12 25 03.74 &  +12 53 13.1  & 6.222 & 10.8 &  0.015 & E1	         & 0.00353   &  Sy2-L \\
NGC4435  & 12 27 40.60 &  +13 04 44.4  & 7.297 & 11.9 &  0.011 & SB(s) (SB0)     & 0.00267   & -      \\
NGC4459  & 12 29 00.13 &  +13 58 42.5  & 7.152 & 11.6 &  0.017 & SA0         & 0.00403   &  HII L \\
NGC4473  & 12 29 48.95 &  +13 25 46.1  & 7.157 & 11.2 &  0.010 & E5	         & 0.00748   & -      \\  
NGC4486  & 12 30 49.42 &  +12 23 28.0  & 5.812 & 10.4 &  0.008 & cD,E0	 & 0.00436   & NLRG-Sy\\  
NGC4486a & 12 30 57.89 &  +12 16 13.7  & 11.2  & 9.01 &  0.009 & E2	         & 0.00050   & -      \\  
NGC4486b & 12 30 31.82 &  +12 29 25.9  & 10.09 & 14.5 &  0.008 & cE0	         & 0.00519   & -      \\  
NGC4552  & 12 35 40    &  +12 33 22    & 6.728 & 11.1 &  0.015 & E1    (S0)      & 0.00113   & HII Sy2-L\\  
NGC4564  & 12 36 27.01 &  +11 26 18.8  & 7.937 & 12.2 &  0.012 & E6    (S0)      & 0.00380   & -      \\   
NGC4596  & 12 39 56.16 &  +10 10 32.4  & 7.463 & 12.4 &  0.008 & SB(r)0$+$ & 0.00623   &  L     \\   
NGC4621  & 12 42 02    &  +11 38 45    & 6.746 & 11.0 &  0.012 & E5	         & 0.00137   & -      \\  
NGC4649  & 12 43 40.19 &  +11 33 08.9  & 5.739 & 10.3 &  0.010 & E2	         & 0.00372   & -      \\   
NGC4697  & 12 48 35.7  &  -05 48 03    & 6.367 & 11.0 &  0.011 & E6	         & 0.00414   &  LLAGN \\   
NGC5576  & 14 21 03.7  &  +03 16 16    & 7.827 & 11.9 &  0.011 & E3	         & 0.00496   & -      \\
NGC5813  & 15 01 11.3  &  +01 42 06    & 7.413 & 12.5 &  0.021 & E1-2	         & 0.00658   &  L     \\  
NGC5845  & 15 06 00.9  &  +01 38 01.4  & 9.112 & 13.8 &  0.020 & E	         & 0.00483   & -      \\  
NGC5846  & 15 06 29.4  &  +01 36 19    & 6.935 & 11.9 &  0.020 & E0	         & 0.00571   &  HII L \\   
NGC7052  & 21 18 33.1  &  +26 26 48    & 8.574 & 14.0 &  0.046 & E	         & 0.0241    & -      \\
UGC9799  & 15 16 44.6  &  +07 01 16.3  & 9.548 & 14.8 &  0.014 & cD;E	         & 0.0345    &  Sy2   \\  
\noalign{\smallskip}
\hline
\end{tabular}
\label{table:first} 
\end{table*}

\subsection{Sky}
\label{sec:2.1}
The accuracy to which we can determine the sky background dictates the depth to which we can profile 
each galaxy. We measure the sky background by manually placing 40 - 50 boxes (10 x 10 pixels) at 
locations around each galaxy using the \texttt{IRAF} task imexamine. The sky value we then adopt is the 
mean of the median values from each box (see Figure~\ref{fig:sky} for details). The boxes are selected to 
lie away from stars, the faint halo of the galaxy, neighbouring galaxies that may exist and to be uniformly 
distributed around each image.

Some images (NGC2778, NGC3245, NGC4258, NGC7052) have a noticeable background gradient. We 
correct the gradient with the use of a \texttt{SExtractor} (\citealt{tex:BA96}) fully resolved background map. 
To ensure that the subtraction of the background map will not deteriorate the galaxy flux we first subtract 
the galaxy with the help of a model constructed with  \texttt{IRAF - ELLIPSE BMODEL}. After we have 
removed the model we create a fully resolved background map which we subtract from the initial image. 
The sky background values derived from the method described above can be found in Table~\ref{table:info} 
and are used later as an input to  \texttt{GALFIT3}. All background values have been independently checked using \texttt{STARLINK - ESP - HISTPEAK} and agree within the quality errors. Nevertheless we do note in particular the extensive structure in the background of the NGC4486 possibly due to the UKIDSS reduction pipeline. The NGC4486 case will be discussed further in Section \ref{sec:index}.

\begin{figure*}
\centering
\subfigure{(a)\includegraphics[height=5cm,width=5cm]{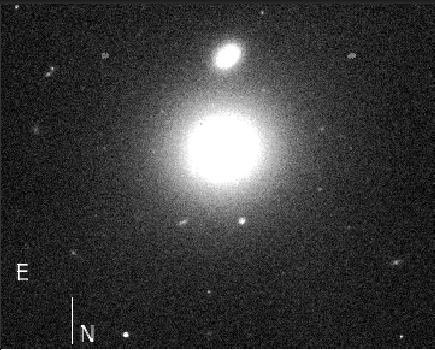}} 
\subfigure{(b)\includegraphics[height=5cm,width=5cm]{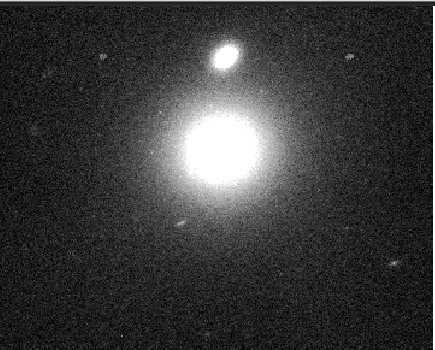}} 
\subfigure{(c)\includegraphics[height=5cm,width=5cm]{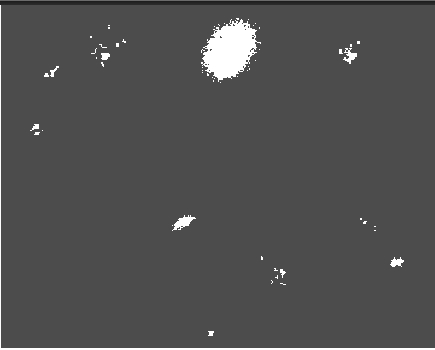}}
\caption{(a)K-band image of NGC5846 (b)The same image cleaned of background stars 
(c)The segmentation map. See Section~\ref{sec:2} for details.}
\label{fig:clean}
\end{figure*}

\subsection{PSF}
\label{sec:2.2}
Two PSFs are created for each galaxy, based on stars taken from the same data frame and using the package \texttt{IRAF DAOPHOT}. The PSF model is described by a penny2 function. Penny2 has a Gaussian core and Lorentzian wings which are free to be tilted in different directions. We construct the PSF from a sample of 10-15 stars selected from each galaxy/image in interactive mode. A different set of stars is used for each PSF.  Saturated stars, or stars very close to the galaxy with unclear background levels, are excluded from the sample. After the creation of the \texttt{DAOPHOT} PSF we use it to subtract all stars from the original image. The left panel and the middle panel of Figure~\ref{fig:clean} show an example galaxy image before and after removing the stars.

\subsection{Image Masks}
\label{sec:2.3}
In some images the main galaxy  is surrounded by satellite galaxies (e.g. UGC9799), bad pixels and saturated stars (e.g. NGC4459). The light distribution from the neighbouring galaxies and the area that the bad pixels cover cannot be cleaned with the same technique we used for the stars. In these cases we use an image mask that indicates to \texttt{GALFIT3} which areas of the image should not be used. We create these maps using \texttt{SExtractor} segmentation maps. The right panel of Figure~\ref{fig:clean} shows the \texttt{SExtractor} segmentation maps with pixels having a non zero value in the map being excluded from the fitting process.

\begin{table*} 
\caption{Galaxy Sample. The background is the mean value of fifty median sky values. Each median value has been estimated for a box of 100 pixels. The stddev is the standard deviation of the mean background value.}
\smallskip
\centering
\begin{tabular}{ccccccc}
\hline
\noalign{\smallskip}
Galaxy    &  Seeing  &background  &  stddev  &  Exposure time  &  Survey  & Telescope/ \\
Name      &  (pixel)   &                        &               &   (sec)	          &	           & Instrument \\
 (1)           &   (2)        &	(3)           &    (4)      &    (5)                     & (6)           &  (7)       \\        
\noalign{\smallskip}
\hline
\hline
\noalign{\smallskip}
NGC 221    & 4.6 & 6540   & 8  & 10   & SERV/1652  & UKIRT/WFCAM  \\ 
NGC 863    & 2.1 & 6606   & 3  & 10   & LAS   &  UKIRT/WFCAM		\\
NGC 1068  & 2.7 & 5945   & 5   & 10   & LAS   & UKIRT/WFCAM  \\ 
NGC 2778  & 2.3 & 10723 & 1.2  & 10   & B2    & UKIRT/WFCAM  \\ 
NGC 2960  & 1.9 & 4740   & 2 & 10   & LAS   & UKIRT/WFCAM   \\ 
NGC 3245  & 1.9 & 10210 & 2   & 10   & B2    & UKIRT/WFCAM   \\ 
NGC 4258  & 2.5 & 10635 & 2   & 10   & B2    & UKIRT/WFCAM   \\ 
NGC 4261  & 1.4 & 5453  & 3   & 10   & LAS   & UKIRT/WFCAM  \\ 
NGC 4303  & 1.7 & 6036  & 3   & 10   & LAS   & UKIRT/WFCAM   \\ 
NGC 4342  & 1.1 & 5407  & 2  & 10   & LAS   & UKIRT/WFCAM  \\ 
NGC 4374  & 2.8 &  5829 & 1.4   & 10   & B2    & UKIRT/WFCAM  \\ 
NGC 4435  & 1.5 & 5782  & 5  & 10   & LAS   & UKIRT/WFCAM   \\ 
NGC 4459  & 1.8 & 5402  & 3   & 10   & LAS   & UKIRT/WFCAM  \\ 
NGC 4473  & 1.9 & 5368  & 4   & 10   & LAS   & UKIRT/WFCAM   \\  
NGC 4486  & 1.6 & 5580    & 9  & 10   & LAS    & UKIRT/WFCAM   \\  
NGC 4486a & 1.2 & 5615  & 4   & 10   & LAS   & UKIRT/WFCAM   \\  
NGC 4486b & 1.2 & 5581  & 5  & 10   & LAS   & UKIRT/WFCAM  \\ 
NGC 4552  & 1.6 & 5612  & 4   & 10   & LAS   & UKIRT/WFCAM   \\ 
NGC 4564  & 1.5 & 5123  & 4   & 10   & LAS   & UKIRT/WFCAM   \\ 
NGC 4596  & 2.1 & 3316    & 2   & 10   & LAS   & UKIRT/WFCAM  \\ 
NGC 4621  & 2.1 & 5117    & 3   & 10   & LAS   & UKIRT/WFCAM   \\ 
NGC 4649  & 3.1 & 5258    & 5   & 10   & LAS   & UKIRT/WFCAM  \\ 
NGC 4697  & 2.9 & 7449  & 3   & 10   & B2    & UKIRT/WFCAM  \\ 
NGC 5576  & 2.6 & 5465    & 3   & 10   & LAS   & UKIRT/WFCAM  \\ 
NGC 5813  & 1.9 & 7477  & 5   & 10   & LAS   & UKIRT/WFCAM  \\ 
NGC 5845  & 2.0 & 8722  & 4  & 10   & LAS   & UKIRT/WFCAM   \\   
NGC 5846  & 1.8 & 8730    & 6   & 10   & LAS   & UKIRT/WFCAM   \\  
NGC 7052  & 2.1 & 6630  & 1.3   & 10   & B2    & UKIRT/WFCAM   \\  
UGC 9799  & 3.3 & 7377  & 2   & 10   & LAS   & UKIRT/WFCAM  \\ 
\noalign{\smallskip}     
\hline
\end{tabular}
\label{table:info} 
\end{table*}

\section{Photometric Decomposition} 
\label{sec:3}
We obtained the structural parameters of the host galaxies by performing \textbf{2D} fitting with \texttt{GALFIT3} (\citealt{tex:PH10}). \texttt{GALFIT3} constructs analytic fits to galaxy images, allowing multi-component functions representing bulge, disk, bar, point-source and sky background components.  \texttt{GALFIT3} uses the Levenberg-Marquardt technique to find the best fit.  \texttt{GALFIT3} algorithm uses this nonlinear least- square technique to  minimise the $\chi^2$ residual between the galaxy image and the model by modifying all the free parameters and accepting them when the $\chi^2$ is reduced.  The normalised $\chi^2$ is in the form:
\begin{equation}
\chi^2 = \frac{1}{n}  \sum_{1}^{nx}  \sum_{1}^{ny} \frac{(f_{d}(x,y) - f_{m}(x,y))^2}{\sigma(x,y)^2}
\label{eq:chi2}
\end{equation}
where $n$ is the number of degrees of freedom in the fit, $nx$ and $ny$ are the $x$ and $y$ image dimensions, $f_{d}(x,y)$ is equal to the image flux at pixel $(x,y)$, and $f_{m}(x,y)$ is the sum of all the functions flux at the same pixel. The term $\sigma(x,y)$ is the Poisson error at each pixel position. The $\sigma(x,y)$ value can be estimated internally by \texttt{GALFIT3} based on the gain and read-noise values found in the header of each galaxy image or provided separately as a FITS image.  Pixels contained in the mask image are not included in the $\chi^2$ calculation. 

The background is kept fixed to the value derived in Section \ref{sec:2.1}. \texttt{GALFIT3} convolves the model with the point-spread function (PSF, see~\ref{sec:2.2}) to account for the results of atmospheric seeing. The free parameters for each component are the magnitude, the scalelength ($r_{s}$)/effective radius ($r_{e}$), the concentration index $n$ for the S{\'e}rsic models (S{\'e}rsic index), the axis ratio, and the position angle. 

We modelled the radial light distribution of each galaxy using combinations of the following analytic functions: a S{\'e}rsic function (see Equation~\ref{eq:sersic}) to model elliptical galaxies, the bulge and/or the bar of lenticular and spiral galaxies; an Exponential function (Equation~\ref{eq:sersic} where n=1) to model the disk of the galaxy and a Moffat function to model/mask the central part of one elliptical galaxy. For some bar-less disk galaxies the combination of a S{\'e}rsic plus an exponential component was insufficient to model the galaxy. In these cases we modelled the galaxies with a combination of two S{\'e}rsic functions.  
The \citet{tex:S68} function is given by:
\begin{equation}
I(r) ~ = ~ I_{e} \exp \left\{ -b_{n}  \left[  \left(  \frac{r}{r_{e}} \right)^{1/n} -1 \right]  \right\} 
\label{eq:sersic}
\end{equation}
where $I_{\rm e}$ is the intensity at the effective radius ($r_{\rm e}$), $n$ is the S{\'e}rsic index and $b_{\rm n}$ is a function of n. The value of b$_{\rm n}$ can be derived from $\Gamma ~ = ~ 2 \gamma(2n,b_{n})$ and is used so that the effective radius encloses half of the total luminosity (see \citealt{tex:GD05}).  When the S{\'e}rsic index is fixed to $n= 4, 1$ or 0.5 the S{\'e}rsic profile is identical to the well known de Vaucouleurs, exponential or Gaussian profile, respectively.

\begin{figure}
\centering
\includegraphics[height=8cm,width=8.0cm]{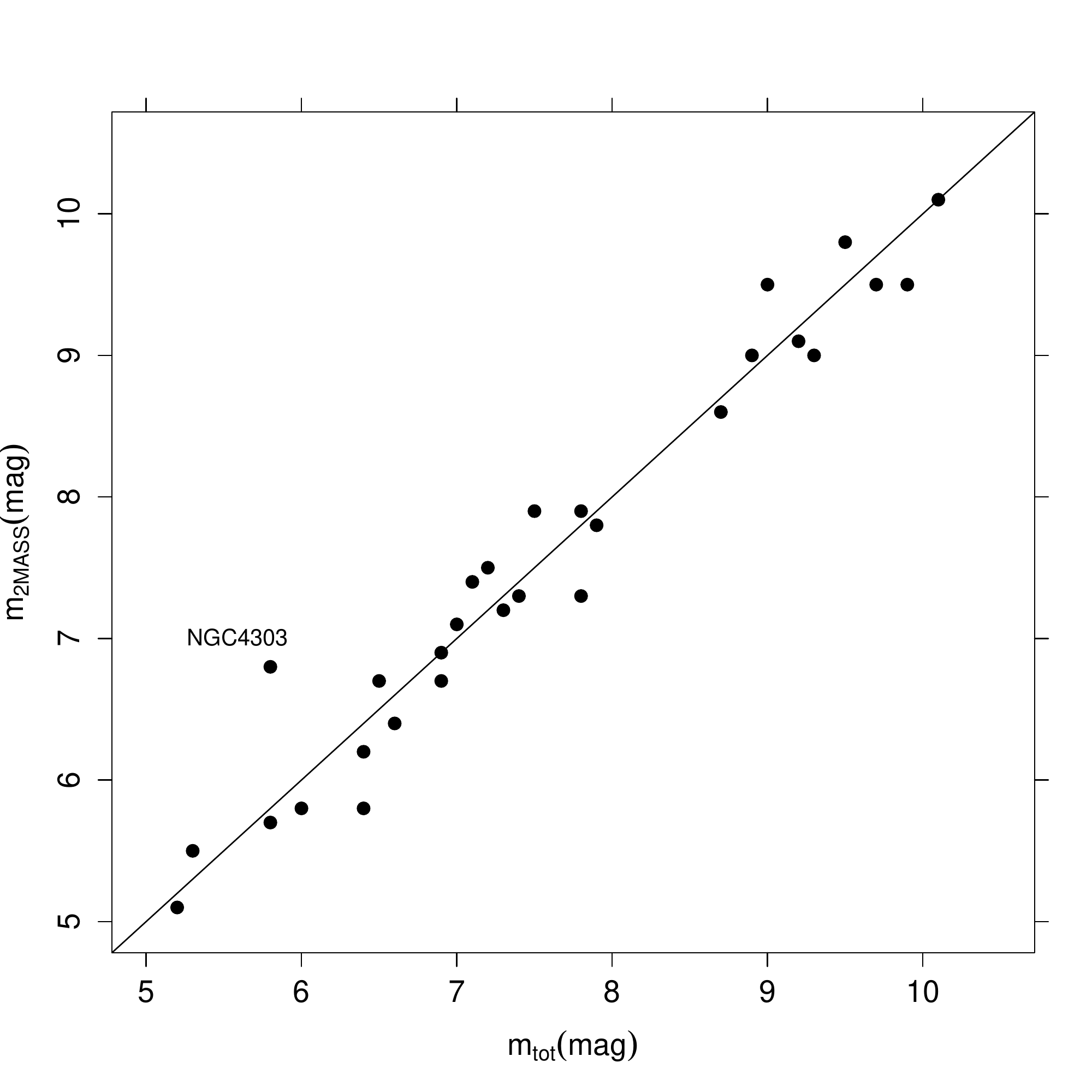}
\caption{Correlation of this study single S{\'e}rsic apparent magnitudes versus the 2MASS apparent magnitudes.}
\label{fig:2mass}
\end{figure}

For the first run of \texttt{GALFIT3}  we performed a single S{\'e}rsic model fit for all galaxies assuming that we can describe the distribution of light with a single component. The first run used initial values for the free parameters as implied by \texttt{SExtractor}. The magnitudes derived from the single S{\'e}rsic model were compared with the 2MASS magnitudes. The 2MASS magnitudes agreed with our single component magnitudes except for two galaxies, NGC4343 and NGC4486A (see Figure~\ref{fig:2mass}). For those cases where an additional component was required, either due to a poor S{\'e}rsic fit or an obvious disk in the images, a second run was conducted. The output parameters of the first run were then used as input parameters for the second run of  \texttt{GALFIT3}. The second run used two-component (i.e. S{\'e}rsic bulge - exponential disk) for disk galaxies.  The S{\'e}rsic bulge plus exponential disk model for the lenticular galaxy NGC4564 was still deficient and so a double S{\'e}rsic model was adopted for a third run. Note that the resulting S{\'e}rsic index for the disk of this galaxy was found to be 1.3 which is plausibly close to the value of the exponential function ($n=1$).

In one case, the disk galaxy NGC4459 the S{\'e}rsic plus disk fit was not sufficient to model the galaxy. After trying to model the galaxy by applying a single S{\'e}rsic model or combining extra functions we conclude that NGC4459 can been modelled better with a combination of a S{\'e}rsic function and a Moffat function. The Moffat function in  \texttt{GALFIT3} has five free parameters: the total magnitude, the FWHM, the beta powerlaw, the axis ratio and the position angle. We fixed the FWHM and the beta parameter to values that we derived through the task \texttt{IRAF}/\texttt{psfmeasure}. 
 
As described above, the outputs of the second run were used as input for the third run for galaxies whose residuals implied the existence of a bar. We applied a third run to galaxies with a bar component by using a three component model, i.e., S{\'e}rsic bulge - Exponential disk - S{\'e}rsic bar. 

An additional run was also applied to active galaxies (as indicated by X-ray or Radio observations) that have a bright nucleus in near-IR (corresponding to a point source at our resolution). These galaxies are NGC863, NGC4258, NGC4303 and NGC4435. In these cases we model the nucleus as a PSF. It is not always clear whether introducing the PSF component improves the model or not and for this reason we provide two models for some active galaxies, one with and one without the PSF.  Table \ref{table:properties} lists the main profile for each galaxy, which is used to derive the $M_{\rm bh}$--$L,n$ relations while Table \ref{table:secproperties} shows the alternative profile information.

Massive elliptical and bulge galaxies often exhibit partially depleted cores (i.e. deviations of the profile in the inner regions), this phenomena is well known and the innermost regions (1-5 per cent of the effective radius) often deviate, see for example \citet{tex:KF09} and \citet{tex:GF11}. While there is relatively little flux involved their presence can cause  particular difficulties in measuring an accurate S{\'e}rsic index. In Section \ref{sec:IG} we present different methods used by previous studies to fit the core galaxies. \texttt{GALFIT3} does not provide a function to model the depleted cores but if we ignore the existence of the core structure and model the core galaxies with a single S{\'e}rsic model we will erroneously weight the fit  to model the inner high signal-to-noise core. In these galaxies where the original profiles showed distinct departures in the inner regions we elect to mask. For these systems, indicated in Table~\ref{table:mask}, we implement a mask and re-profile and gradually increase the mask size until a stable outcome is found. Table~\ref{table:mask} Col~3 shows the final mask sizes in units of the effective radius for each of our galaxies and no obvious correlation or rule of thumb is seen. We conclude that masking is critical for the recovery of an accurate S{\'e}rsic index but actually affects the bulge luminosity relatively little as the majority of the flux lies outside the core region. Also in the case of some bright galaxies the center of the galaxy has been saturated and as a result an artificial drop of counts appears. Here too we overcome the problem of the saturated area by masking the data. The galaxies to which a core mask has been applied are NGC1068, NGC221, NGC4261, NGC4374,  NGC4473, NGC4486, NGC4486A, NGC4621, NGC4649, NGC5576, NGC5813, NGC5846 and UGC9799.

\begin{figure*}
\centering
\includegraphics[height=9cm,width=15.0cm]{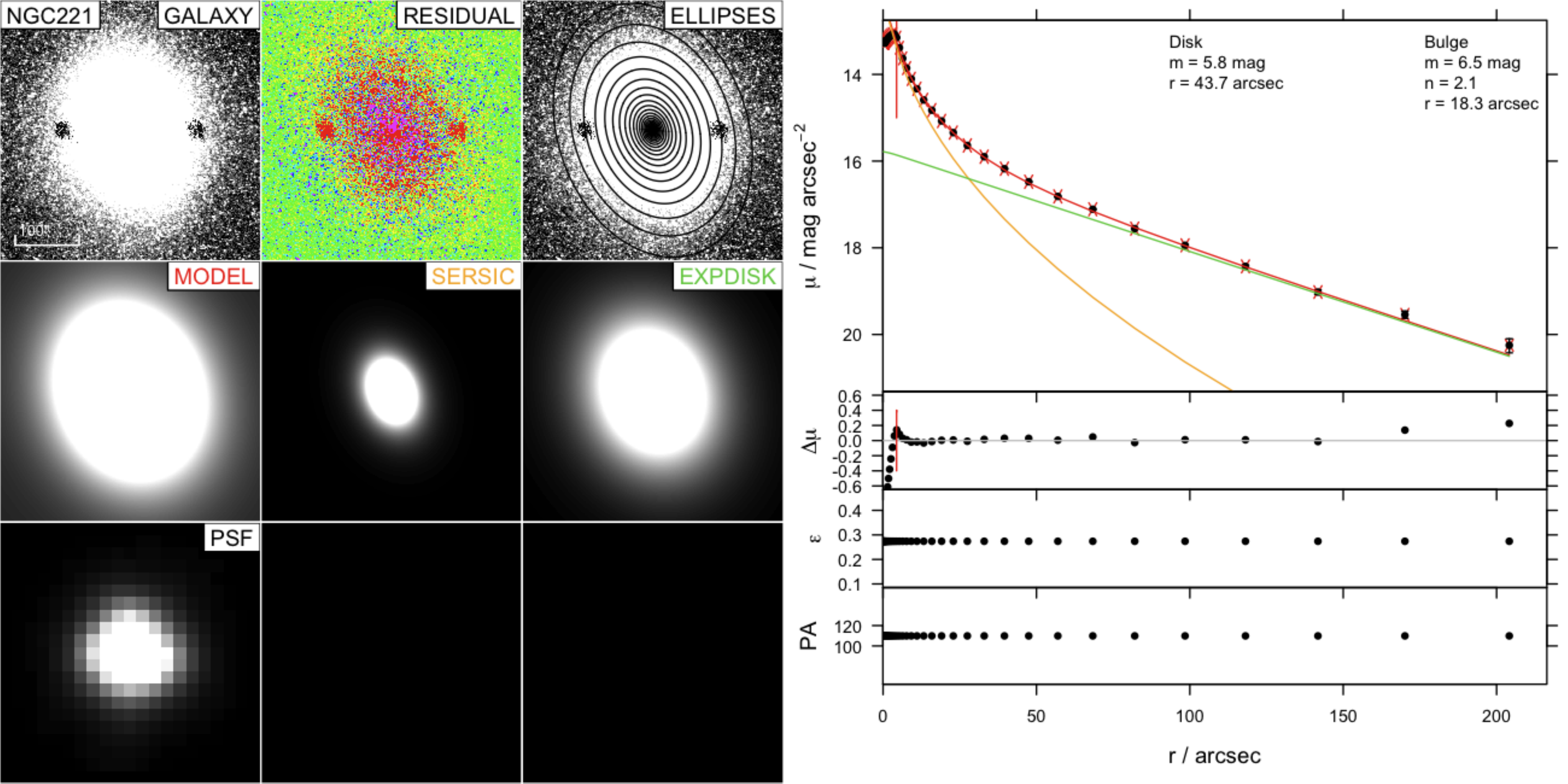}
\caption{Two-dimension decomposition for NGC221. 
Left Panels: We display the original image, the residuals, the original image with \texttt{IRAF ELLIPSE} ellipses on top, the final model, the sub-components and the PSF which \texttt{GALFIT3} used to convolve the original image. 
Right Panels from top to bottom: surface brightness $\mu$, the deviation of the galaxy surface brightness (red line) from the model surface brightness, the ellipticity $\epsilon$ and the position angle PA of the bulge. The black circles indicate the surface brightness profiles of the original image while the red line the surface brightness from the model as ellipse measure them. The rest of the lines correspond to each of the sub-components. The colour of each line corresponds to the colour of the legend found on the left panel. The black line error bars show the uncertainty of estimating the surface brightness due to uncertainty of measuring the sky value. The red line error bars indicate the uncertainty of \texttt{IRAF ELLIPSE} to measure the surface brightness. The red vertical line indicate the use of a mask that prevents \texttt{GALFIT3} from profiling the core of the galaxy. See the properties of the fitting components in Table~\ref{table:properties}.}
\label{fig:221}
\end{figure*}

\begin{figure*}
\centering
\includegraphics[height=9cm,width=15.0cm]{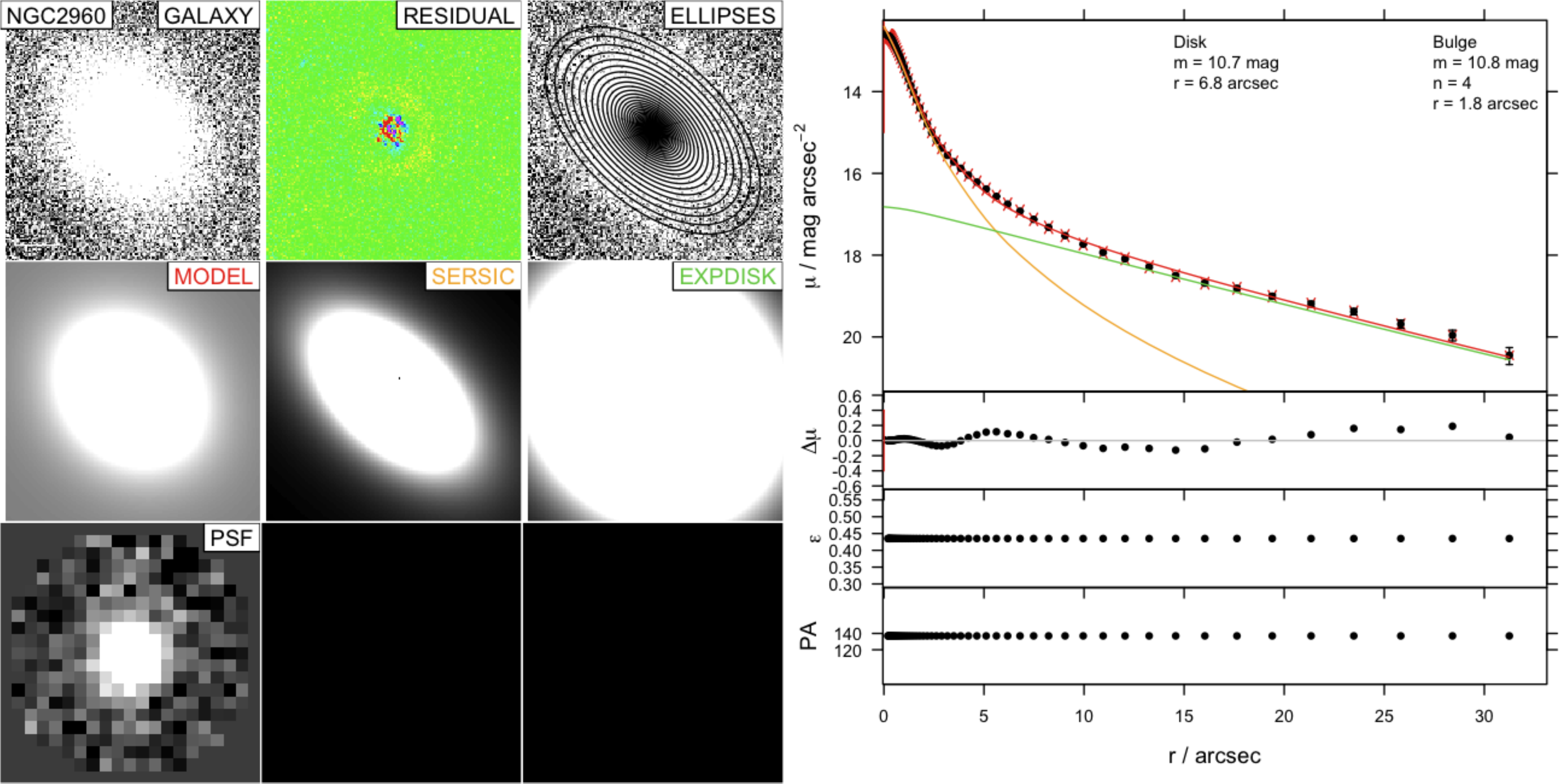}
\caption{The location of the galaxy on the image does not permit \texttt{GALFIT3} to 
profile the galaxy further than 35 arcsec. But even under this limitation \texttt{GALFIT3} output look to be realistic so we trust 
the bulge properties. The layout is as in Figure~\ref{fig:221}.}
\label{fig:2960}
\end{figure*}

\begin{figure*}
\centering
\includegraphics[height=9cm,width=15.0cm]{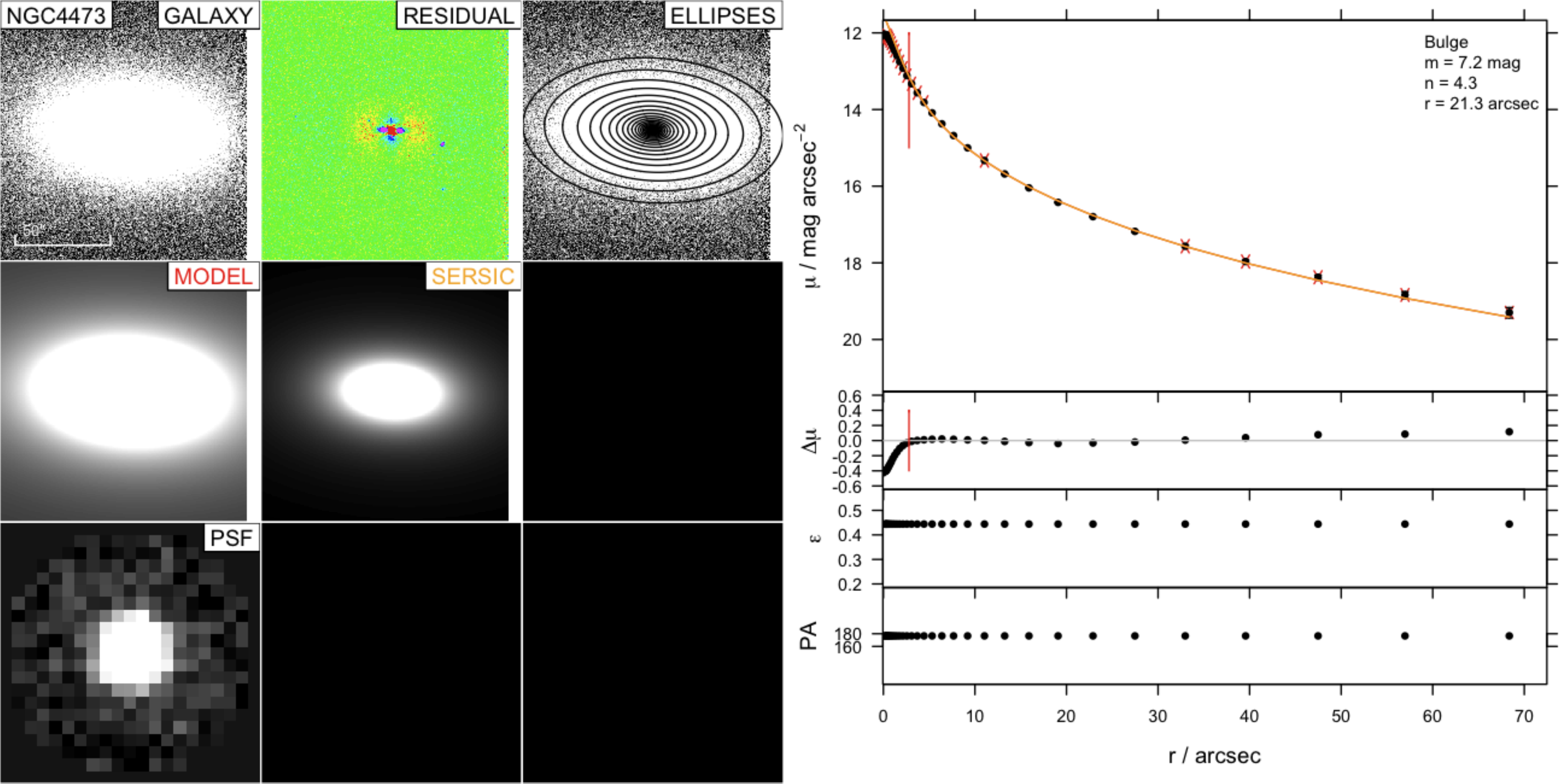}
\caption{NGC4473 has been classified as an elliptical galaxy with an inner rotating disk. 
We masked the inner 2.8 arcsec. The layout is as in Figure~\ref{fig:221}.}
\label{fig:4473}
\end{figure*}

\begin{figure*}
\centering
\includegraphics[height=9cm,width=15.0cm]{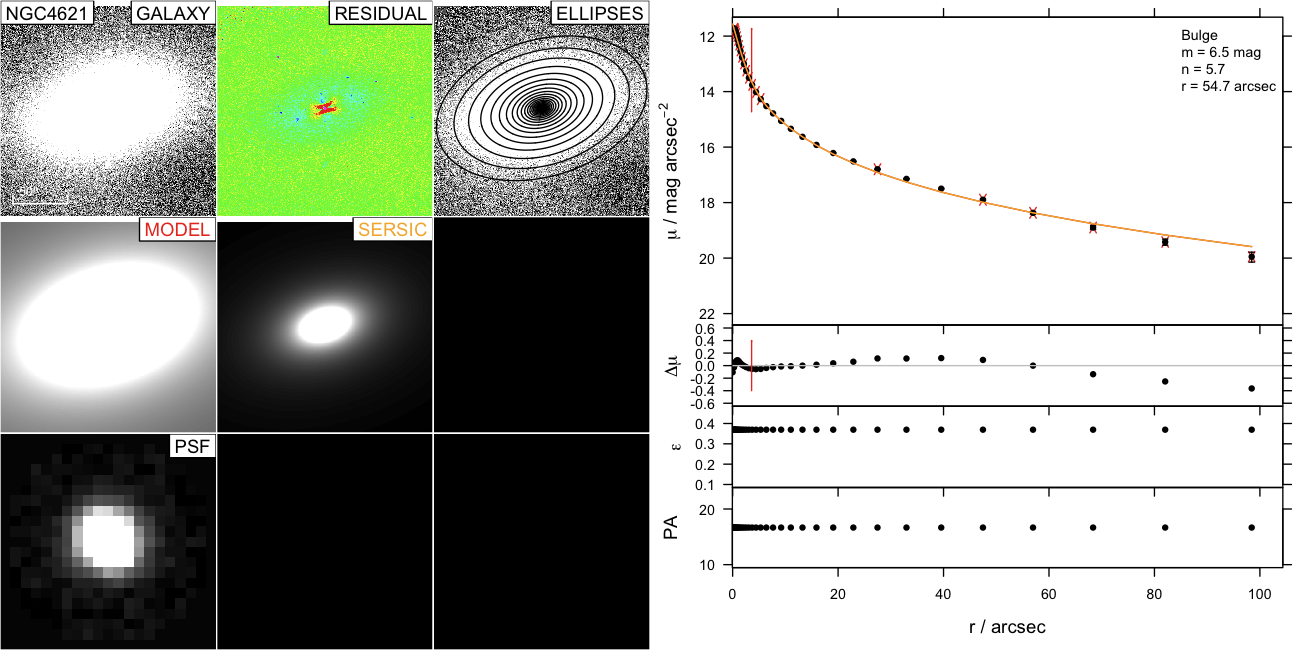}
\caption{NGC4621 has been classified as an S0 and as an elliptical galaxy from different studies. We do not find evidences of a disk but the existence of a core which we masked (3.6 arcsec).  The layout is as in Figure~\ref{fig:221}.}
\label{fig:4621}
\end{figure*}

\begin{figure*}
\centering
\includegraphics[height=9cm,width=15.0cm]{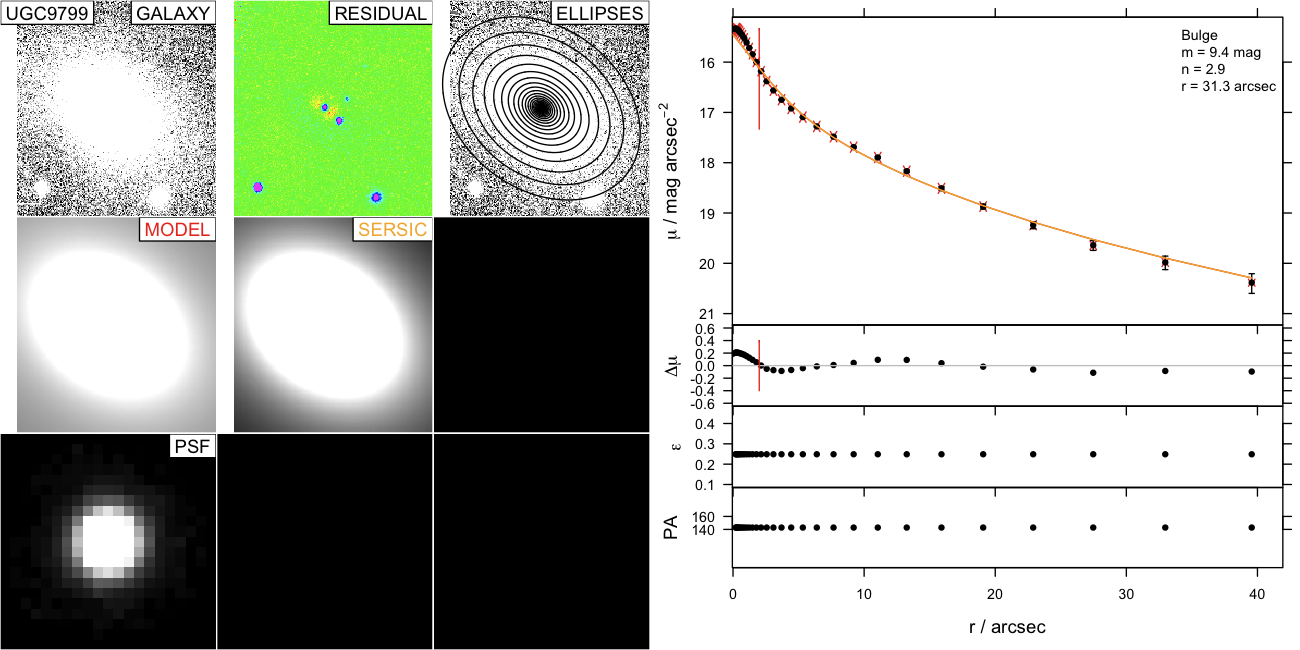}
\caption{The surface brightness profile for UGC9799, we masked the inner 2 arcsec. The galaxy is placed at the edge of the image that makes difficult to constrain the fit. The layout is as in Figure~\ref{fig:221}.}
\label{fig:9799}
\end{figure*}

When the minimisation is complete, \texttt{GALFIT3} produces  FITS files for the original image, the model, the residual and the individual images for each component. To visually examine the goodness of the fit we use \texttt{IRAF ELLIPSE} to produce a \textbf{1D} profile of both the input galaxy, the fit and the each sub-component. This process ensures that the data and models are inspected in an identical manner with the position angle and the axis ratio of the ellipses were fixed to the values that \texttt{GALFIT3} has estimated for the bulge/spheroid. We placed the resulting ellipses onto both the model image and the sub-component images of the model. With this test we can see if the azimuthally measured surface brightness along the major axis of the model is in agreement with the surface brightness profile of the galaxy, and also the contribution to the overall profile from each component. The derived surface brightness profiles are displayed in Figures \ref{fig:221}-\ref{fig:9799} together with the image of the galaxy, the residual, the model, the sub-components and the PSF as indicated. It is important to stress that these profiles are \emph{not} an output of \texttt{GALFIT3} but simply an inspection tool that process the original image and \texttt{GALFIT3} output in an identical manner. 

The apparent magnitudes produced by \texttt{GALFIT3} are converted into absolute magnitudes using the values of distance $(d)$ and extinction ($A_{\rm K}$) as listed in Tables \ref{table:first}  and \ref{table:masses} respectively. 

The faint limit in surface brightness to which our fits are deemed reliable varies for each image from 20.6 mag$~$arcsec$^{-2}$ to 22.7 mag$~$arcsec$^{-2}$. Five galaxies (NGC221, NGC2960, NGC4473, NGC4621 and UGC9799) in our sample  are located at the edge of their cutout. These galaxies cannot be profiled "down to their faint surface brightness limit". After testing the model surface brightness profile with \texttt{IRAF ELLIPSE} following the method described above for these five galaxies we decided that their derived  parameters are robust (see Figures~\ref{fig:221},\ref{fig:2960},\ref{fig:4473},\ref{fig:4621} and \ref{fig:9799}).

\begin{table} 
\caption{Mask size of the inner regions for the galaxies for which their profile deviates from the S{\'ersic} model. In the case of NGC1068 the saturated AGN has been masked while in the case of the NGC4486A the central area of the galaxy has been masked due to the existence of a bright star.}
\smallskip
\centering
\begin{tabular}{ccc}
\hline
\noalign{\smallskip}
Galaxy    &   Mask     & Mask/r$_{\rm eff}$  \\
Name      &  (arcsec) &    \\
 (1)           &   (2)          & (3)     \\        
\noalign{\smallskip}
\hline
\hline
\noalign{\smallskip}
NGC 221     &  2.75 & 15\% \\ 
NGC 1068   &  0.6   & saturated area\\ 
NGC 4261   &  1.0   & 4\% \\ 
NGC 4374   &  1.0   & 3\%\\ 
NGC 4473   &  1.4   & 6\%\\  
NGC 4486   &  3.8    & 10\% \\  
NGC 4486A&  $\sim$5   & saturated area\\ 
NGC 4621   &  1.8   & 3\%   \\ 
NGC 4649   &  2.6   & 5.6\%\\ 
NGC 5576   &  2.0   & 11\% \\ 
NGC 5813   &  1.0   & 1\%   \\ 
NGC 5846   &  2.6   & 5.6\%\\  
UGC 9799   &  1.0   & 3\%   \\ 
\noalign{\smallskip}     
\hline
\end{tabular}
\label{table:mask} 
\end{table}

\subsection{Uncertainties} 
\label{sec:unc}

As we described in Section \ref{sec:2.1} we measured the background sky value as accurately  as the image quality allows. However small variances on the mean sky value can significantly modify the output values.  To calibrate the errors due to sky uncertainty we re-run \texttt{GALFIT3} using the best-fit values but changing the mean sky level by $\pm$1 sigma (where sigma  is the uncertainty to which the mean sky level is known and listed in column 4 of Table~\ref{table:info}). These two additional runs provide us with 1$\sigma$ uncertainties for the magnitudes and the S{\'e}rsic indices required for deriving robust $M_{\rm bh}$--$L,n$ relations (see Section \ref{sec:fit}).

The dominant systematic uncertainty is the validity of the choice of function(s) to describe the light distribution of a galaxy.  Most galaxies in our sample leave residuals structures after removing the model, which may indicate smaller components that have not been modelled. In those cases where we see ambiguity  we refit with/without the ambiguous component and report for completeness the alternative  results in Table \ref{table:secproperties}.

\begin{figure}
\centering
\includegraphics[height=4cm,width=8.0cm]{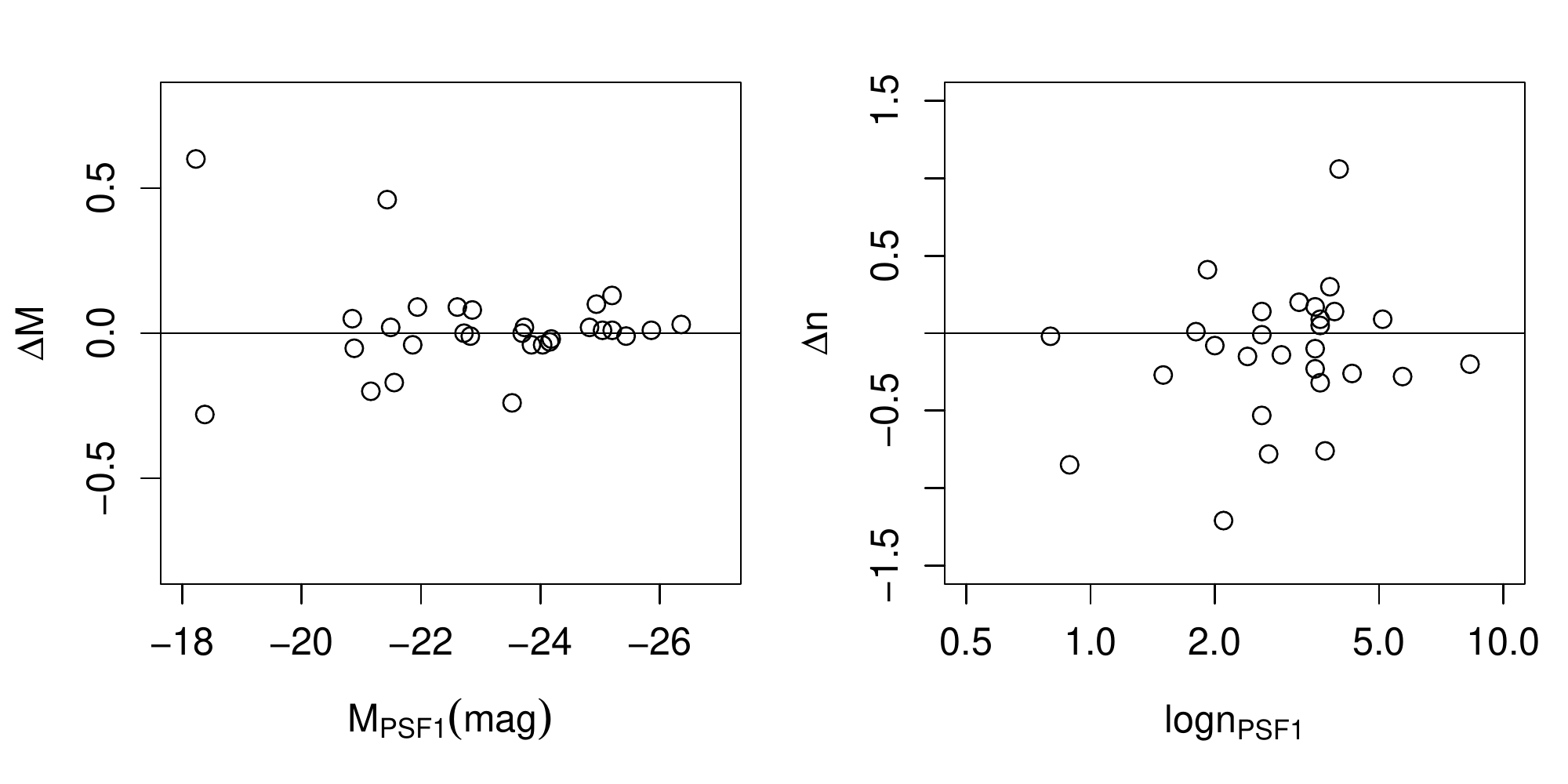}
\caption{ Left panel: Correlation of the spheroid absolute magnitudes of the best fit for two different PSFs. Right panel:  The same as the left panel for S{\'e}rsic indexes. The black line in both panels is the 1-1 relation.}
\label{fig:psfcomp}
\end{figure}

Finally we test the uncertainty introduced by the PSF. To explore this we re-run \texttt{GALFIT3} using the best-fit values but changing the PSF that  \texttt{GALFIT3} use for convolution (see Figure \ref{fig:psfcomp} and Section \ref{sec:2.2}). We find that the uncertainty introduced by the PSF ($\Delta M=0.02$, $\Delta n=0.2$) is small compared with the sky value uncertainty. 

\subsection{Notes on Modelling for Individual Galaxies}
\label{sec:IG}
\textbf{NGC221} 
(M32), M31s closest, satellite has a contaminated brightness profile due to the M31 disk. The contaminated light has a gradient from north-west, where it takes the maximum value, to south-east. Previous studies have excluded the inner 10$~$arcsec from their studies (\citealt{tex:K87,tex:G02,tex:CG02}). \citet{tex:G02} found that NGC221 can be best profiled with a bulge/disk model ($n_{\rm bulge}=1.51$) while H09 found $n_{\rm bulge}=4.00$ plus disk\footnote{H09  use the two dimensional bulge/disk decomposition program \texttt{BUDDA} and K-band images while \citet{tex:G02} use the \textbf{1D} algorithm \texttt{UNCMND} and R-band images.}.

\noindent{\textbf{NGC863}
(Mkr590) is a Seyfert 1 galaxy with a broad line spectrum.}

\noindent{\textbf{NGC1068}
(M77) is one of the most well studied barred spiral galaxies. The main bar was first observed in the near-IR by \citet{tex:SM88} while other studies have observed multiple bars (e.g. \citealt{tex:E04}). The nucleus hosts a Seyfert1-2 source with double jet observed in the radio (\citealt{tex:GB96}) making the mid- and near- IR nuclei (inner 4$~$arcsec) appear extremely red (\citealt{tex:HS98,tex:BN00}). \citet{tex:DF07} identify the existence of a pseudobulge based on the nuclear structure of the galaxy. H09 found $n_{\rm bulge}=1.51$,  $n_{\rm bar}=0.7$ plus an exponential disk.}

\noindent{\textbf{NGC2778} 
has been classified as an elliptical galaxy but GD07\footnote{GD07 derive their major-axis surface brightness profiles via fitting elliptical isophotes using \texttt{IRAF ELLIPSE}.} showed that it can be described better with a S{\'e}rsic bulge $n_{\rm bulge}=1.6$ plus a exponential disk which indicate a lenticular galaxy. This conclusion is also supported also from kinematical studies (\citealt{tex:RC99}).}

\noindent{\textbf{NGC2960}
 (Mrk1419) No previous information.}

\noindent{\textbf{NGC3245} 
Kinematical studies show circularly rotating disks (\citealt{tex:WB08}). H09 found $n_{\rm bulge}=3.9$ plus an exponential disk.}

\noindent{\textbf{NGC4258} 
(M106) A barred-spiral Seyfert galaxy that has been studied extensively over a broad band of wavelengths. The nucleus contains an edge-on warped   accretion disk with radio jet (\citealt{tex:HM97}) and strong maser emission (\citealt{tex:CH84}). Both GD07 and H09 used a bulge/disk model and found  $n_{\rm bulge}=2.04$, $n_{\rm bulge}=2.6$ respectively. \citet{tex:FD10}, found that  $n_{\rm bulge}=2.8$ in mid-infrared and presented evidence of pseudobulge characteristics.}

\noindent{\textbf{NGC4261} 
is the main elliptical galaxy in a group of 33 galaxies located behind the Virgo cluster (\citealt{tex:HD83}). The galaxy corresponds to the radio source (3C 270) which contains a pair of highly symmetric kpc-scale jets (\citealt{tex:BD85}) and an edge-on nuclear disk of gas and dust in the optical (\citealt{tex:FF96a,tex:JW00,tex:FC06a}). In X-ray it is possible that the galaxy hosts a heavily obscured AGN (\citealt{tex:ZB05}). The isophotal analysis shows boxy isophotes at large radii both in the optical and near-IR bands (\citealt{tex:BF94,tex:FF96a,tex:QB00}). GD07 derived a S{\'e}rsic index fit of $n=7.3$.}

\noindent{\textbf{NGC4303}     		     
 (M61) is a double-barred AGN galaxy (\citealt{tex:E04}) with bright star-forming regions in a ring around the nucleus and in the spiral arms in the UV (\citealt{tex:CV97}) and also visible in near-IR images (\citealt{tex:MH01}). \citet{tex:WJ09} performed a two dimensional bulge-disk-bar decomposition using \texttt{GALFIT3} on H-band images and derived $n=1.55,1.0,0.55$. \citet{tex:FD10} showed that the spheroid component of the galaxy is a pseudobulge with $n_{\rm bulge}=1.7$.}

\noindent{\textbf{NGC4342} 
is an S0 elongated galaxy with disky isophotes (\citealt{tex:BF94}). \citet{tex:BJ98} discovered the existence of a nuclear disk in addition to the outer disk through analysis of the rotation curve in HST/WFPC2 U-, V- and I-band imaging.}

\noindent{\textbf{NGC4374} 
(M84, 3C272.1) is a radio elliptical galaxy. GD07 found $n_{\rm bulge}=4.97$ while \citet[hereafter KF09]{tex:KF09} found $n_{\rm bulge}= 7.9$ after excluding the inner 4.2$~$arcsec\footnote{KF09 derive their major-axis surface brightness profiles via fitting elliptical isophotes allowing a boxy/disky parameter to vary (\texttt{MIDAS/ESO}).}.}

\noindent{\textbf{NGC4435}
has boxy isophotes in the inner region and disky isophotes at large radii (\citealt{tex:FC06a}).}

\noindent{\textbf{NGC4459}
 is a Virgo lenticular galaxy. H09 modelled NGC4459 with two components (bulge/disk $n_{\rm bulge}=2.5$). KF09 classified NGC4459 as an elliptical galaxy and estimated a S{\'e}rsic index of $n=3.16$.}
        
\noindent{\textbf{NGC4473} 
is an elliptical galaxy with primarily disky isophotes (\citealt{tex:BD88,tex:BF94}). Its unusual surface profile brightness has aroused plenty of interest, e.g.,  \citet{tex:BG96,tex:FC06a,tex:KC06}. The distribution of light in the inner part of the galaxy is dominated by a counter-rotating stellar disk (\citealt{tex:CE07}). KF09 modelled the galaxy by excluding the inner 23$~$arcsec and measured a  S{\'e}rsic index of $n=4.0$ in agreement with the \textbf{2D} profiling of H09. GD07 found a S{\'e}rsic index of $n=2.73$.} 

\noindent{\textbf{NGC4486}
 (M87) is the second brightest elliptical Virgo galaxy and classified as a cD due to extra halo light originating from the cluster. \citet{tex:FC06a} and GD07 agreed on their S{\'e}rsic index measurements 6.1 (using S{\'e}rsic-core model) and 6.8 respectively while KF09 profiled with the extra halo and found $n=11.86$. A big discrepancy in the S{\'e}rsic values appears when we transferred to two dimensional profiling where both \citet{tex:D01} and  H09 found a significantly low S{\'e}rsic index $n=3.0$.}

\noindent{\textbf{NGC4486A}
has extra light from the nuclear disk visible in almost all wavelengths. In combination with a very bright star next to the centre it is difficult to provide a reliable fit at the centre.  \citet{tex:FC06a,tex:KF09,tex:H09} found S{\'e}rsic indices values 2.7, 2.04 and 4.2 respectively.}

\noindent{\textbf{NGC4486B} 
is a low-luminosity dwarf galaxy with extra light near the centre and characterised by a double core (\citealt{tex:LT96}) which flattens the profile close to the nucleus. Due to its orbit around NGC4486, a fraction of the light from NGC4486b is actually halo light from its companion and consequently affects the surface brightness of NGC4486b. KF09 estimated a S{\'e}rsic index of $n=2.2$ after modelling the additional light from NGC4486 and masking the inner 1.3$~$arcsec. \citet{tex:SG06} found $n=2.7$.}

\noindent{\textbf{NGC4552} 
(M89) \citet{tex:CC93} first profiled this S0 galaxy  with a single S{\'e}rsic index of $n=13.87$ excluding the inner 2$~$arcsec while KF09 applied the same model and found $n=9.1$. \citet{tex:FC06a} applied a core-S{\'e}rsic model and found $n_{\rm g}=7.1$. Finally in two-dimension modelling \citet{tex:D01} and H09 used a bulge/disk model ($n_{\rm bulge}=4.2,4.6$ ) yielding a much lower $n$ value.}
	
\noindent{\textbf{NGC4564} 
has been classified as elliptical in RC3 while \citet{tex:TE04} classified it as an S0 galaxy. The existence of the disk is also verified by GD07 and KF09 with a bulge S{\'e}rsic index value of 3.15 and 4.69 respectively. Also H09 found $n_{\rm bulge}=3.6$ plus an exponential disk.}
	
\noindent{\textbf{NGC4596}
bar properties have been studied by \citet{tex:K90} and \citet{tex:E05}. H09 found $n_{\rm bulge}=3.3$, $n_{\rm bar}=1.0$ plus an exponential disk.}

\noindent{\textbf{NGC4621}
(M59) \citet{tex:FC06a} and KF09 profiled NGC4621 with a single S{\'e}rsic index and found $n_{\rm bulge}=6.8$ and 5.36 respectively. While H09 applied a bulge/disk model and found $n_{\rm bulge}=4.1$.}

\noindent{\textbf{NGC4649}
\citet{tex:FC06a} found $n_{\rm bulge}=$4.7 plus core while GD07 and H09 fitted single S{\'e}rsic model and derived $n_{\rm bulge}=6.04$ and 3.4 respectively.}

\noindent{\textbf{NGC4697}
GD07 measured $n=4.0$ while H09 found $n_{\rm bulge}=3.0$ plus an exponential disk.}

\noindent{\textbf{NGC5576}
is an elliptical galaxy which \citet{tex:TE04} profiled with both a single S{\'e}rsic model  ($n=4.47$) and a S{\'e}rsic plus core model ($n=4.89$).}

\noindent{\textbf{NGC5813}	
is an elliptical galaxy with a core (\citealt{tex:LA95,tex:RB01}). H09 profiled with a bulge/disk model $n_{\rm bulge}=4.6$.}

\noindent{\textbf{NGC5845}
 is a dwarf elliptical galaxy in the group of NGC5846 with an unusually high central surface brightness. It hosts a nuclear disk and dust disk that extends to 15$~$arcsec on the major axis (\citealt{tex:QB00,tex:MT05}). GD07 and \citet{tex:TE04} performed \textbf{1D} profiling of the V-band images and derived a S{\'e}rsic index of $n=3.22$ (single S{\'e}rsic fit) and $n=2.88$ (S{\'e}rsic fit plus core fit) respectively. H09 found $n=4.6$.}

\noindent{\textbf{NGC5846}
 is the main galaxy in an isolated group of 250 galaxies (\citealt{tex:MT05}) with a compact radio core at the optical centre (\citealt{tex:MH92}).  \citet{tex:FB97} found slightly boxy isophotes in central regions while \citet{tex:RB01} argued that due to strong dust filaments it is not possible to study the nucleus and have reliable information about the isophotal shape. H09 modelled the galaxy light profile with a single S{\'e}rsic and found $n=3.1$.}

\noindent{\textbf{NGC7052} 
is an isolated radio galaxy with an edge-on dust ring along the major axis (\citealt{tex:NM90}) and nuclear disk of gas/dust (\citealt{tex:JC96}) which affects the boxy/disky measurements of isophotes in the outer regions of the galaxy  (\citealt{tex:QB00}). GD07 and H09 modelled the galaxy light profile and estimated S{\'e}rsic indices of  $n=4.55$ and $=3.4$ respectively.}

\noindent{\textbf{UGC9799} 
(3C 317) is the central elliptical galaxy (cD morphological class; cf. \citealt{tex:SG07}) of the cooling flow cluster, Abell 2052.  It features a central X-ray excess and is host to an active galactic nucleus (AGN) evident as a bright, steep-spectrum radio source (e.g. \citealt{tex:ZS93}) with a compact optical counterpart (\citealt{tex:RG03,tex:SG07}).  \citet{tex:SG07} and \citet{tex:DC07} also identify a distinct outer halo to UGC 9799 (consistent with its classification as a cD morphological type) in Jacobus Kapteyn Telescope $R$-band and HST NICMOS $H$-band (F160W) imaging respectively.  After masking the inner $\sim$1-2 arcsec (to exclude the central AGN) and modelling the outer halo with an exponential light distribution, these authors recovered S{\'e}rsic index fits of $n=1.2$ and 2.3 to its major-axis surface brightness profile respectively\footnote{Both \citet{tex:SG07} and\citet{tex:DC07} derived their major-axis surface brightness profiles via fitting of `perfect' (i.e., non-boxy/disky) elliptical isophotes to their galaxy images using \texttt{IRAF ELLIPSE}.}.}

\begin{figure*}
\centering
\includegraphics[height=9cm,width=15.0cm]{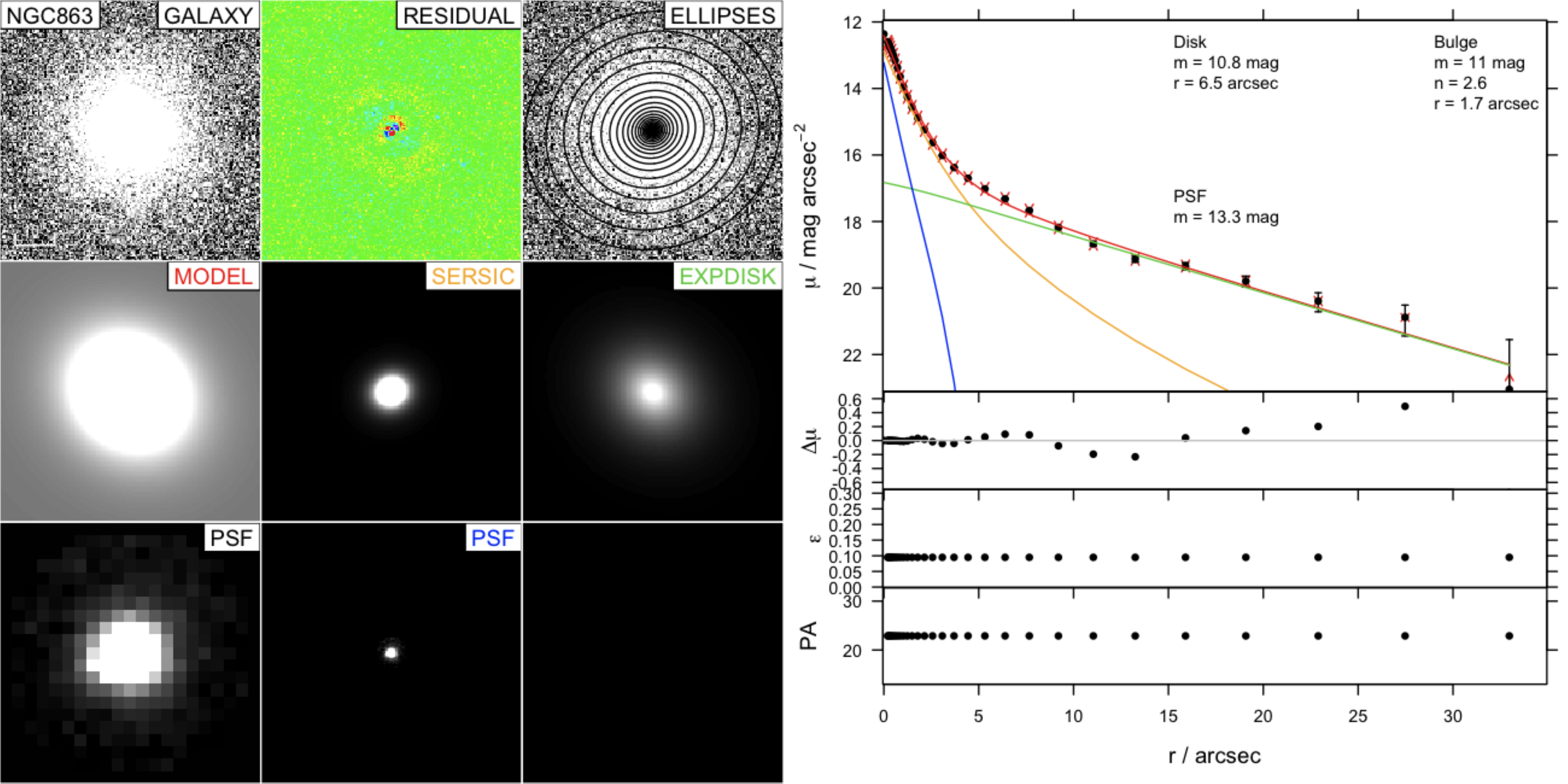}
\caption{NGC863 has been classified as a spiral galaxy. The layout is as in Figure~\ref{fig:221}.}
\label{fig:863}
\end{figure*}

\begin{figure*}
\centering
\includegraphics[height=9cm,width=15.0cm]{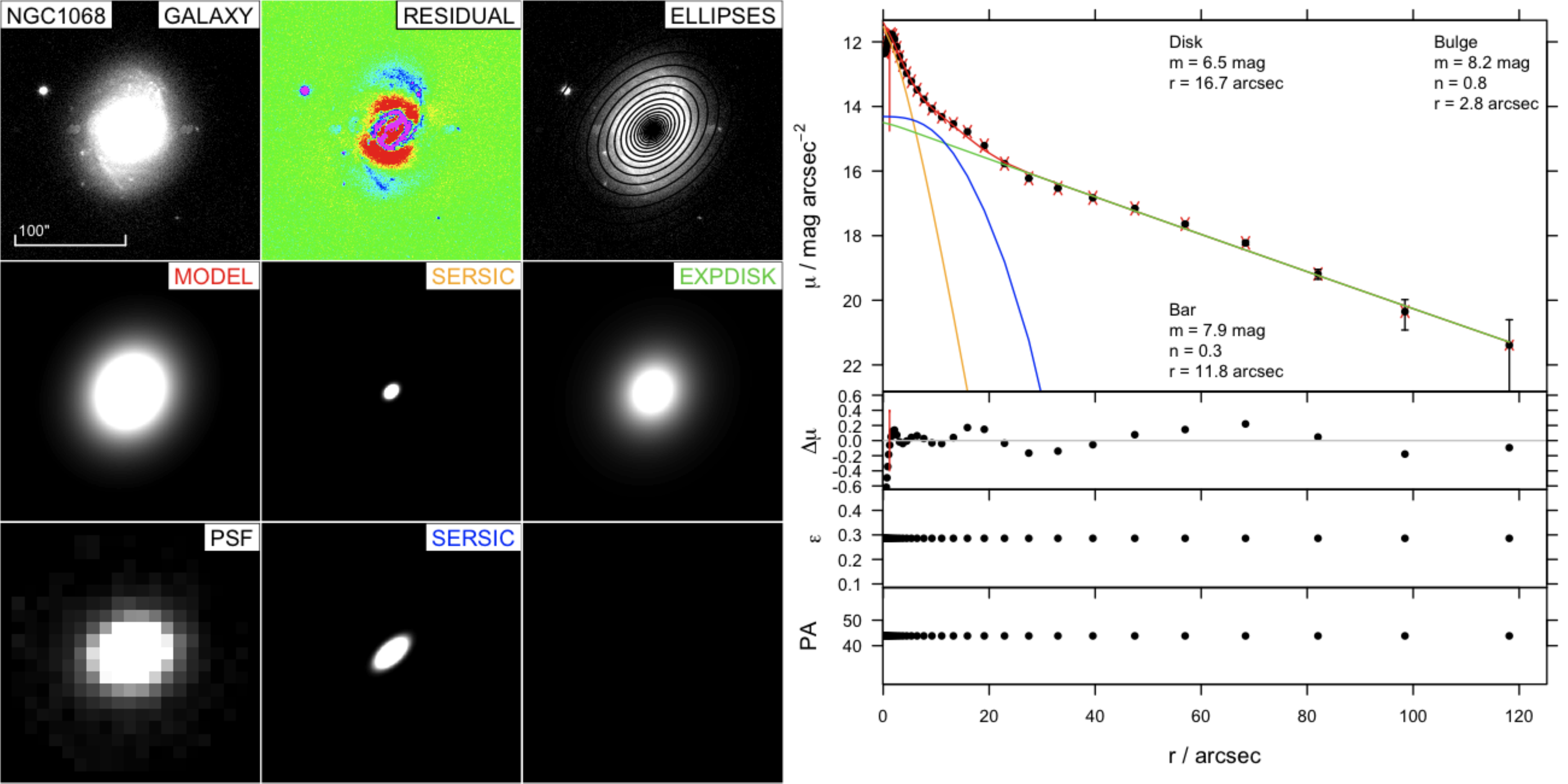}
\caption{At the nuclei of the galaxy arise an artificial drop of counts. To prevent 
\texttt{GALFIT3} profiling this inner part we masked the nuclei with a box of 
1.5 arcsec$^2$. The layout is as in Figure~\ref{fig:221}.}
\label{fig:1068}
\end{figure*}

\begin{figure*}
\centering
\includegraphics[height=9cm,width=15.0cm]{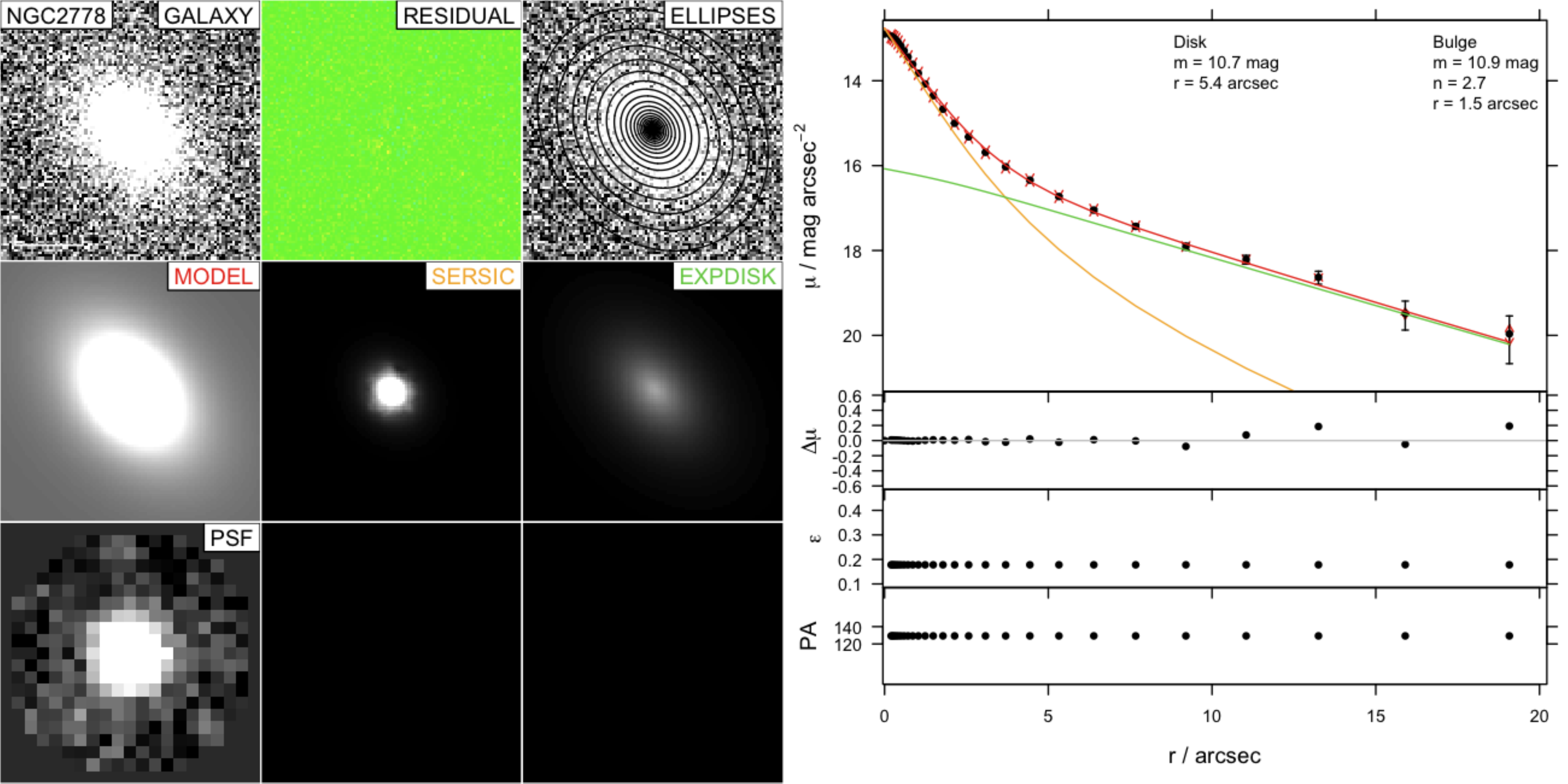}
\caption{The surface brightness profile for NGC2778. The layout is as in Figure~\ref{fig:221}.}
\label{fig:2778}
\end{figure*}

\begin{figure*}
\centering
\includegraphics[height=9cm,width=15.0cm]{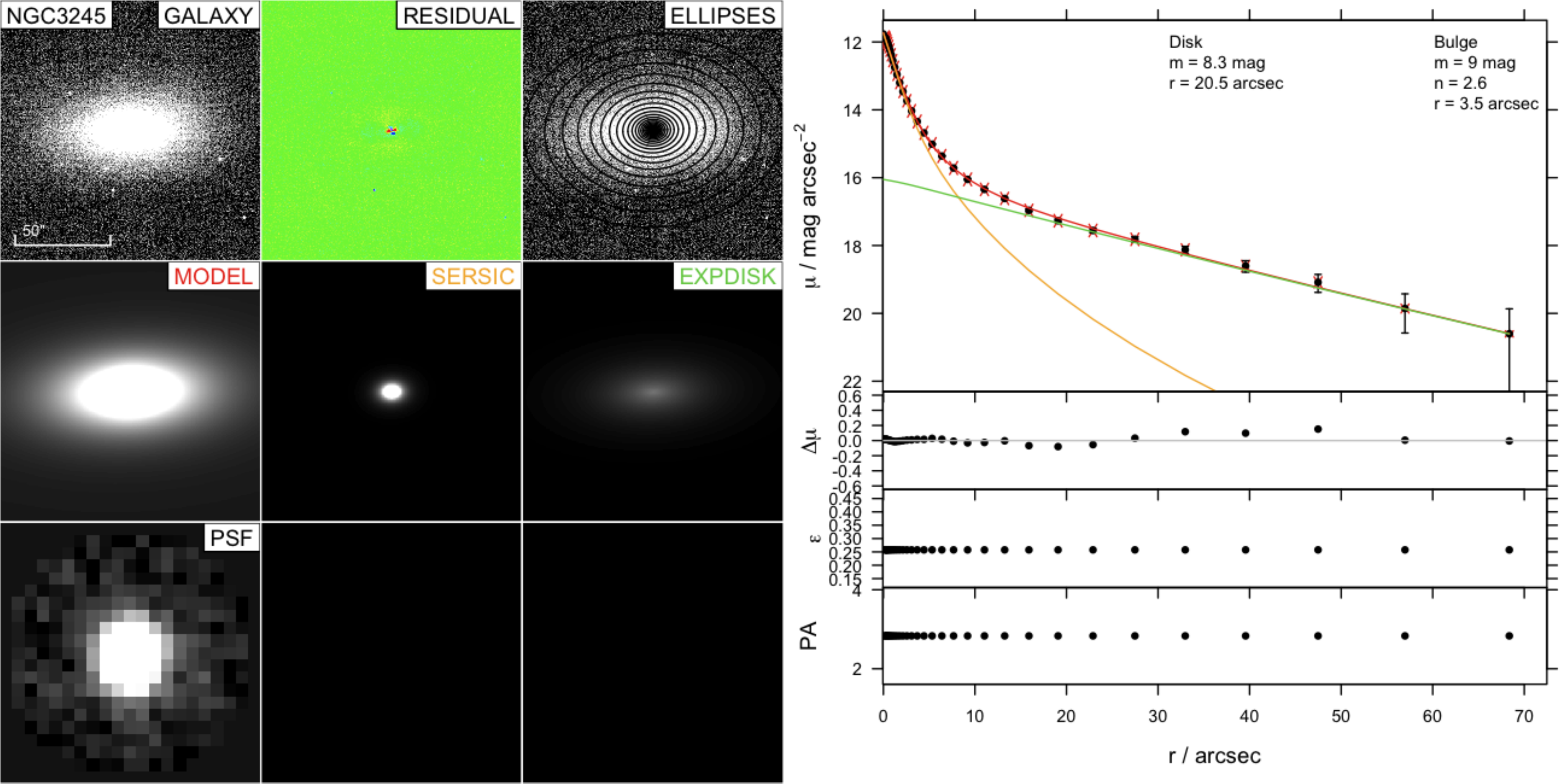}
\caption{\texttt{GALFIT3} results are in agreement with the kinematics studies that show existence  
of disk. The layout is as in figure~\ref{fig:221}.}
\label{fig:3245}
\end{figure*}

\begin{figure*}
\centering
\subfigure{\includegraphics[height=9cm,width=15.0cm]{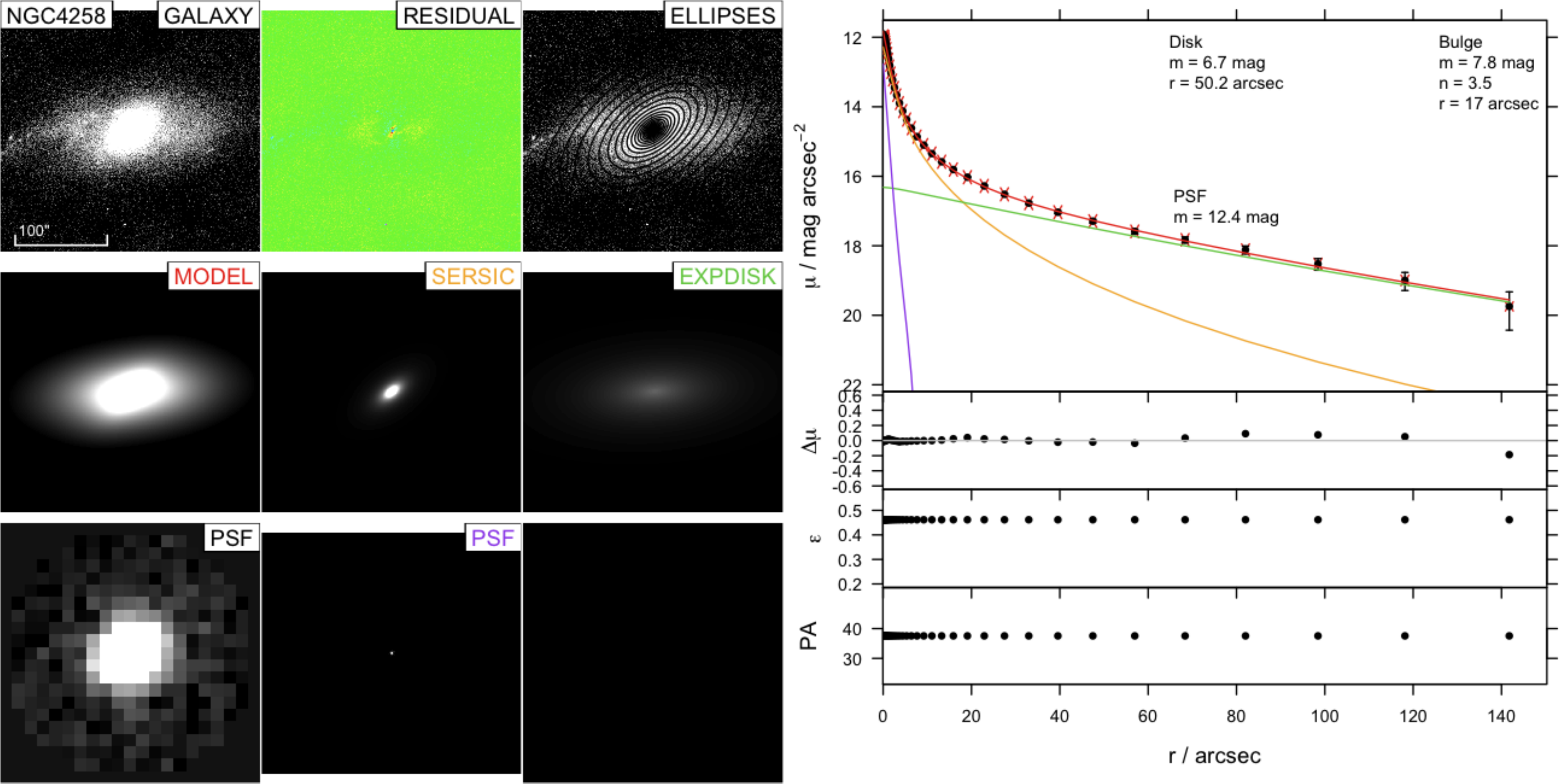}} 
\subfigure{\includegraphics[height=9cm,width=15.0cm]{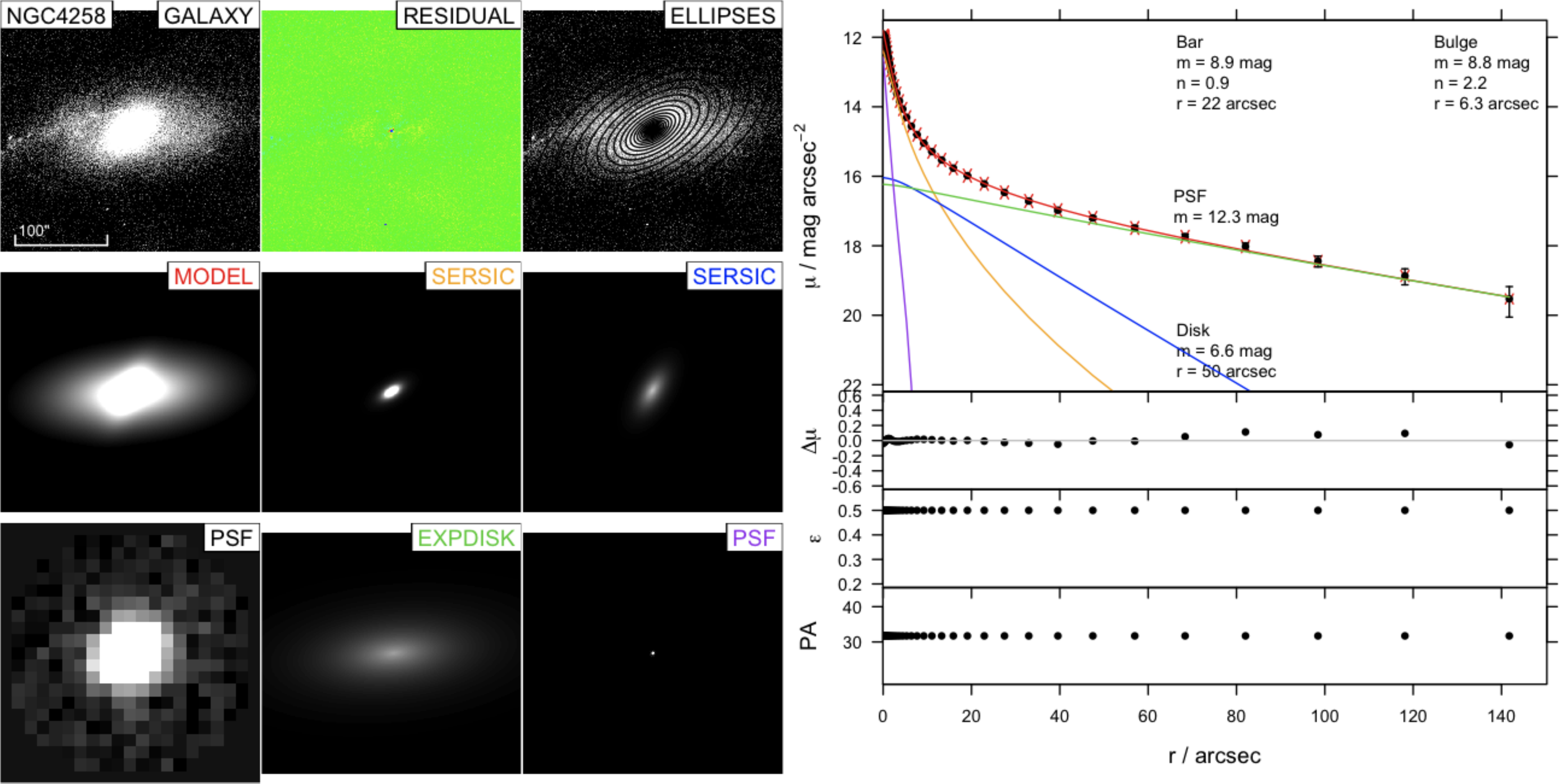}} 
\caption{The surface brightness profile for NGC4258 for two different fits: (a) S{\'e}rsic 
+ exponential model + PSF nuclei and (b) double S{\'e}rsic + exponential model + PSF nuclei. 
The layout is as in Figure~\ref{fig:221}.}
\label{fig:4258}
\end{figure*}

\begin{figure*}
\centering
\includegraphics[height=9cm,width=15.0cm]{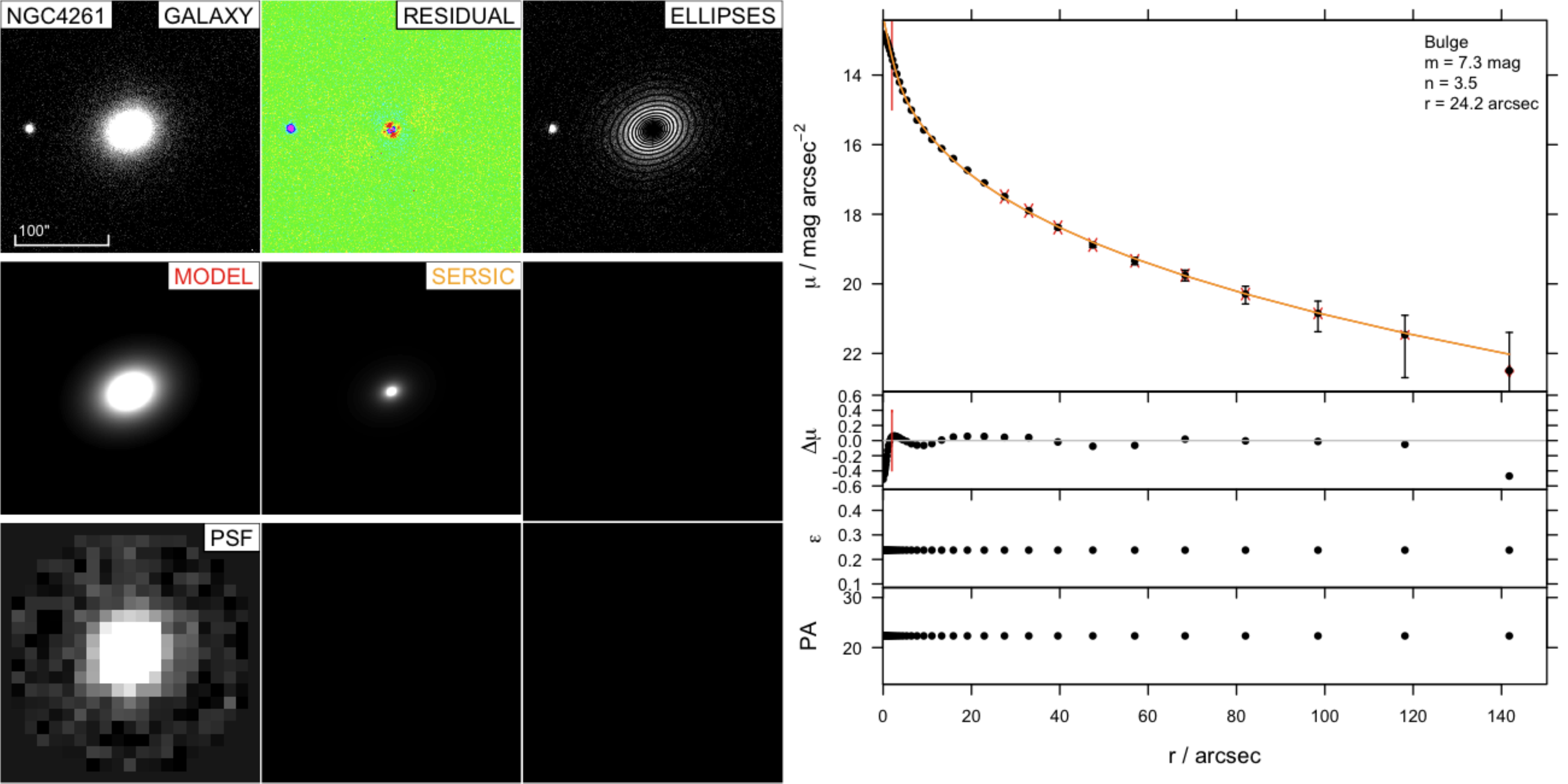}
\caption{For the NGC4261 galaxy \texttt{GALFIT3} generate low S{\'e}rsic index model compared to the one dimensional pre-existing models (see Subsection~\ref{sec:IG}). We masked the inner 2 arcsec. The layout is as in Figure~\ref{fig:221}.}
\label{fig:4261}
\end{figure*}

\begin{figure*}
\centering
\includegraphics[height=9cm,width=15.0cm]{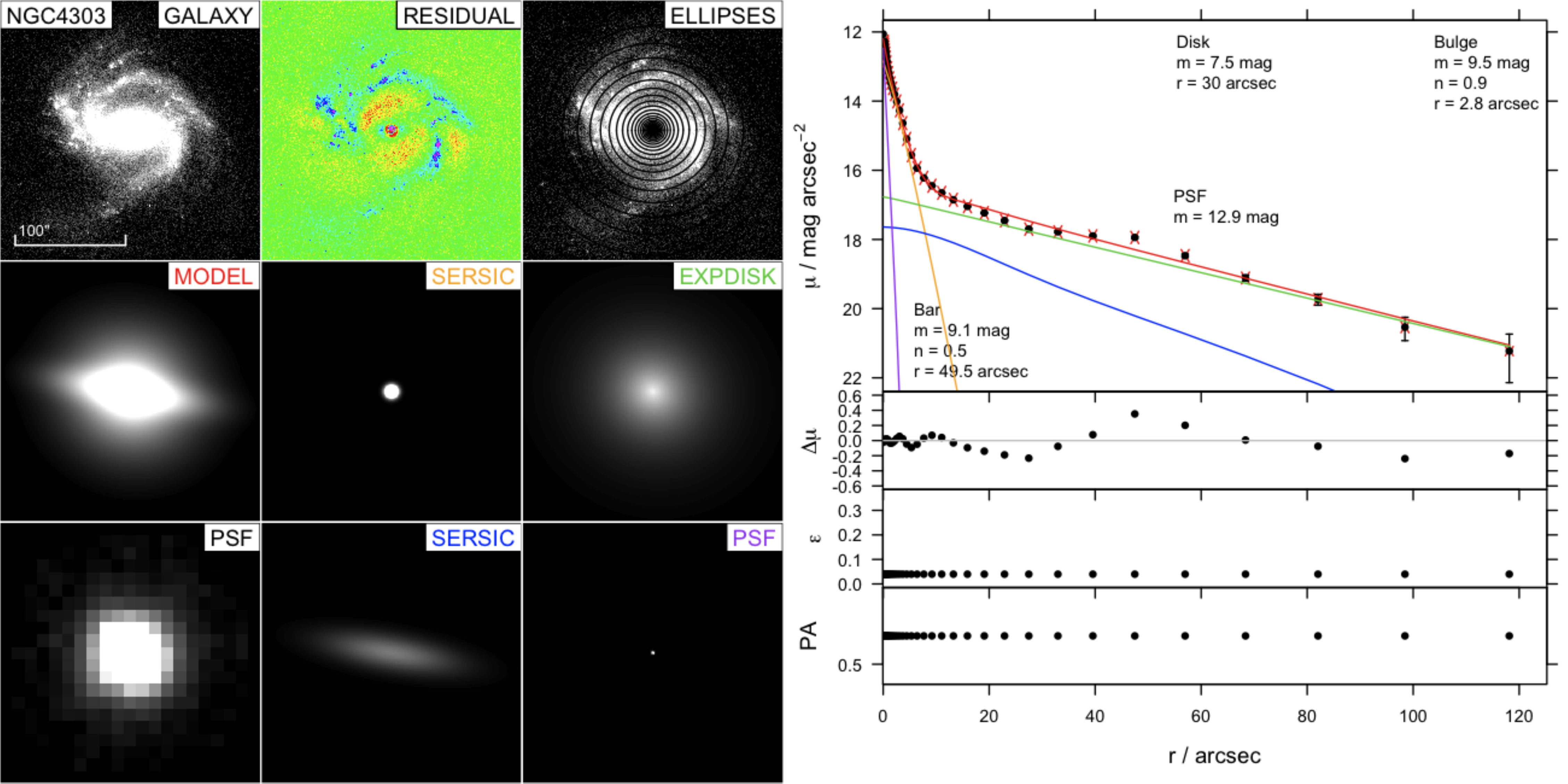} 
\caption{The surface brightness profile for NGC4303. The layout is as in Figure~\ref{fig:221}.}
\label{fig:4303}
\end{figure*}

\begin{figure*}
\centering
\includegraphics[height=9cm,width=15.0cm]{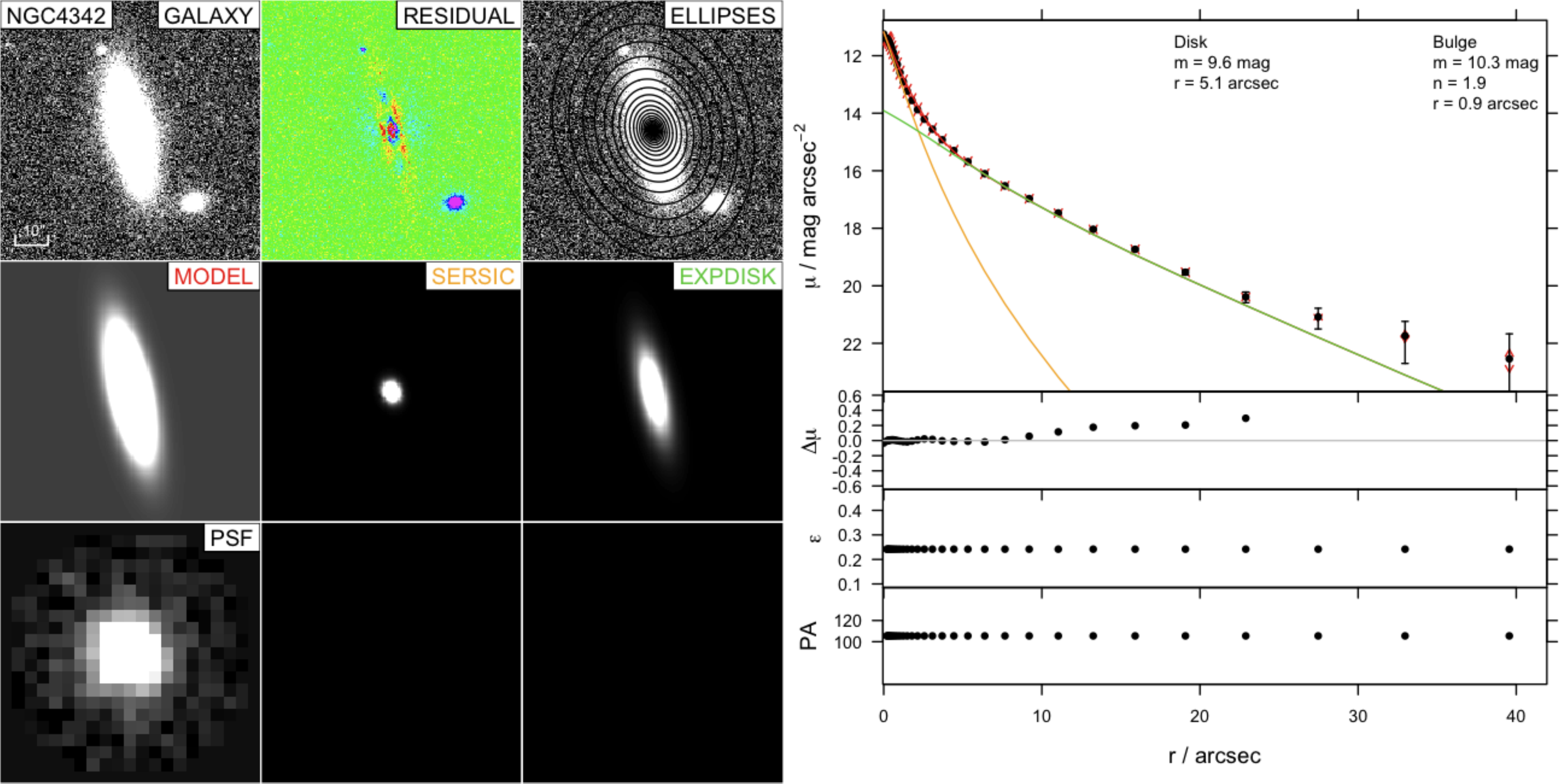}
\caption{The layout is as in Figure~\ref{fig:221}. \texttt{IRAF ELLIPSE} outer ellipses are contaminated with
extra light from the satellites as a result the model surface brightness mismatch  \texttt{IRAF ELLIPSE}  points.  }
\label{fig:4342}
\end{figure*}

\begin{figure*}
\centering
\subfigure{\includegraphics[height=9cm,width=15.0cm]{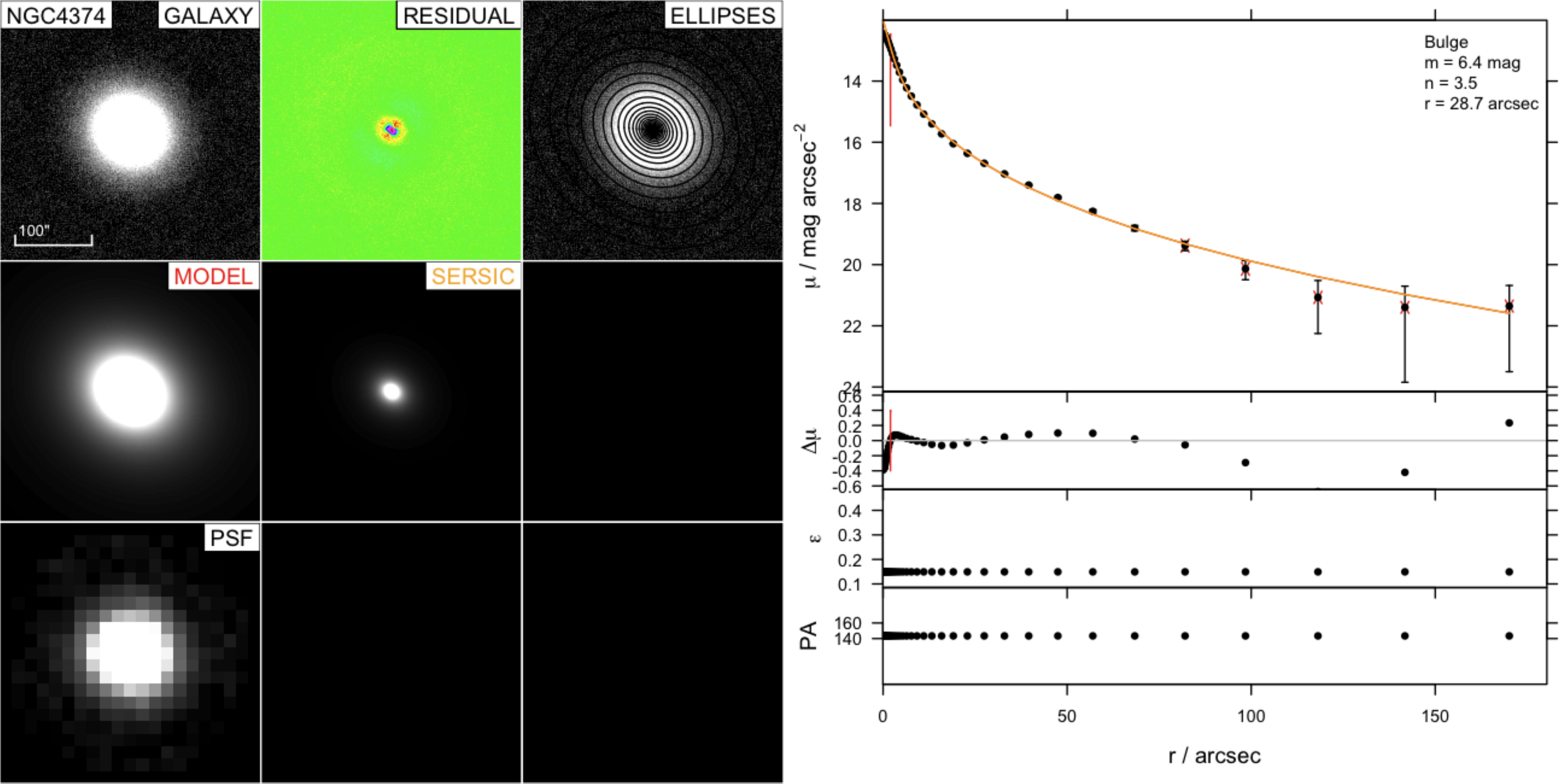}} 
\subfigure{\includegraphics[height=9cm,width=15.0cm]{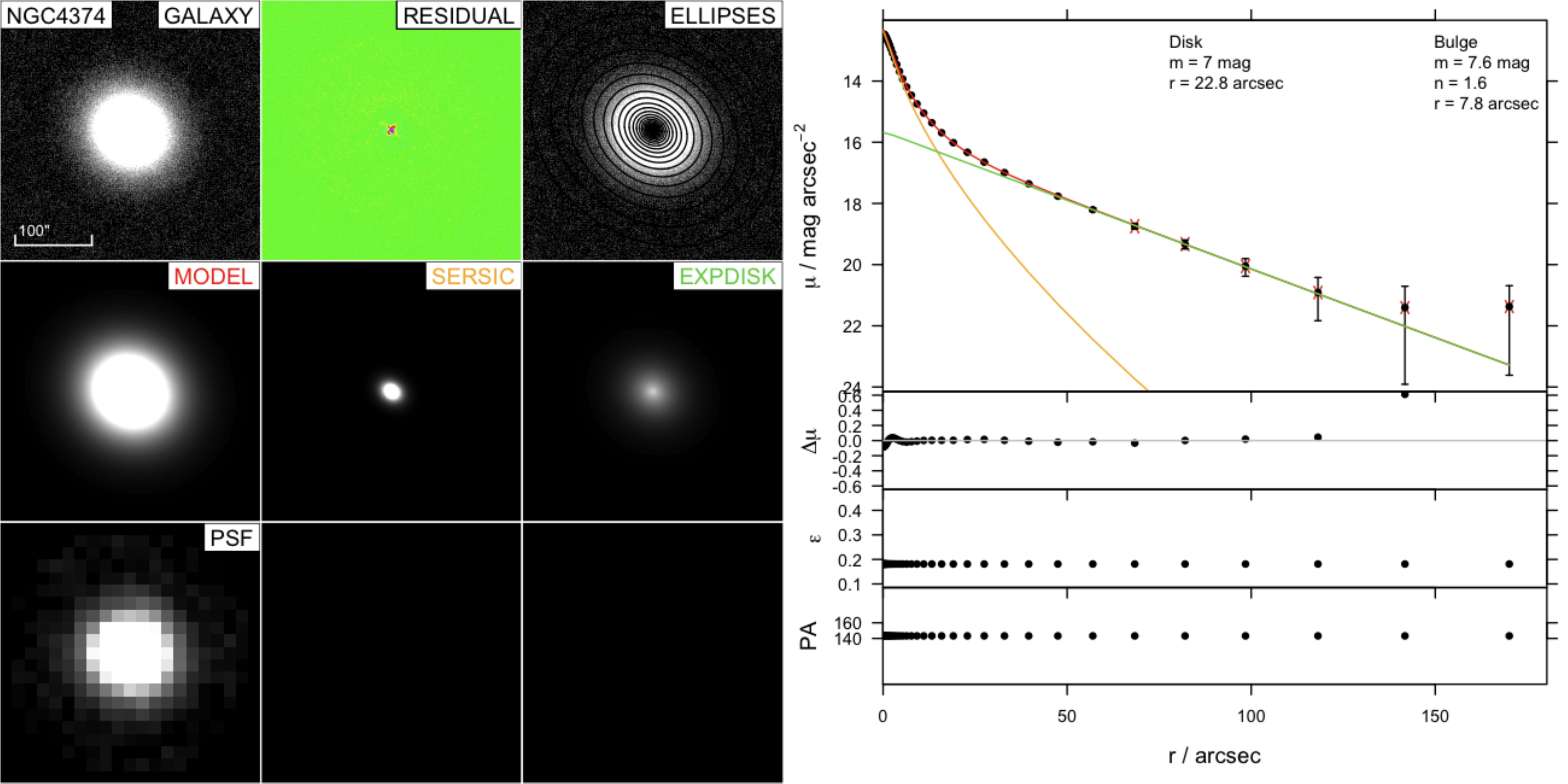}} 
\caption{NGC4374 has been classified as an high S{\'e}rsic index elliptical galaxy from 
previous studies. We believe that NGC4374 can also be profiled with a two components model. 
The layout is as in Figure~\ref{fig:221}.}
\label{fig:4374}
\end{figure*}

\begin{figure*}
\centering
\subfigure{\includegraphics[height=9cm,width=15.0cm]{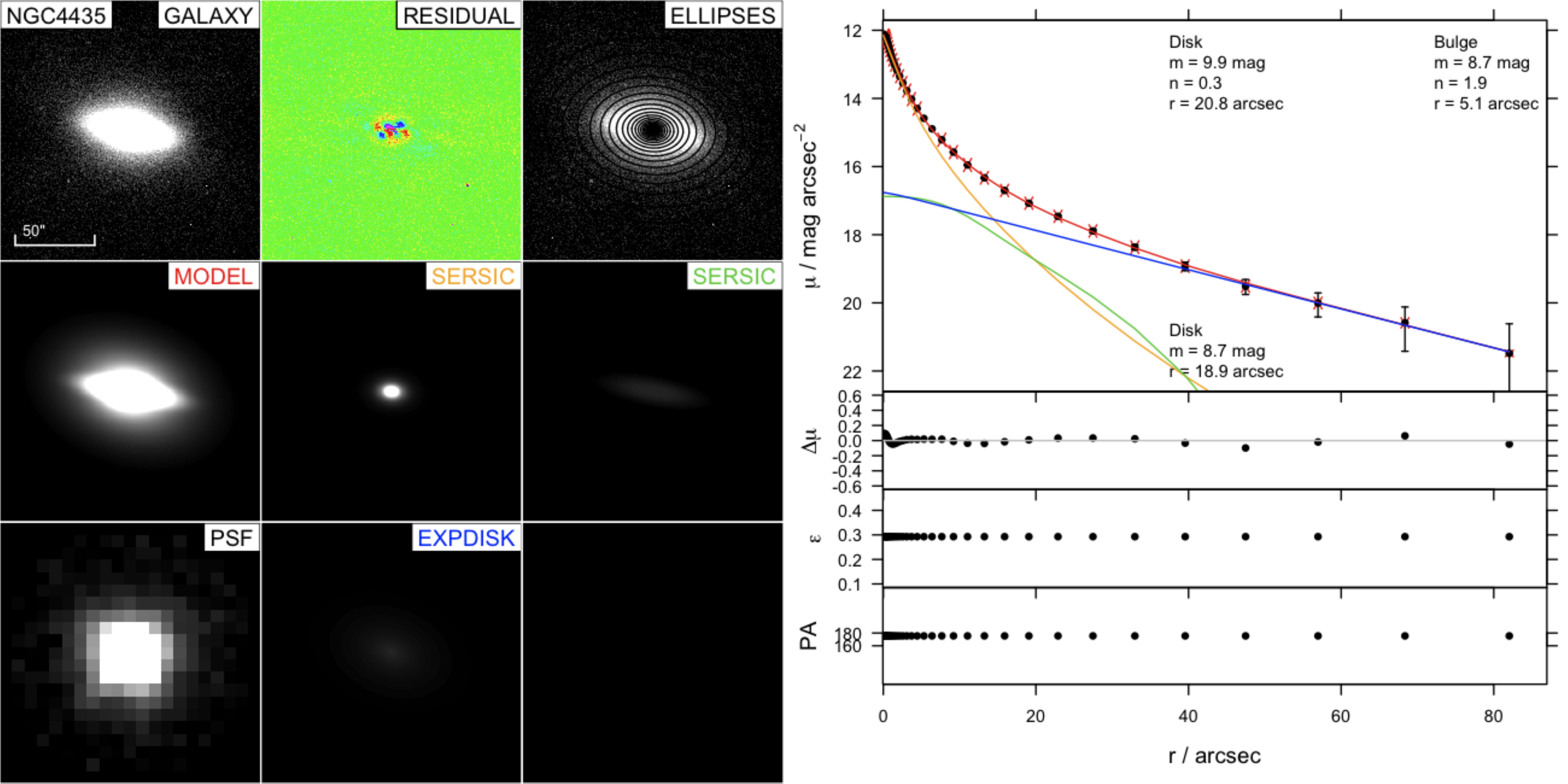}} 
\subfigure{\includegraphics[height=9cm,width=15.0cm]{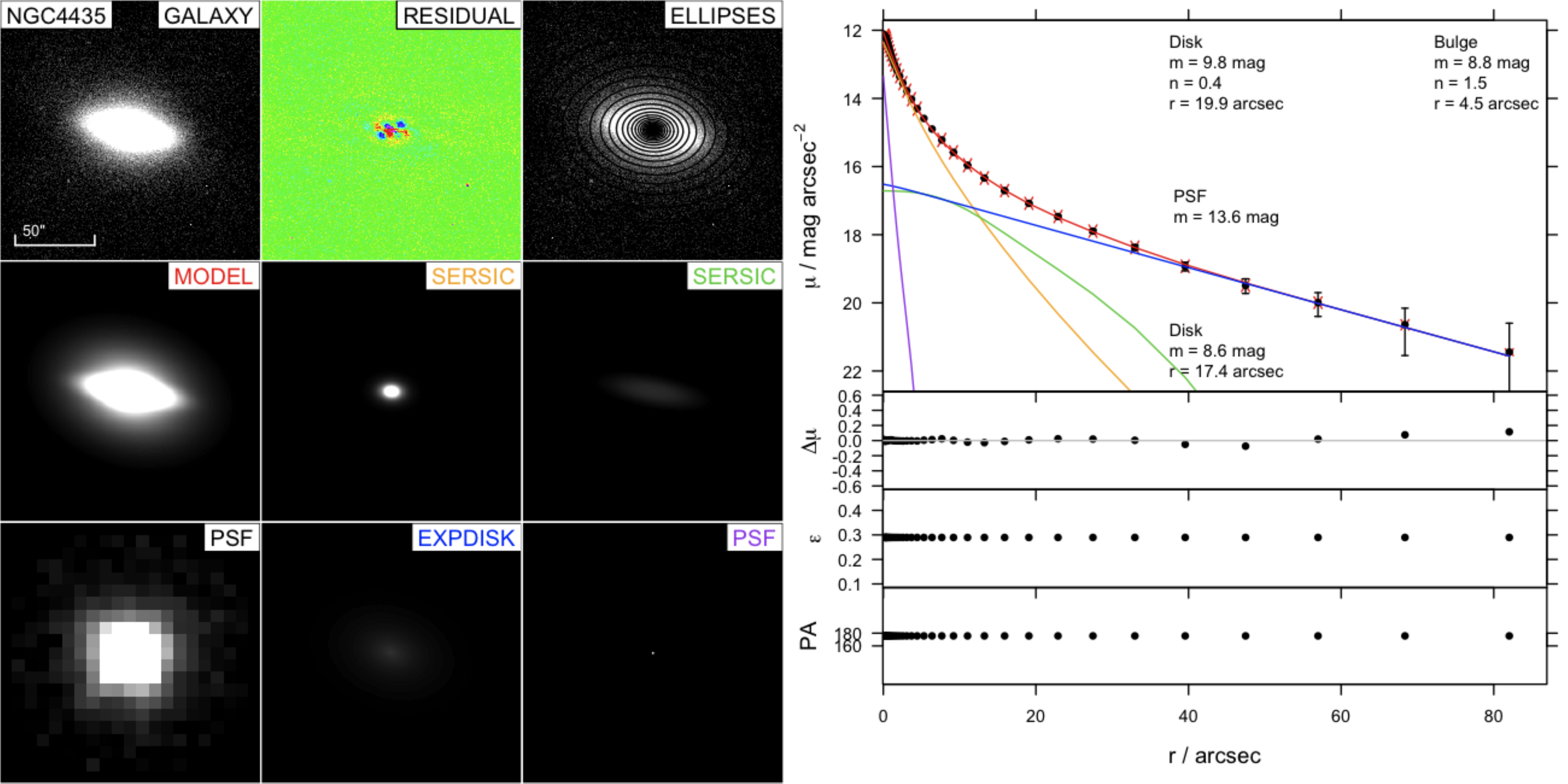}} 
\caption{The surface brightness profile for NGC4435 for different fits: (a) double 
S{\'e}rsic + exponential model and (b) double S{\'e}rsic + exponential model + PSF nuclei.
The layout is as in Figure~\ref{fig:221}.}
\label{fig:4435}
\end{figure*}

\begin{figure*}
\centering
\includegraphics[height=9cm,width=15.0cm]{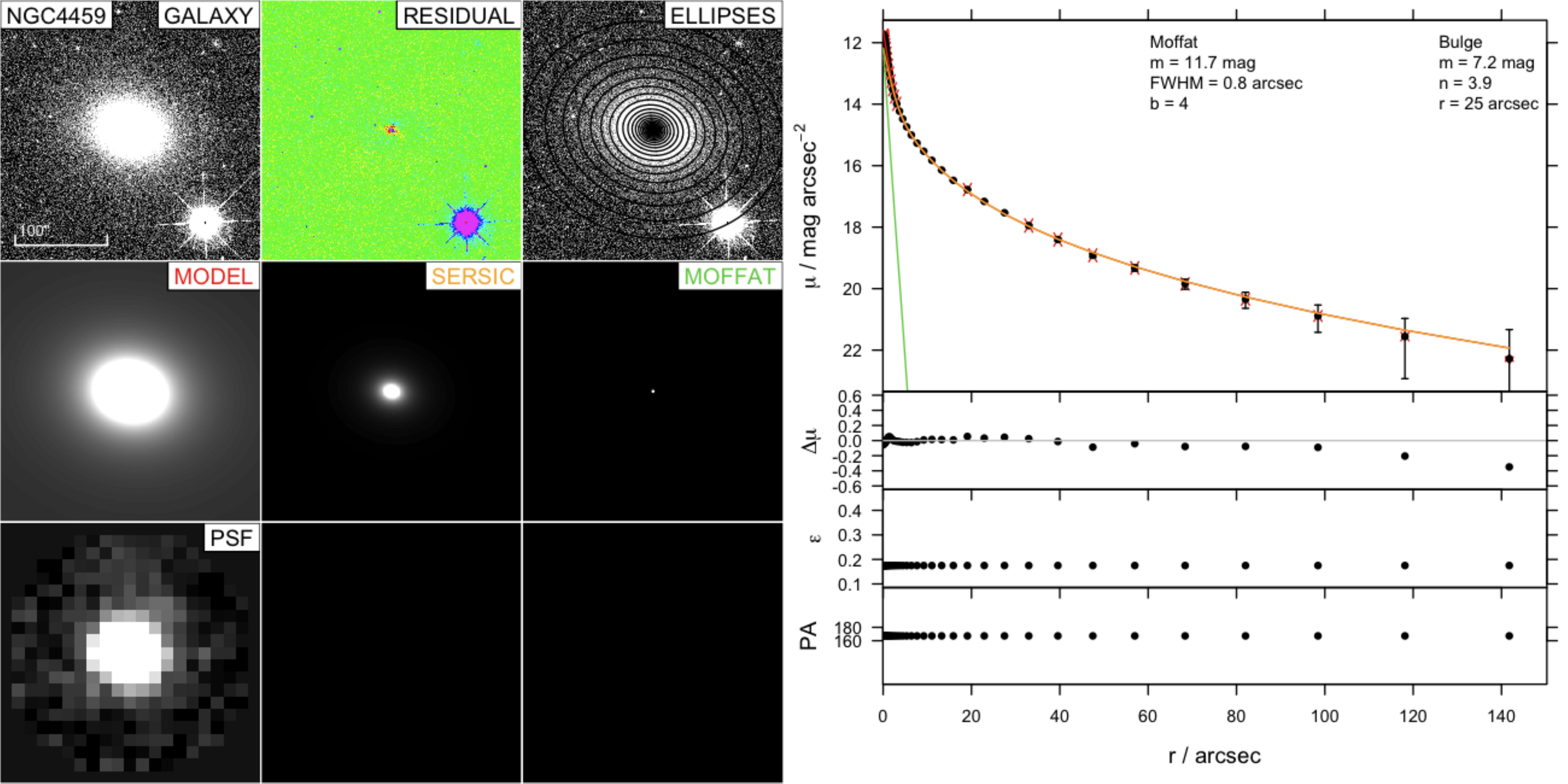}
\caption{NGC4459 has been classified as an S0 or as an elliptical galaxy from different 
studies. Our profiles show that is an elliptical galaxy with an extra light in the 
center that need to be mask or profiled. We found that the function that describes better 
this core light is the Moffat function. For more details about how we apply the Moffat 
function see Section~\ref{sec:3}. The layout is as in Figure~\ref{fig:221}.}
\label{fig:4459}
\end{figure*}

\begin{figure*}
\centering
\includegraphics[height=9cm,width=15.0cm]{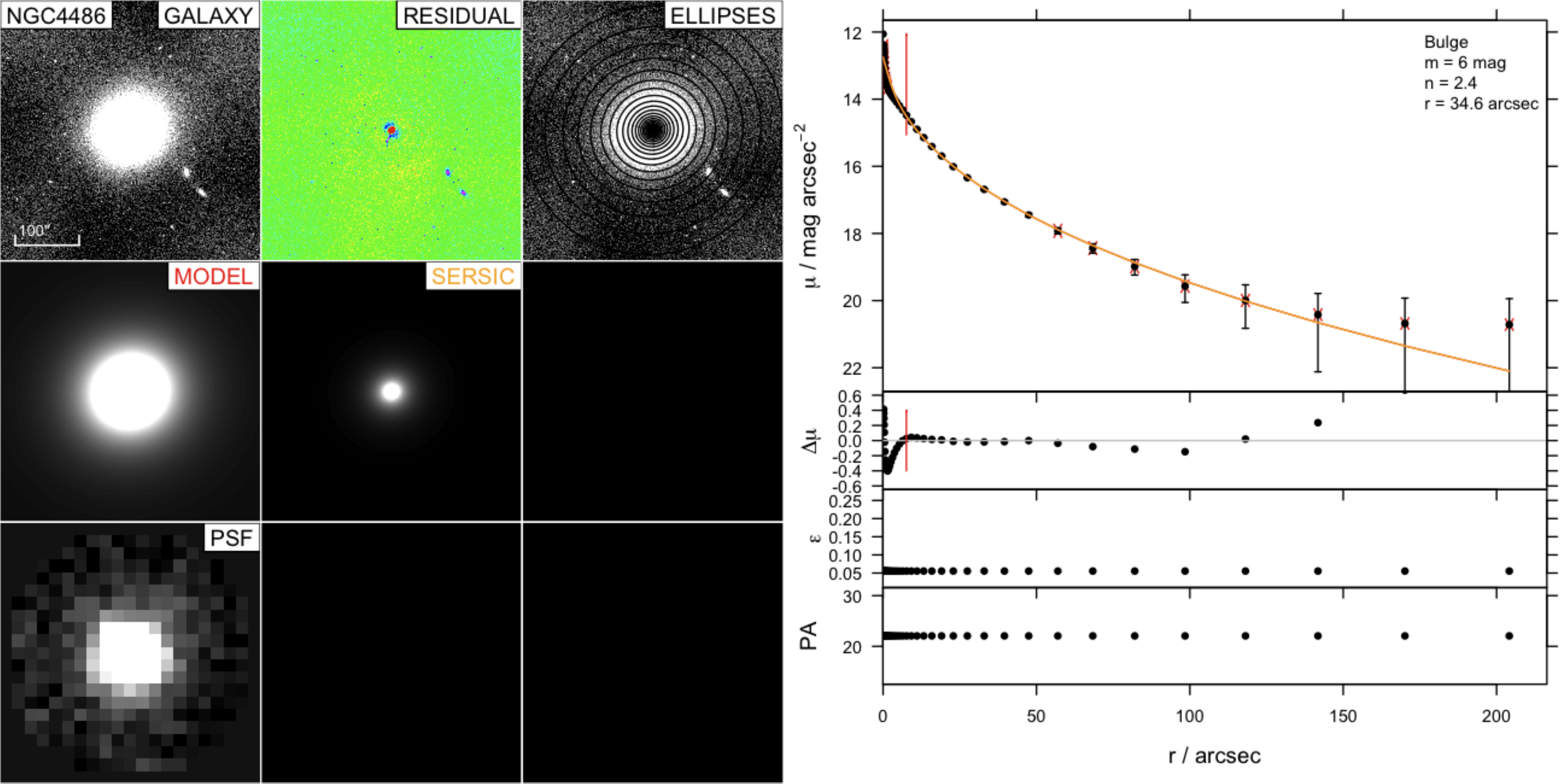}
\caption{M87 is an elliptical galaxy with a recognisable jet in most of the wavelengths. We masked the inner 7.5 arcsec. Notice the S{\'e}rsic index variance between one dimensional and two dimensional profiles (see Section~\ref{sec:IG}). 
The layout is as in Figure~\ref{fig:221}.}
\label{fig:4486}
\end{figure*}

\begin{figure*}
\centering
\includegraphics[height=9cm,width=15.0cm]{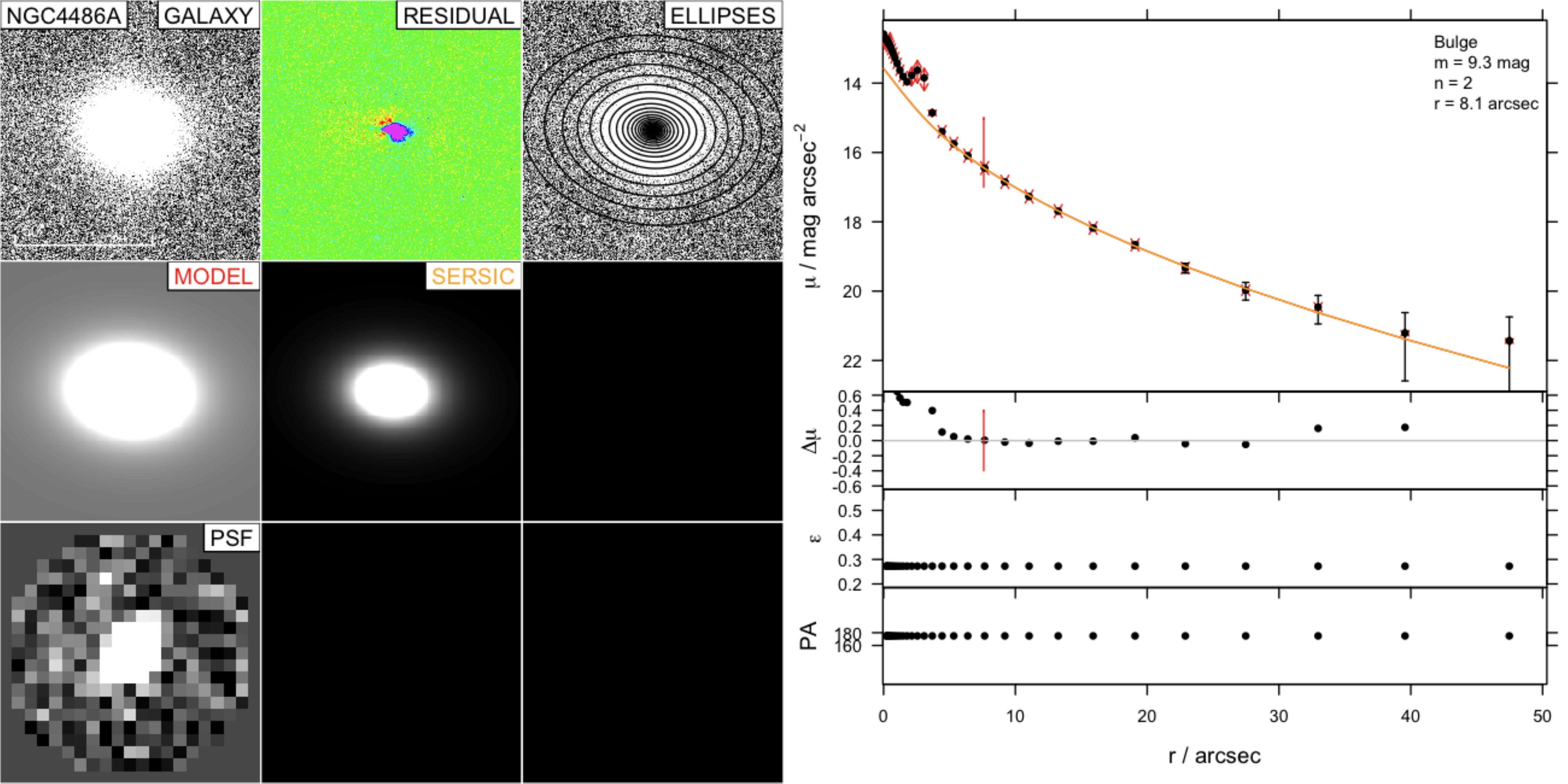}
\caption{For profiling NGC4486A we masked the inner 7 arcsec where the flux is contaminated by a star placed next to the core of the galaxy. The layout is as in Figure~\ref{fig:221}.}
\label{fig:4486A}
\end{figure*}

\begin{figure*}
\centering
\subfigure{\includegraphics[height=9cm,width=15.0cm]{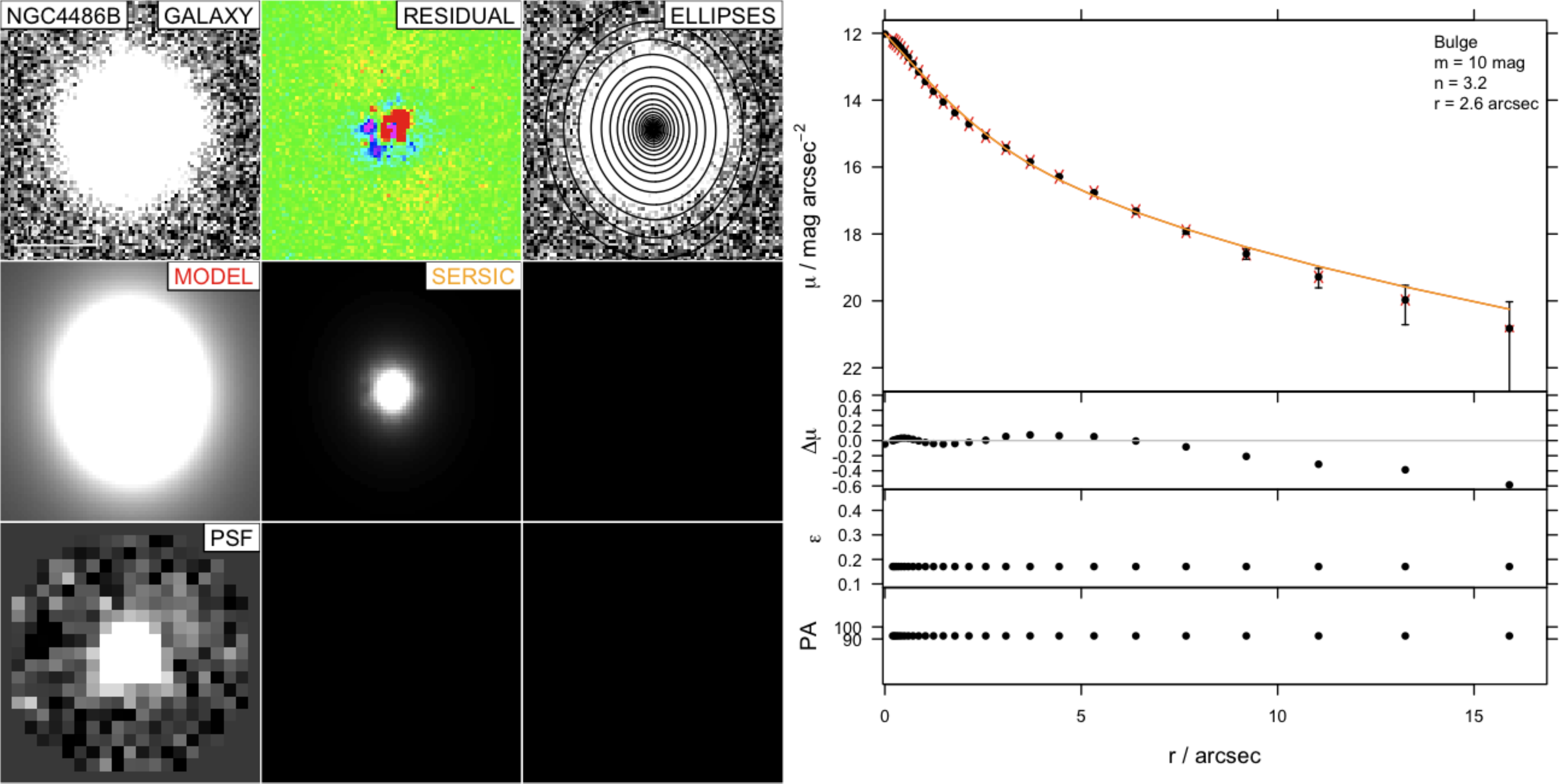}} 
\subfigure{\includegraphics[height=9cm,width=15.0cm]{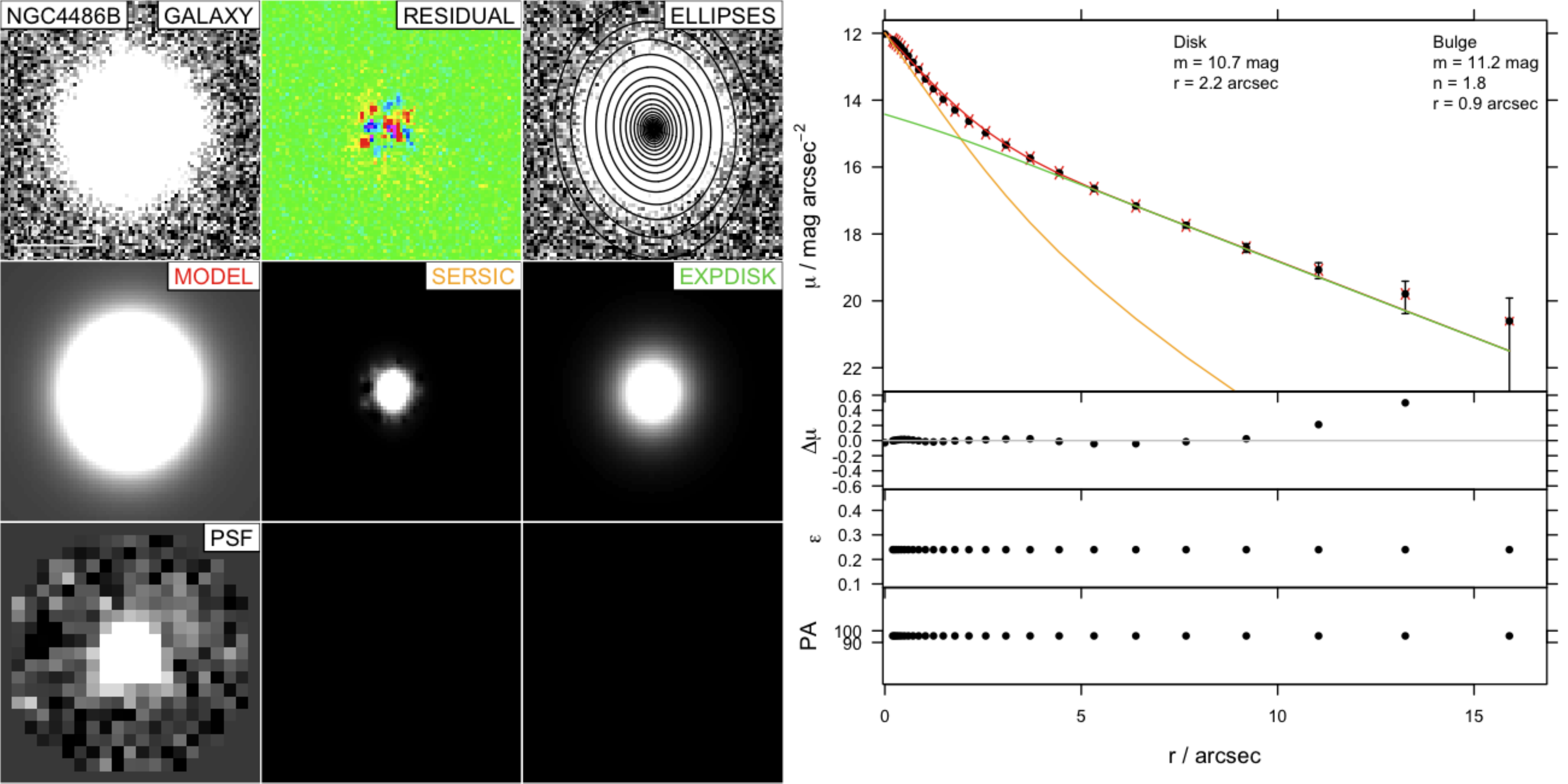}} 
\caption{NGC4486B has been characterised as a low luminosity cE0 with extra light in the center. Our surface brightness profile for NGC4486B for two different fits: (a) single S{\'e}rsic  and (b) S{\'e}rsic + exponential. The layout is as in Figure~\ref{fig:221}.}
\label{fig:4486B}
\end{figure*}

\begin{figure*}
\centering
\includegraphics[height=9cm,width=15.0cm]{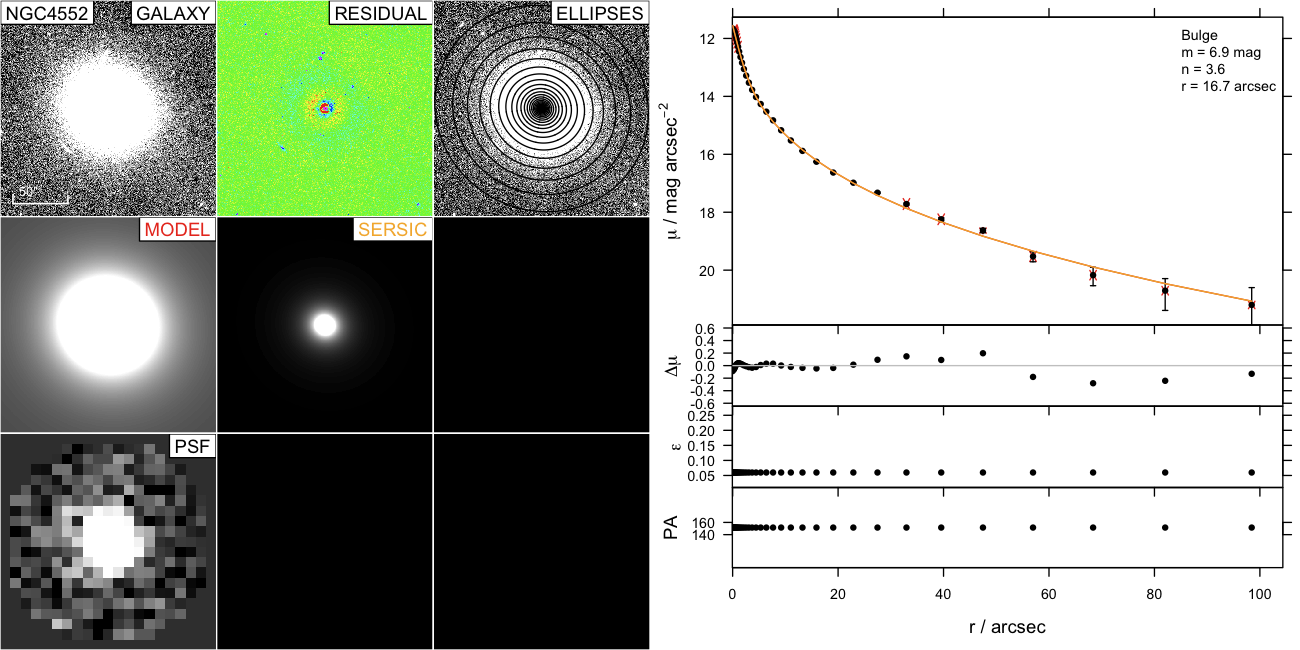}
\caption{Notice the S{\'e}rsic index variance between one dimensional and two dimensional profiles (see Section~\ref{sec:IG}). The layout is as in Figure~\ref{fig:221}.}
\label{fig:4552}
\end{figure*}

\begin{figure*}
\centering
\includegraphics[height=9cm,width=15.0cm]{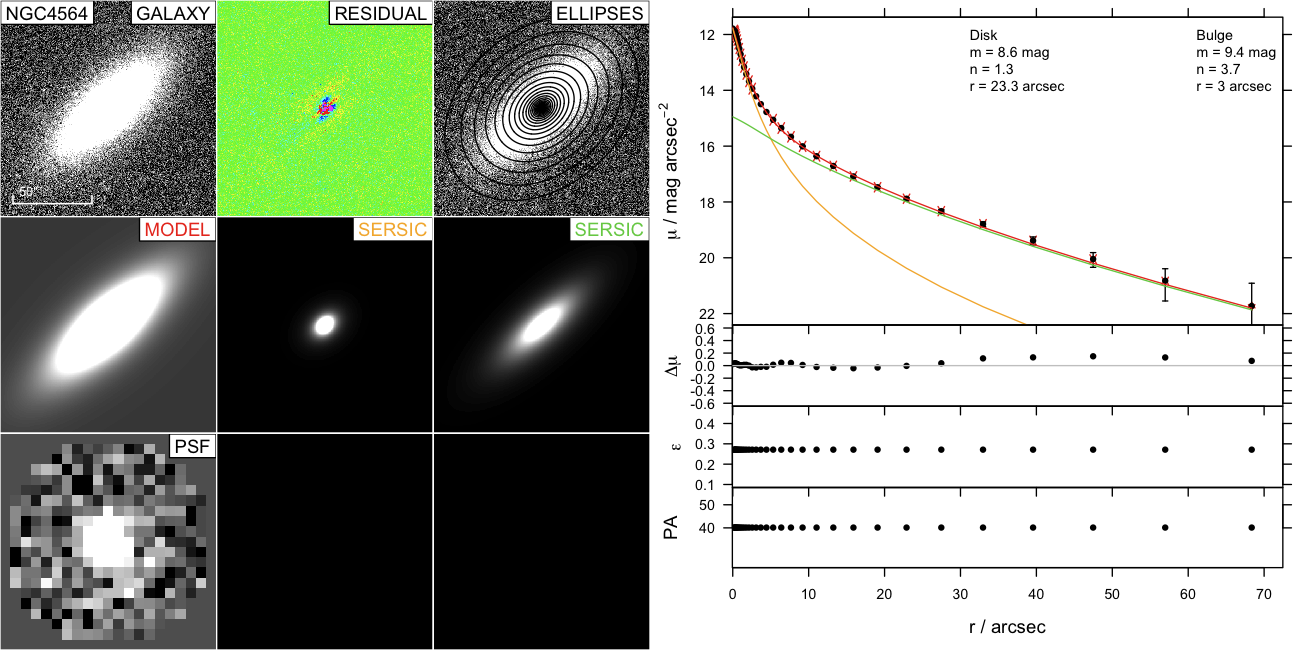}
\caption{The NGC4564 best fit is with a double S{\'e}rsic model. The layout is as in Figure~\ref{fig:221}.}
\label{fig:4564}
\end{figure*}

\begin{figure*}
\centering
\includegraphics[height=9cm,width=15.0cm]{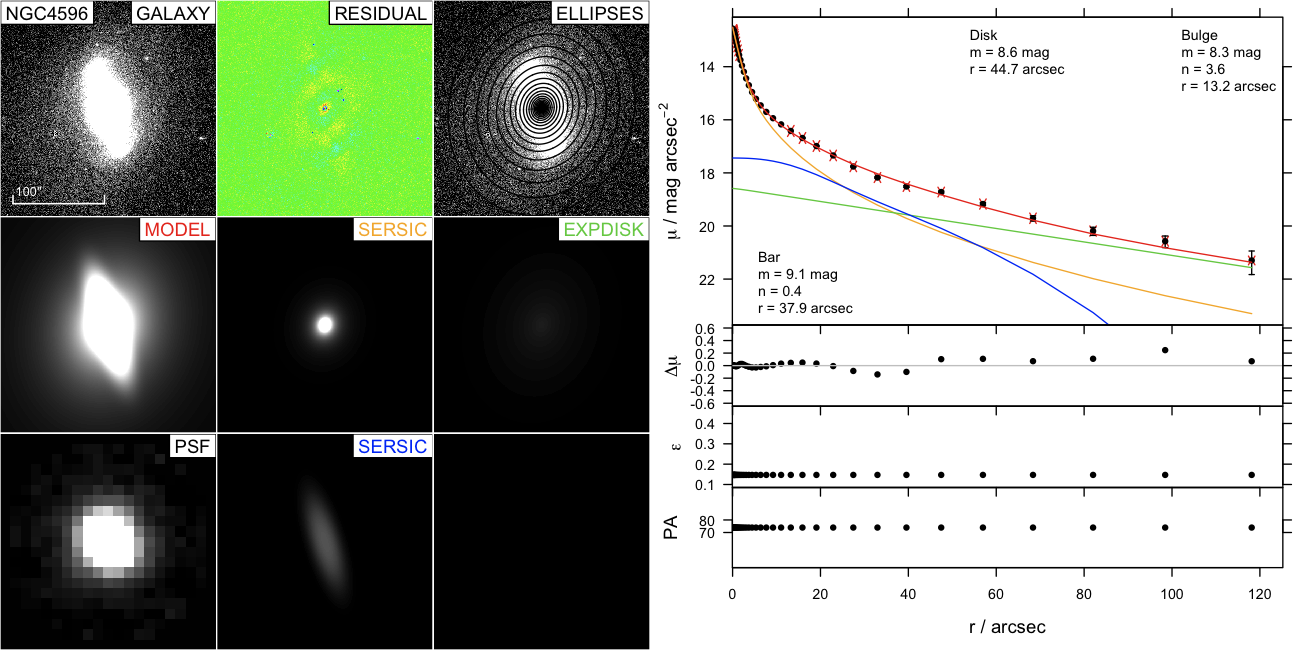}
\caption{The surface brightness profile for NGC4596. The layout is as in Figure~\ref{fig:221}.}
\label{fig:4596}
\end{figure*}

\begin{figure*}
\centering
\includegraphics[height=9cm,width=15.0cm]{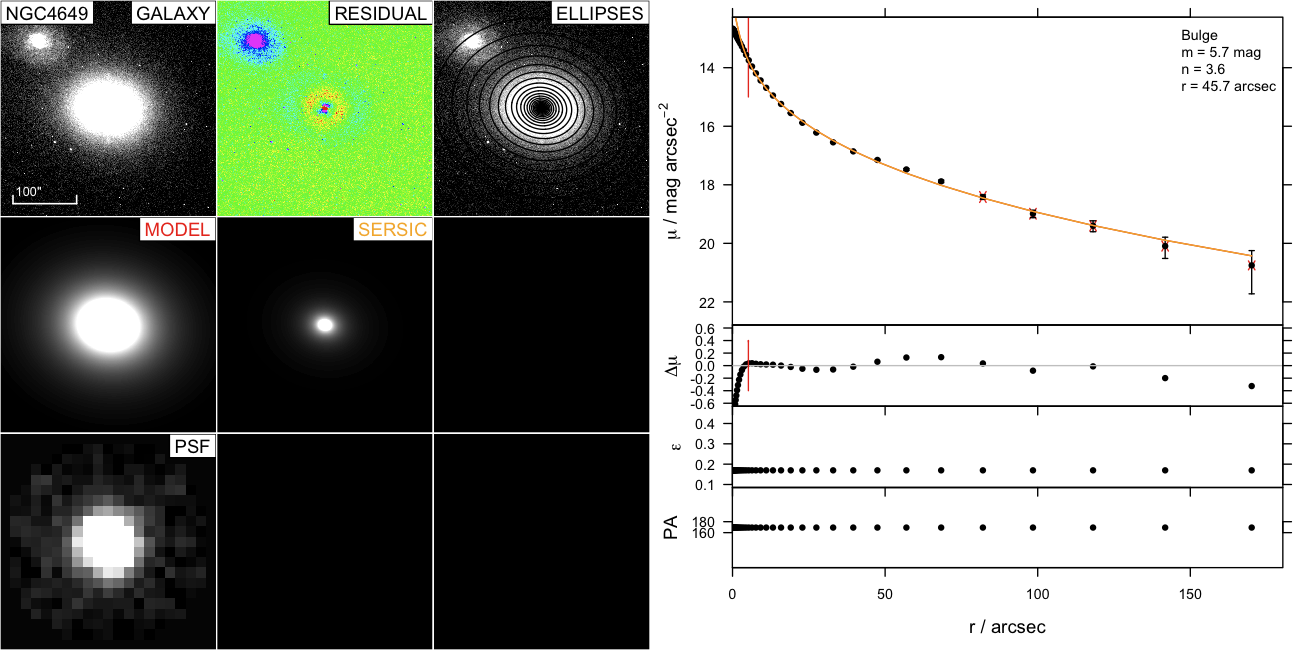}
\caption{The surface brightness profile for NGC4649, we masked the inner 5.2 arcsec. The layout is as in Figure~\ref{fig:221}. }
\label{fig:4649}
\end{figure*}

\begin{figure*}
\centering
\subfigure{\includegraphics[height=9cm,width=15.0cm]{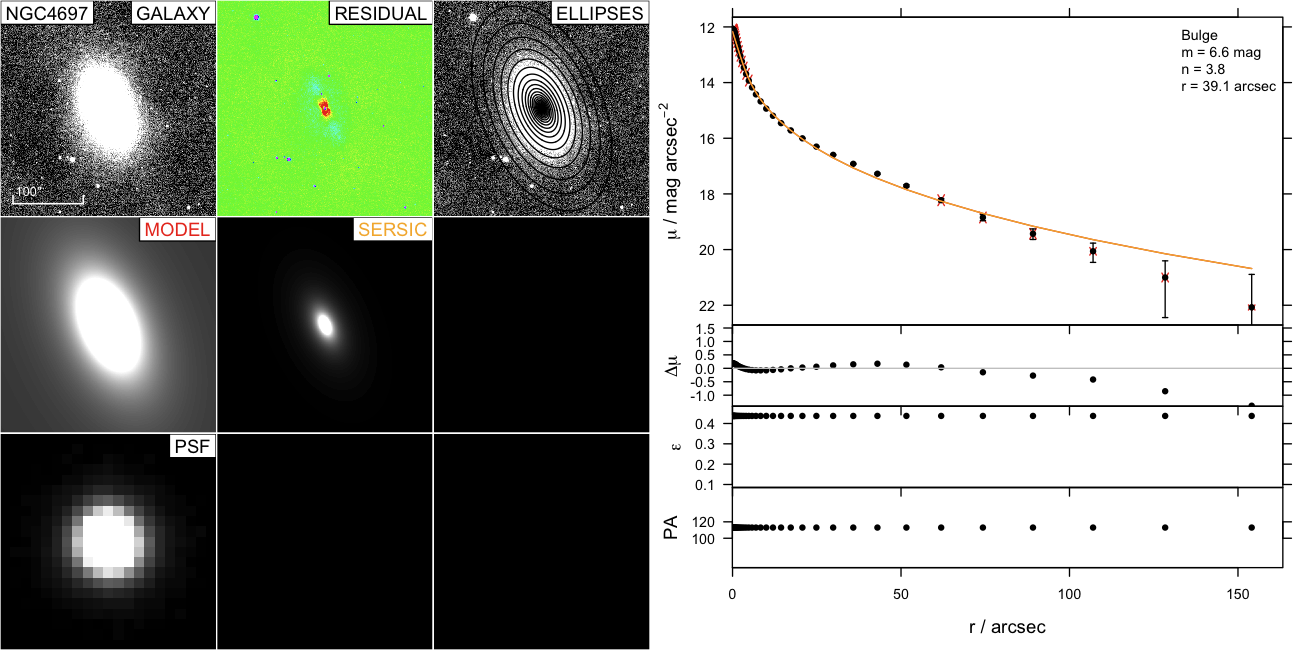}}
\subfigure{\includegraphics[height=9cm,width=15.0cm]{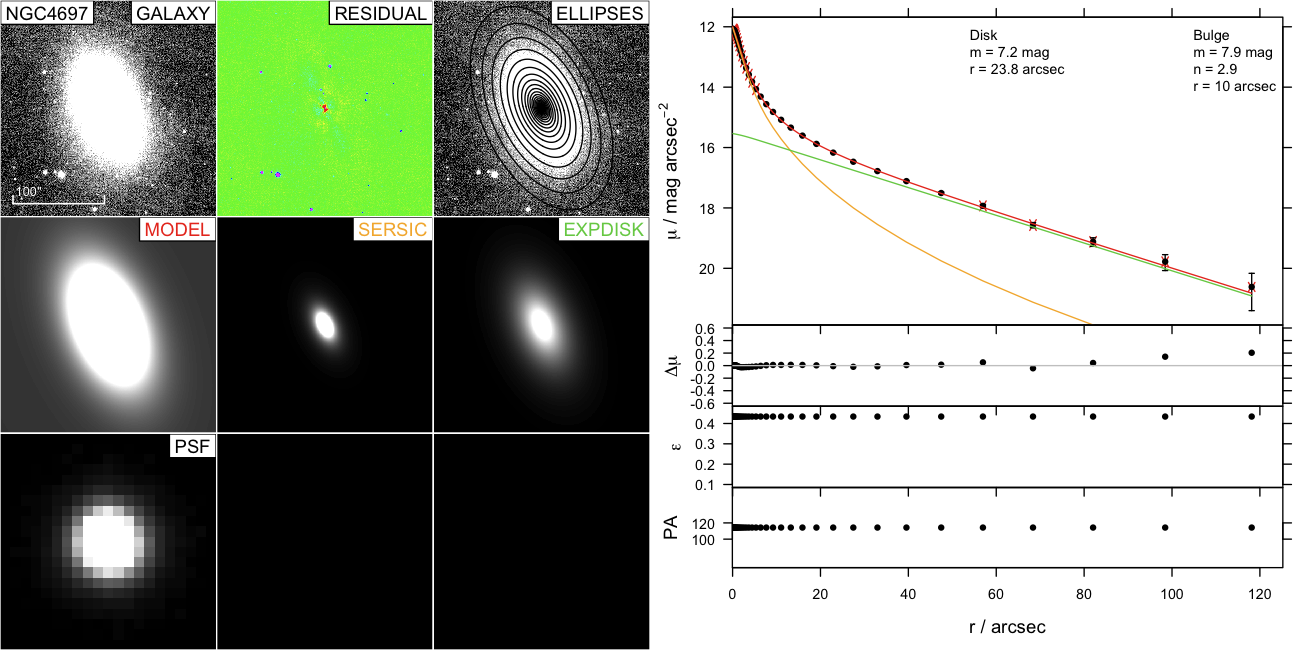}}
\caption{The surface brightness profile for NGC4697. The layout is as in Figure~\ref{fig:221}.}
\label{fig:4697}
\end{figure*}

\begin{figure*}
\centering
\includegraphics[height=9cm,width=15.0cm]{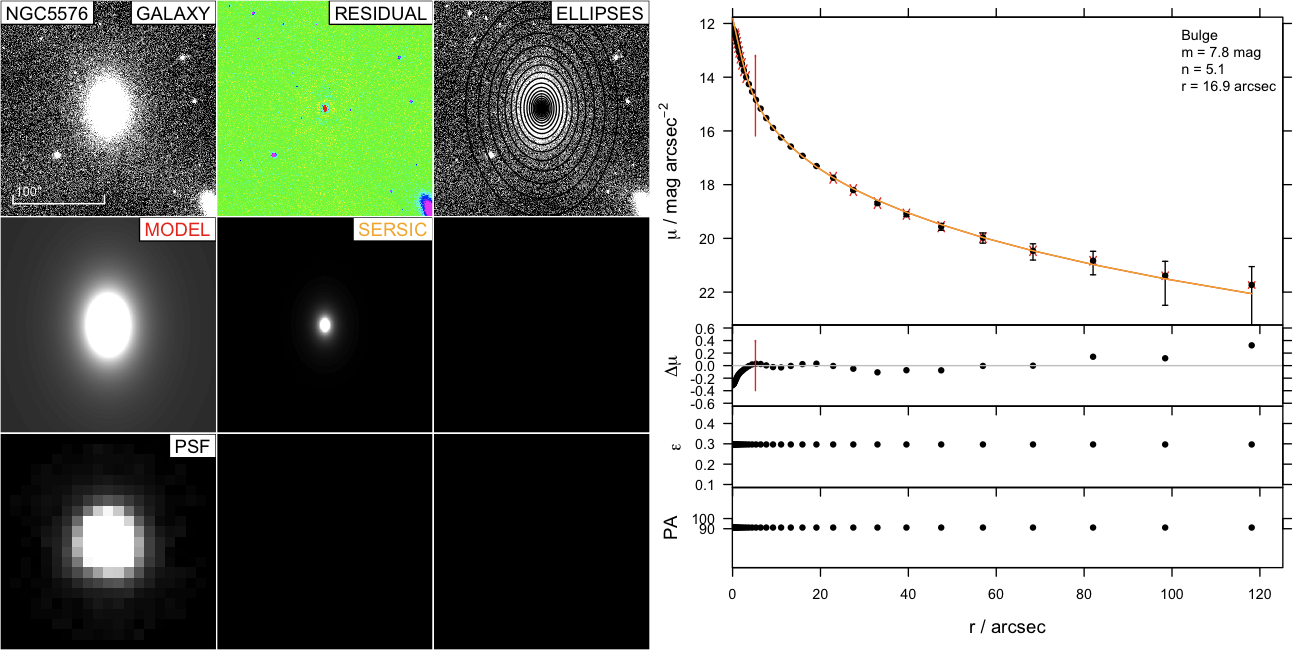}
\caption{The surface brightness profile for NGC5576, we masked the inner 5.2 arcsec. The layout is as 
in Figure~\ref{fig:221}.}
\label{fig:5576}
\end{figure*}

\begin{figure*}
\centering
\includegraphics[height=9cm,width=15.0cm]{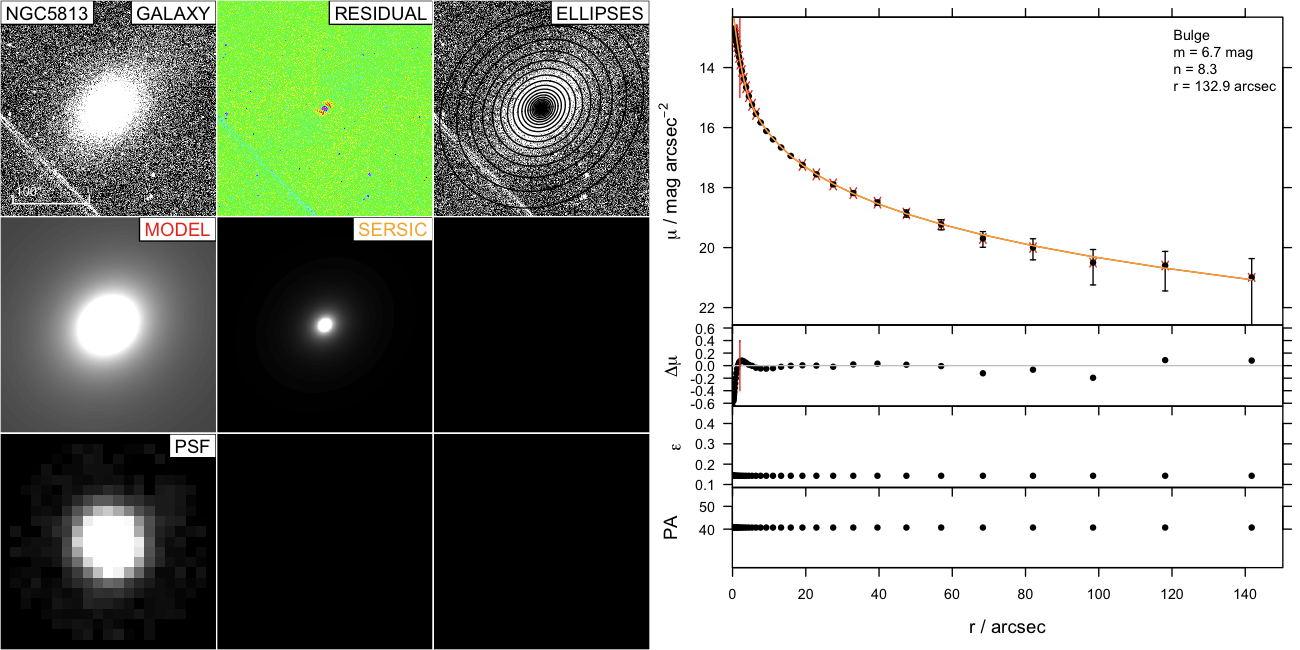} 
\caption{The surface brightness profile for NGC5813, we masked the inner 2 arcsec. The layout is as in Figure~\ref{fig:221}.}
\label{fig:5813}
\end{figure*}

\begin{figure*}
\centering
\includegraphics[height=9cm,width=15.0cm]{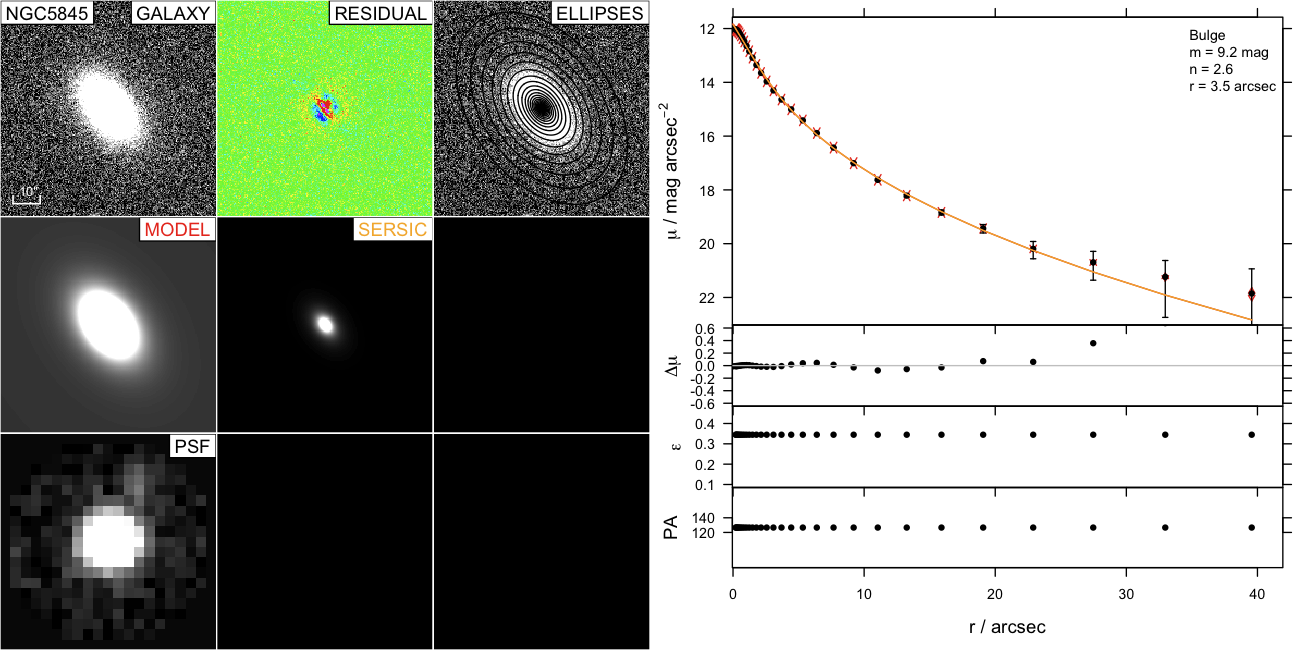}
\caption{The surface brightness profile for NGC5845. The layout is as in Figure~\ref{fig:221}.}
\label{fig:5845}
\end{figure*}

\begin{figure*}
\centering
\includegraphics[height=9cm,width=15.0cm]{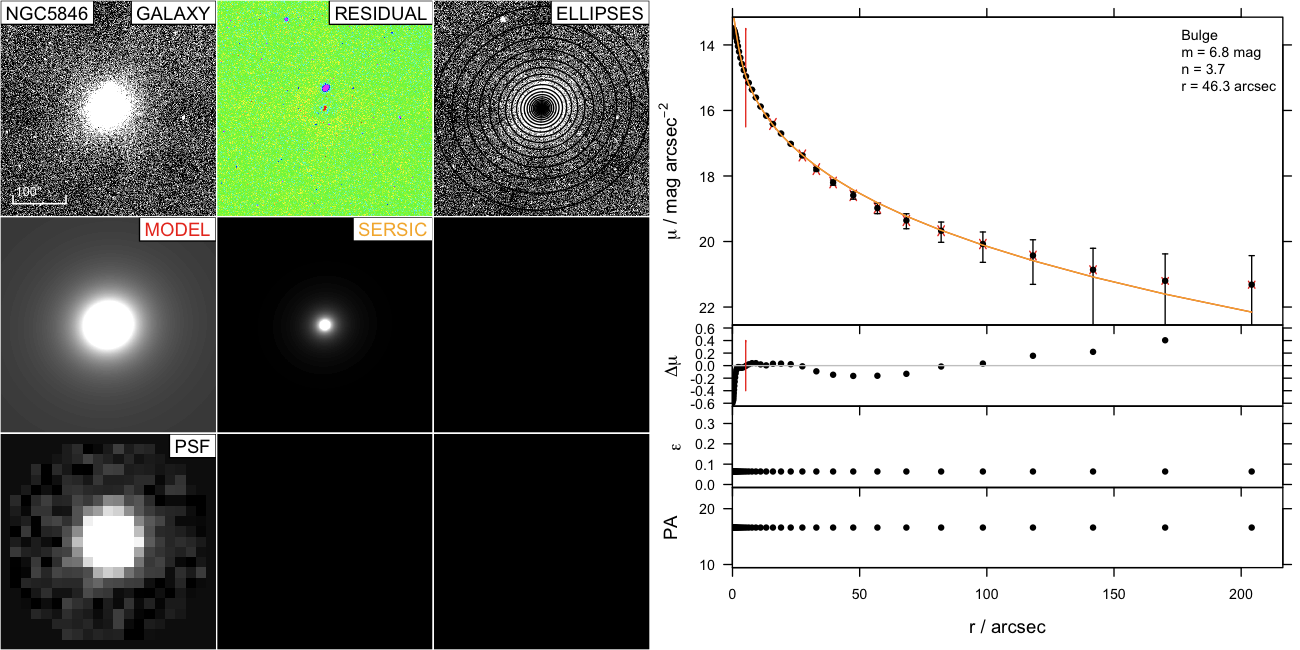}
\caption{The surface brightness profile for NGC5846, we masked the inner 5.2 arcsec. The layout is as in Figure~\ref{fig:221}.}
\label{fig:5846}
\end{figure*}

\begin{figure*}
\centering
\subfigure{\includegraphics[height=9cm,width=15.0cm]{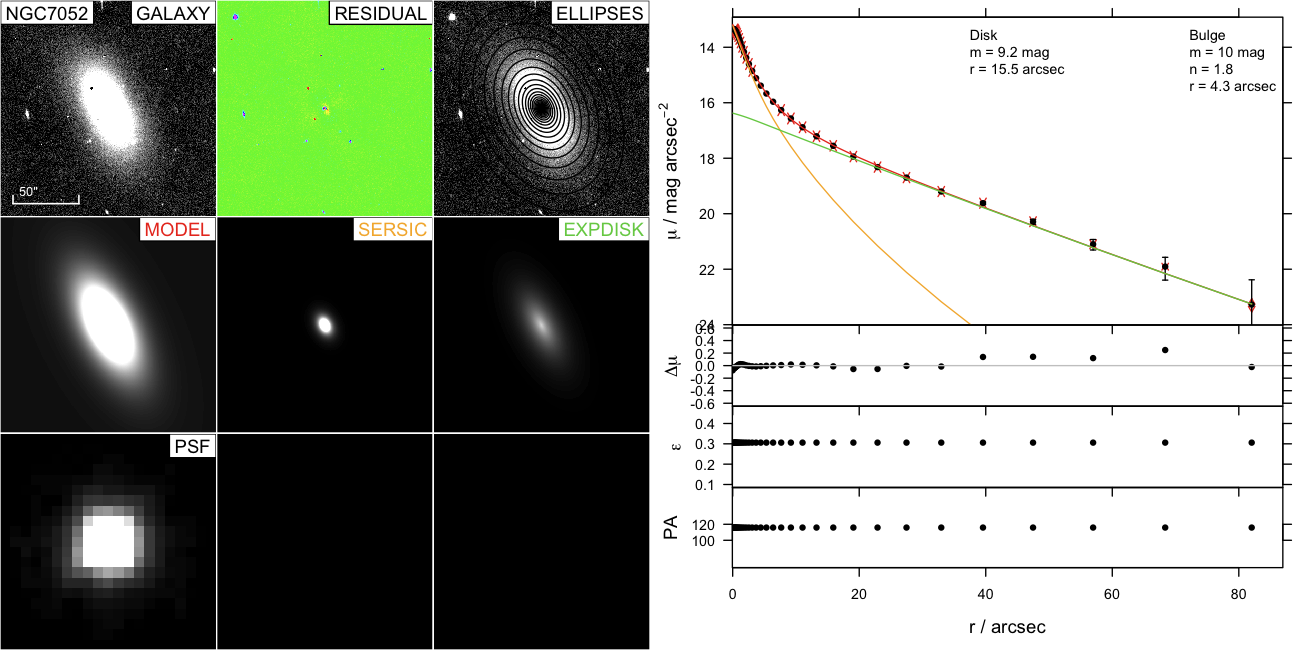}}  
\subfigure{\includegraphics[height=9cm,width=15.0cm]{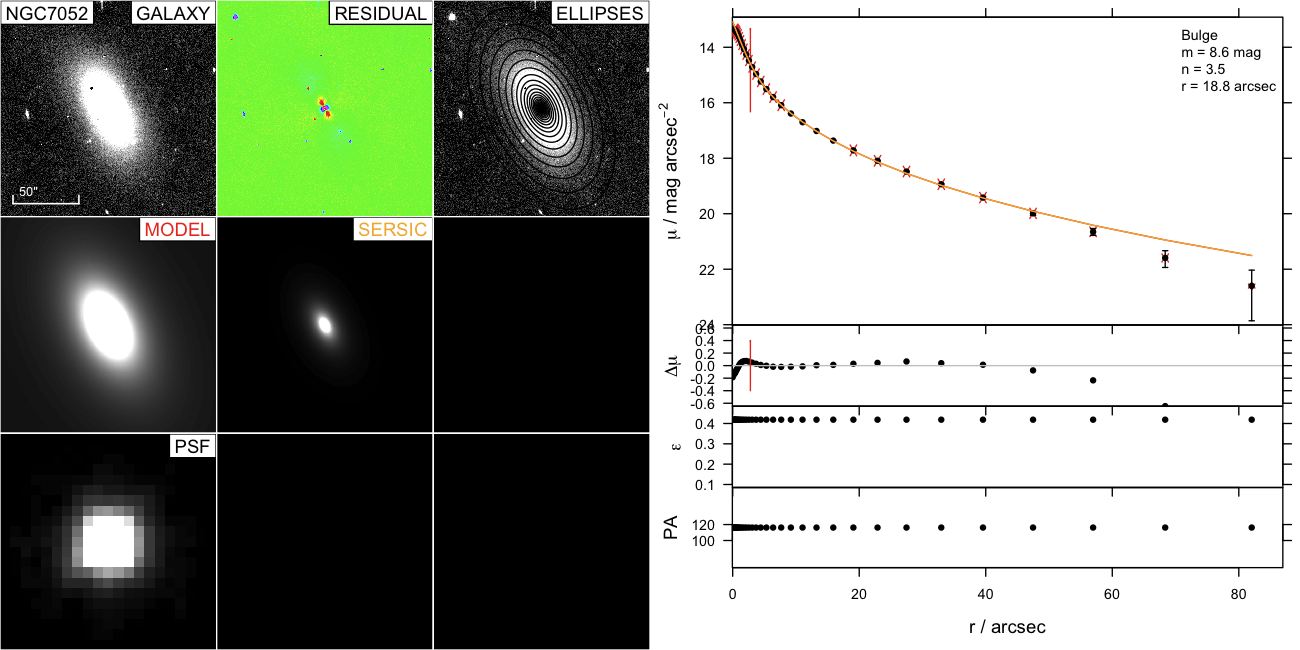}} 
\caption{Previous studies model NGC7052 with a single S{\'e}rsic model. We found the galaxy can be profiled accurate by adding an additional component. The layout is as in Figure~\ref{fig:221}.}
\label{fig:7052}
\end{figure*}

Our fits to all these galaxies are shown in Figures \ref{fig:221} - \ref{fig:9799} and the results of our profiles are tabulated in Table~\ref{table:properties}. Where the number of components is ambiguous we show the alternative profiles and tabulate its results in Table ~\ref{table:secproperties}.

\section{Results}
\label{sec:4}

\subsection{Fitting Methodology}
\label{sec:fit}
In Figure \ref{fig:ML} we plot the black hole masses versus the absolute bulge/elliptical K-band magnitude of the host galaxies.  The fitting algorithm used to estimate the linear $M_{\rm bh}$--$L$ relation and the log-linear $M_{\rm bh}$--$n$ relation is the regression analysis given in \citet{tex:TG02}
\begin{equation}
 \chi^2=\sum_{i=1}^N\frac{(y_{i}-\alpha-\beta x_{i})^2}{\delta y_{i}^2 +\beta^2 \delta x_{i}^2}
\label{eq:fitting}
\end{equation}
where $x=$logn or $x=M_{\rm K,sph}$, $y= \log(M_{\rm bh}/M_{\odot})$ and $\delta y_{i}$,  $\delta \chi_{i}$ are the errors of the $x$ and $y$ measurements. \citet{tex:TG02} inserted the intrinsic scatter $\epsilon_{0}$ of the $M_{\rm bh}$--$L,n$ correlations by replacing $\delta y_{i}$ with  $(\delta y_{i}^2 + \epsilon_{0}^2)^{1/2}$. Where $\epsilon_{0}$ is computed by repeating the fit until  $\chi^2 / (N-2)$ = 1. The uncertainty on the $\epsilon_{0}$ is estimated when $\chi^2 / (N-2)$ = 1$\pm \sqrt{2/N}$.

Here we assume that the errors cited in the literature are 1$\sigma$ uncertainties if not clearly stated. We now implement a Monte Carlo method to derive the errors. To do this we randomly perturb each SMBH mass and the galaxy magnitude in each case assuming a Normal error distribution. We repeat the fit 1001 times, each time applying equation \ref{eq:fitting}, assuming that the uncertainty is zero, consequently  $\delta y$ and  $\delta x$ are zero.

The final values for the intercept $\alpha$ and the slope $\beta$ are then derived from the median value of the 1001 individual sets of $\alpha$ and $\beta$ values while the  $\delta y$ and the $\delta x$ are the standard deviations.  We can see an illustration of the method in the bottom of the right panel  of Figure \ref{fig:ML}.  The 1001 red points shows the measurement distribution for NGC221 around the mean value. There are an equal number of points in each quadrant.

\subsection{Robustness of Mass Measurements}
A number of galaxies (NGC863, NGC4435, NGC4486b and UGC9799) have poorly constrained mass measurements. NGC863 is the only galaxy in our sample for which the SMBH mass has been estimated via the method of reverberation mapping.  Reverberation mapping masses use the local $M_{\rm bh}$--$\sigma$ relation to  normalise their values. Figure \ref{fig:ML} shows that  NGC863 is offset from the expected relation by $\sim$1dex while the virial mass estimation appears to be rather consistent with the $M_{\rm bh}$--$L$ relation. NGC4435 and UGC9799 have only upper limit on their SMBH mass estimations. NGC4486b SMBH mass measurement has been characterised as weak while the mass estimation models show a possibility of zero mass black hole. For these reasons the above referred galaxies have been excluded from the following fits.

\subsection{$M_{\rm bh}$--$L$ correlation}
\label{sec:ml}
\begin{figure*}
\centering
\includegraphics[height=6cm,width=12.0cm]{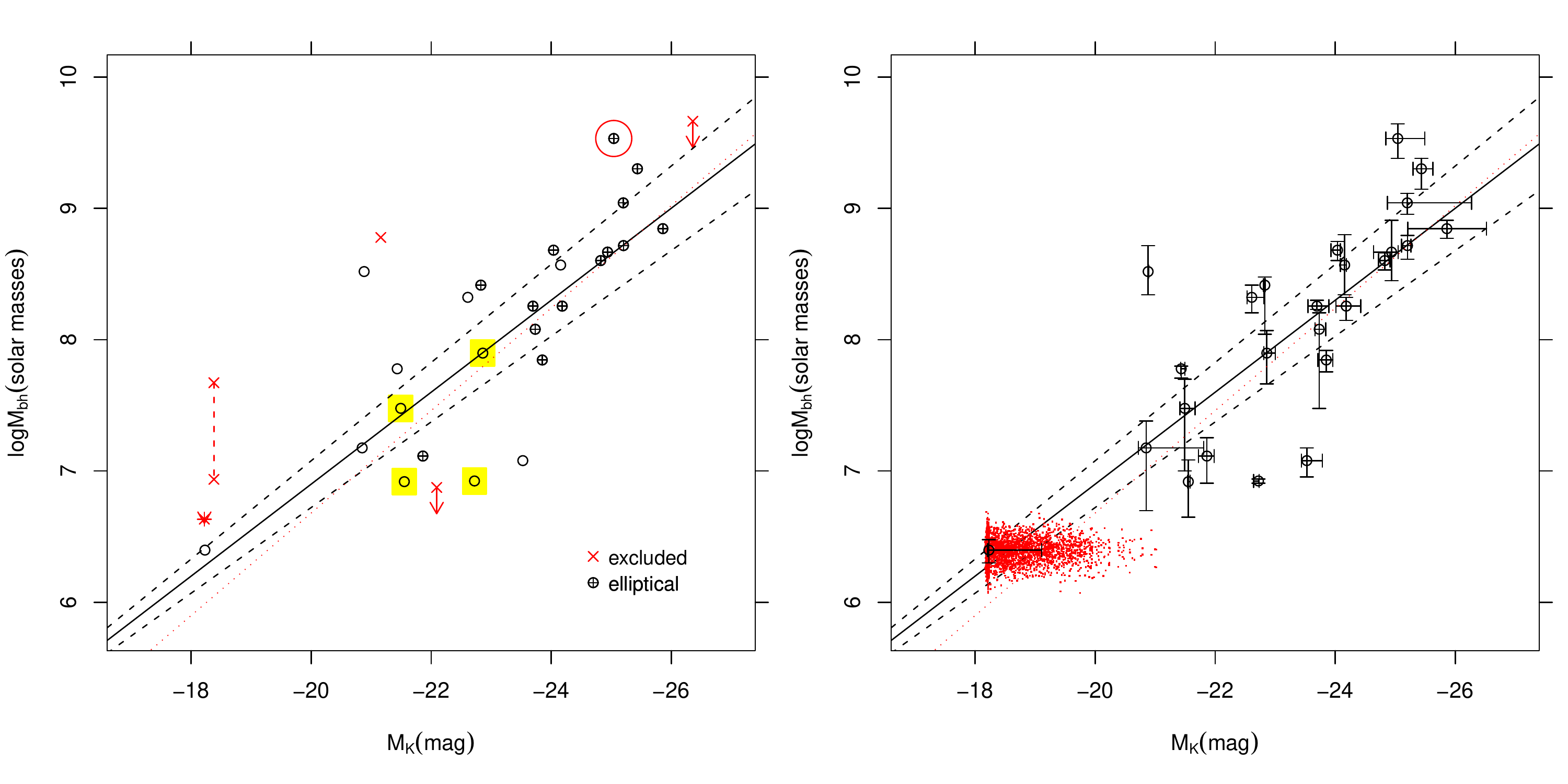}
\caption{Left panel: The $M_{\rm bh}$--$L_{\rm K}$ relation for the full sample (solid line) with 1$\sigma$ uncertainty (dashed line).  The circles with a cross indicate elliptical galaxies. The yellow rectangular symbol denote galaxies with bar. The red dotted line comes from \citet[eq.13]{tex:GD07}. The four galaxies excluded for the fitting (red crosses) are NGC863,  NGC4435, UGC9799 and NGC4486B. The red star indicate the position of the Milky Way which has not been included in the line fit but shown merely for reference. The red dashed line connects the two possible black hole masses for the galaxy NGC863. The red circle indicate the galaxy NGC4486.  Right Panel:  Same plot as the left panel with the addition of the individual error bars. See \S  \ref{sec:ml} for details.}
\label{fig:ML}
\end{figure*}

Figure \ref{fig:ML} shows the $M_{\rm bh}$--$L$ distribution for our full sample with the error bars shown in the right panel and various symbols indicating the morphological characteristics in the left panel. Applying Equation \ref{eq:fitting} to our trustworthy sample of 25 galaxies we derive:
\begin{equation}
 \log(M_{\rm bh}/M_{\odot})=~-0.35(\pm~0.024) (M_{\rm K} + 18)~ + ~ 6.2(\pm~0.13)
\end{equation}
with an intrinsic scatter of 0.52$^{+0.1}_{-0.06}$ dex in $ \log M_{\rm bh}$. This level of intrinsic scatter is relatively high and may arise from the varied morphological mix of our sample (see Table 1).

If we exclude the barred galaxies (NGC1068, NGC4258 , NGC4303 and NGC4596) for which bulge fluxes are consider the most uncertain, and the extreme outlying galaxy NGC4342 from the regression analysis we derive:

\begin{equation}
 \log(M_{\rm bh}/M_{\odot})=~-0.37(\pm~0.03) (M_{\rm K} + 18)~ + ~ 6.1(\pm~0.18)
\end{equation}
with an intrinsic scatter of 0.43$^{+0.09}_{-0.06}$dex in log$M_{\rm bh}$ for a subsample of 20 galaxies.

Finally, if we additionally exclude the low quality image cD galaxy, NGC4486 (see \S \ref{sec:index}),  we find:
\begin{equation}
 \log(M_{\rm bh}/M_{\odot})=~-0.36(\pm~0.03) ( M_{\rm K} + 18)~ + ~ 6.17(\pm~0.16)
\end{equation}
with an intrinsic scatter of 0.40$^{+0.09}_{-0.06}$dex in logM$_{\rm bh}$ for a secure high quality subsample of 19 galaxies.

When we apply the equation to our elliptical subsample consisting of 13 galaxies (again excluding NGC4486) we obtain:
\begin{equation}
 \log(M_{\rm bh}/M_{\odot})=~-0.42(\pm~0.06) (M_{\rm K} + 22)~ + ~  7.5(\pm~0.15)
\end{equation}
with an intrinsic scatter of 0.31$^{+0.087}_{-0.047}$dex in $\log M_{\rm bh}$.

The red dotted line in Figure \ref{fig:ML}  shows the  \citet{tex:MH03} relation after \citet{tex:G07} corrections have been applied. This relation has been derived from a sample of 26 galaxies (9 of which are within our sample) and has an intrinsic scatter of 0.35$~$dex (the best fit of  \citet{tex:MH03} full sample, consisting of 37 galaxies gives an intrinsic scatter 0.51$~$dex). From Figure \ref{fig:ML} we see that our  $M_{\rm bh}$--$L$ relation is found to be consistent with previous measurements. Previous near-IR $M_{\rm bh}$--$L$ relations are based on 2MASS data. The shallow nature of the 2MASS imaging makes it difficult to identify faint components of a galaxy. For instance \citet{tex:MH03} misclassified NGC221, NGC2778 and NGC4564 as elliptical galaxies.

We noticed that the intrinsic scatter of the $M_{\rm bh}$--$L$ relation is increased when we include barred galaxies. The increased dispersion of the scatter could be the result of the uncertainty introduced by estimating the individual luminosity for each component. Also the barred galaxies in our sample have nuclei activity. Extracting the bulge luminosities for these galaxies is complex because of the contamination of the bulge flux from active nuclei. 

In conclusion we find that our high quality data replicate but do not improve the intrinsic scatter suggesting a genuine spread in the data $\epsilon_{0} = \pm$0.31. We note the significant uncertainty when including multiply component systems $\epsilon_{0} = \pm$0.52.

\subsection{$M_{\rm bh}$--$n$ correlation}
\label{sec:index}

\begin{figure*}
\centering
\includegraphics[height=6cm,width=12.0cm]{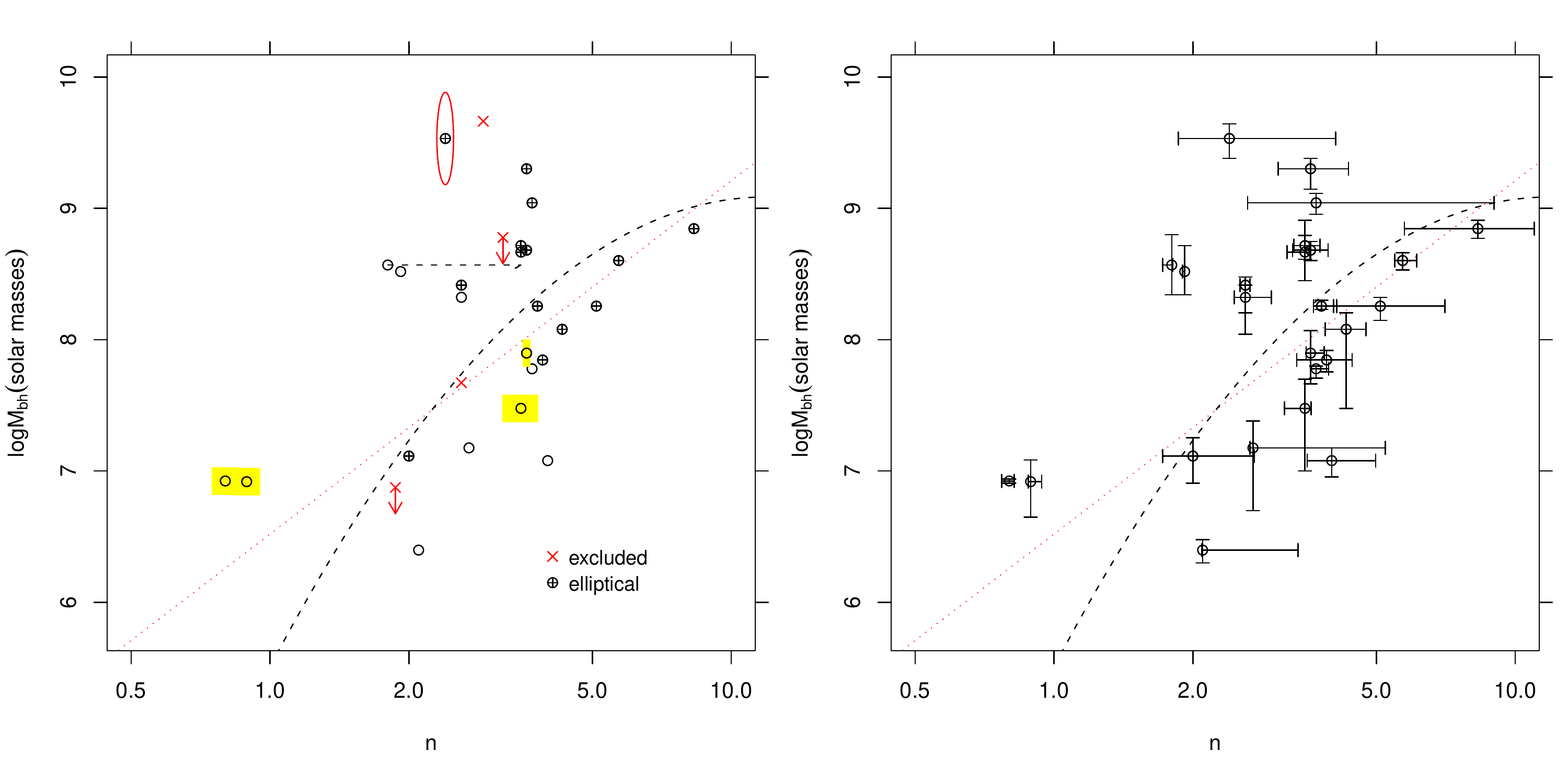}
\caption{The SMBH mass versus the spheroid S{\'e}rsic index in the K-band. Different groups of galaxies are indicated with different symbols as in Figure \ref{fig:ML}.  The black dashed line indicate the S{\'e}rsic index of NGC7052 for a single S{\'e}rsic fit.  The linear and quadratic $M_{\rm bh}$--$n$ relation of GD07 are shown as a red dotted line and a black dashed curve.} 
\label{fig:nL}
\end{figure*}

Figure \ref{fig:nL} shows the $M_{\rm bh}$ as a function of S{\'e}rsic indices in the K-band. The sample selection is the same as that noted is \S \ref{sec:ml}. Contrary to expectations our S{\'e}rsic index values  does not show strong correlation with $M_{\rm bh}$.

\begin{figure*}
\centering
\includegraphics[height=6cm,width=12.0cm]{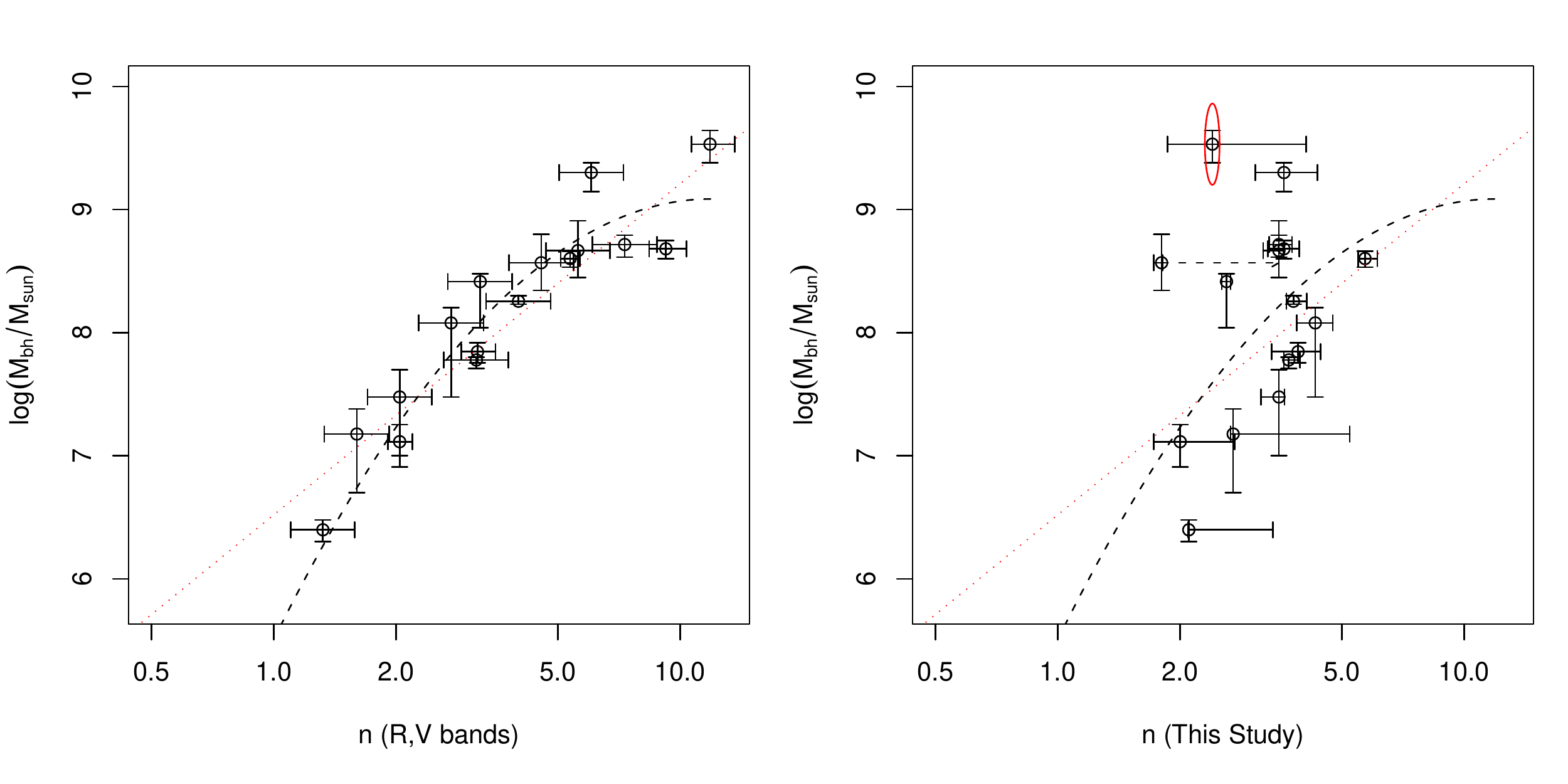}
\caption{Comparison between this study S{\'e}rsic index and values from literature. Left panel: Correlations between supermassive black hole mass and the S{\'e}rsic index of the spheroid for a subset of my sample reported from \textbf{1D} fitting in R,V bands  by \citet{tex:GD07a, tex:KF09, tex:SG07}. Dotted and dashed lines are GD07  best linear and quadratic fits.  Right panel: The same galaxies as left panel but with the S{\'e}rsic indices derived from this study with \textbf{2D} fit. } 
\label{fig:nLcom}
\end{figure*}

In Figure \ref{fig:nLcom} we compare our S{\'e}rsic indices with S{\'e}rsic indices from the literature.  In the left panel of Figure \ref{fig:nLcom} we plot $M_{\rm bh}$ versus the S{\'e}rsic indices found in the literature for 16 galaxies matching our sample. These measurements are derived for  R,V- band images using one dimensional profiling (listed in Table \ref{table:masses}). The right panel shows the $M_{\rm bh}$ versus this study's S{\'e}rsic indices. Individual comparisons for each galaxy's S{\'e}rsic index can be found in \S \ref{sec:IG}.

The tight relation between the $M_{\rm bh}$ and S{\'e}rsic index found in \textbf{1D} optical analysis disappears in our \textbf{2D} near-IR study. We notice that  \textbf{2D} analysis appear to have an upper limit of $n\sim$5. Especially, massive elliptical galaxies like NGC4261, NGC4486 and NGC452 appear to have small S{\'e}rsic indexes. In the case of NGC4486 we suspect that the sky gradient of the image obstruct us from fitting the halo of the galaxy. The discrepancy between the different studies therefore appear to be a result of either the method used to model the galaxy (i.e. 1D v 2D) and/or the transfer from optical wavelength to K-band images. 

Kelvin et al.  (in preparation) perform multi-wavelength \textbf{2D} profiling in nine bands, from u-band to K-band, with \texttt{GALFIT3}, finding no important change of  S{\'e}rsic indices for early-type systems in moving from r-band to near-IR. However disks and disk components are noted to show an increase in S{\'e}rsic index with wavelength and lowering in half-light radii. It appears then that the distinction between disk and bulge is less pronounced in the near-IR data compared to the optical. 

We believe that a further cause of the mismatch could be the different  pixel weighting adopted by \textbf{1D} v \textbf{2D} studies.  \texttt{GALFIT3} weights pixels using a sigma (weight) map. The sigma map shows the one standard deviation of counts at each pixel. Such maps can be created by the program itself or supplied by the user. We chose to follow \texttt{GALFIT3} manual suggestion and let the program create the maps internally (See Appendix A for details). The only input values are required for the sigma images to be produced internally are the background sky value and the root-mean-square (RMS) scatter of the background sky value. 

Another known source of uncertainty of the S{\'e}rsic index is the switch from major axis fitting to minor axis fitting (\citealt{tex:CC93}). \citet{tex:FD04} showed that major and minor axis  S{\'e}rsic index mismatch occur when there are radial variations of the isophotal eccentricity.

Further work is required to investigate what causes the breakdown of the $M_{\rm bh}$--$n$ correlation. Vika et al. (in preparation) will pursue this by exploring the photometric properties derived from \textbf{1D} and \textbf{2D} fits for a larger sample of $\sim$200 elliptical galaxies and the contribution of different pixel weighs.

Our conclusion is that the $M_{\rm bh}$--$n$ relation is no longer clearly apparent in the high quality near-IR data. We believe that this is caused by a combination of the use of  \textbf{2D} fitting in conjunction with the difficulty to distinguish the bulge and disk components in near-IR data. While we cannot rule out minor errors in the profiling process we have explored a variety of alternative fits with extensive masking. As the motivation was to derive an $M_{\rm bh}$--$n$ relation suitable for application to automated  \texttt{GALFIT3} analysis we must conclude that the $M_{\rm bh}$--$n$ relation is unsuitable for such use either because of a breakdown of the relation when \textbf{2D} fitting is used or the excessive care required to mitigate the core deviations.

\section{Summary}  
\label{sec:5}
One of the main motivations of this study was to derive $M_{\rm bh}$--$L,n$ relations using high quality near-IR data and using the same methods that we will apply to the GAMA survey in order to derive the SMBH mass function (e.g. \citealt{tex:VD09}). In this paper, we tested the  $M_{\rm bh}$--$L,n$ relations using updated SMBH masses and provide new estimations of the galaxy component luminosities and light profile shapes. We made use of K-band galaxy images for a sample of 29 galaxies taken by WFCAM as part of the UKIDSS-LAS. 

We used \texttt{GALFIT3}  to produce \textbf{2D} surface-brightness photometry on K-band images and decomposed the different components of the host galaxy. We carefully modelled all the components of each galaxy and derived estimates of the various structural parameters for each galaxy along with a concise discussion of each galaxy previous studies at optical wavelengths. 

We have used 21 elliptical and disk systems with classical bulge galaxies to derive the $M_{\rm bh}$--$L$ relation with intrinsic scatter of 0.41. We confirm a strong correlation between the central $M_{\rm bh}$ and its host galaxy's spheroid luminosity found from a number of previous studies.  Overall, we see that the scatter of the $M_{\rm bh}$--$L$ relation is much larger when we include bar galaxies and galaxies with active nuclei.  We found no improvement of the intrinsic scatter for the $M_{\rm bh}$--$L$ relation by using higher quality data which may indicate that we have reached the physical limit to which one we can constrain the $M_{\rm bh}$--$L$ relation.

Using the same sample of galaxies we failed to find a clear $M_{\rm bh}$--$n$ correlation but we noticed that the S{\'e}rsic index can vary significantly from study to study. The available data are inadequate for deriving accurate outcomes for the different S{\'e}rsic index values. Our best explanation is that the mismatch arises from the different weighting of pixel during the fit that each study uses. Further comparison of \textbf{1D} analysis versus \textbf{2D} analysis is required to fully understand this result. 

In conclusion we have established an $M_{\rm bh}$--$L$ relation based on  \texttt{GALFIT3} \textbf{2D} profiling of near-IR data which we will shortly apply to the GAMA dataset (Vika et al in preparation). 

\section*{Acknowledgments}

This work is based in part on data obtained as part of the UKIRT Infrared Deep Sky Survey. The UKIDSS project is defined in \citealt{tex:LW07}. UKIDSS uses the UKIRT Wide Field Camera (WFCAM; \citealt{tex:CA07}) and a photometric system described in \citealt{tex:HW06}. The pipeline processing and science archive are described in Irwin et al (2010, in preparation) and \citealt{tex:HC08}. We thank the referee for helpful suggestions that improve this paper. 

\bibliography{references}

\begin{table*} 
\caption{\texttt{GALFIT3} Derived Parameters. Column(1): name of the galaxy; Column(2-4): the spheroid apparent
magnitude, effective radius along the semi-major axis and S{\'e}rsic index; Column (5-6): the disk 
apparent magnitude and scale-length; Column (7-9): the bar apparent magnitude, effective radius and 
S{\'e}rsic index; Column (10-11): the spheroid and total absolute magnitude; Column(12): the best fit chosen.}
\smallskip
\centering
\begin{tabular}{lccccccccccl}
\hline
\noalign{\smallskip}
Galaxy   &m$_{\rm sph}$  & r$_{\rm eff}$ &$n_{\rm sph}$ & m$_{\rm d}$  & r$_{\rm s}$  & m$_{\rm bar}$ & r$_{\rm bar}$ & $n_{\rm bar}$ & M$_{\rm sph}$ & M$_{\rm tot}$  & best fit\\
Name     &   (mag)	 & (arcsec)      & Index        &    (mag)     & (arcsec)     & (mag)	      & (mag)         & Index	      &	(mag)         &  (mag)         &         \\
 (1)     &   (2)  	 &   (3)         &  (4) 	&     (5)      &  (6)         & (7)	      & (8)           & (9)	      & (10)          &   (11)         &  (12)	 \\	   
\noalign{\smallskip}
\hline
\hline
\noalign{\smallskip}             
NGC221     &     6.5 $^{+0.88    }_{- 0.02  }$     &  18.3    &    2.1 $^{+0.01    }_{-1.28  }$  &	 5.8 &   43.7 &  /	  &  /   &   /                & -18.2  &  -19.5     &bulge$+$disk$+$mask centre \\
NGC863    &    10.9 $^{+0.2     }_{-  0.32 }$     &  1.66    &    2.6 $^{+ 0.83   }_{- 0.001 }$ &	10.8 &    6.5 &  /	 &  / & /                     & -18.4  &  -19.7     &bulge$+$disk$+$psf \\	  
NGC1068  &     8.2  $^{+0.01    }_{- 0.08  }$    &  2.8     &    0.8   $^{+0.03    }_{-0.02  }$ & 6.5        &   16.7 &   7.9 & 11.8 &  0.3 & -22.7  & -24.9     &bulge$+$disk$+$bar$+$mask centre\\      
NGC2778  &    10.9$^{+0.96   }_{-  0.13 }$      &  1.5     &    2.7   $^{+0.04   }_{- 2.53 }$   &	10.7 &    5.4 &  /	&  /   &   /		      & -20.8  &  -21.9    &bulge$+$disk	      \\	  
NGC2960  &    10.8$^{+0.26    }_{- 0.09  }$      &  1.8     &    4.0  $^{+0.46    }_{-0.98  }$  &	10.7 &    6.8 &  /	&  /   &   /		      & -23.5  &  -24.8    &bulge$+$disk	      \\  
NGC3245  &     9.0 $^{+ 0.2    }_{- 0.08 } $     &  3.5     &    2.6   $^{+ 0.14   }_{- 0.36 }$  &	 8.3 &   20.5 &  /	&  /   &   /		      & -22.6  &  -24.1    &bulge$+$disk	      \\	  
NGC4258  &     7.8 $^{+ 0.17   }_{-  0.08 }$     & 17.0     &    3.5  $^{+0.34    }_{-0.11  }$  &	 6.7 &   50.2 &  /	&  /   &   /	   	      & -21.5  &  -24       &bulge$+$disk$+$psf         \\      
NGC4261  &     7.3 $^{+ 0.06  }_{-  0.1  }$      & 24.2     &    3.5  $^{+ 0.19  }_{-  0.27}$   &	 /   &   /    &  /	&  /   &   /		      & -25.2  &  -25.2    &elliptical$+$mask centre \\	
NGC4303  &     9.5  $^{+0.03    }_{- 0.01  }$    &  2.8     &    0.9   $^{+0.01    }_{-0.05  }$ &	 7.5 &   30.0 &   9.1	& 49.5 &  0.5  & -21.6  &  -25.2   &bulge$+$disk$+$bar$+$psf \\ 
NGC4342  &    10.3 $^{+0.008   }_{-0.004  }$     &  0.86    &    1.9 $^ {+0.02   }_{- 0.001 }$  &	 9.6 &    5.1 &  /	&  /   &   /          & -20.9  &  -22.3   &bulge$+$disk	       \\
NGC4374  &     6.4  $^{+0.02    }_{- 0.01  }$    &  28.7     &    3.5   $^{+0.02    }_{-0.03  }$ &	 / &  / &  /	&  /   &   /	                         &-24.9   &  -24.9   &elliptical$+$mask centre	       \\		
NGC4435  &     8.8  $^{+0.03   }_{-  0.08 }$     &  4.5     &    1.5   $^{+0.05   }_{- 0.05 }$  &	 8.7 &   18.9 &   9.9	& 20.8 &  0.3  & -22      &  -22.9   &bulge$+$disk$+$bar     \\
NGC4459  &     7.2  $^{+0.11    }_{- 0.14  }$    & 25.0     &    3.9  $^{+0.54    }_{-0.53  }$ &	/    &   /    &  /	&  /   &  /		      & -23.8  &  -24.1    &elliptical$+$moffat core   \\
NGC4473  &     7.2  $^{+0.11    }_{- 0.07  }$    & 21.3     &    4.3  $^{+0.43    }_{-0.45  }$ &   /    &   /    &  /	&  /   &  /		      & -23.7  &  -23.7    &elliptical$+$mask centre \\  
NGC4486  &     6.0  $^{+0.45    }_{- 0.2   }$    & 34.6     &    2.4  $^{+ 0.54   }_{- 1.68 }$ &	/    &   /    &  /	&  /   &  /		      & -25.0  &  -25.0    &elliptical$+$mask centre   \\ 
NGC4486a &     9.3 $^{+0.12   }_{- 0.14  }$      &  8.1     &    2.0   $^{+0.28   }_{-0.72  }$  &	/    &   /    &  /	&  /   &  /		      & -21.9  &  -21.9    &elliptical$+$mask centre \\
NGC4486b &    10.0$^{+0.12 }_{-0.13  }$    &   2.6        & 3.2$^{+ 0.12 }_{-  0.1}$           &	/    &   /   &  /	&  /   &  /	               & -21.2 & -21.2      &elliptical	   \\   
NGC4552  &     6.9   $^{+0.05    }_{-0.11   }$   & 16.7     &    3.6   $^{+0.31    }_{-0.33  }$  &	/    &   /    &  /	&  /   &  /      	      & -24.0 &  -24.0     &elliptical  	   \\		    
NGC4564  &     9.4   $^{+0.06    }_{-0.01   }$   &  3.0     &    3.7    $^{+0.1     }_{-0.24  }$  &	/    &   /    &   8.6	& 23.4 &  1.3	      & -21.4  &  -23.1    &bulge$+$disk	     \\ 	
NGC4596  &     8.3   $^{+0.14   }_{- 0.05  }$    & 13.2     &    3.6   $^{+0.08   }_{-0.25  }$  &	 8.6 &   44.7 &   9.1	& 37.9 &  0.4 & -22.9  &  -24        &bulge$+$disk$+$bar       \\
NGC4621  &     6.5  $^{+0.09    }_{- 0.1   }$    & 54.7     &    5.7  $^{+ 0.22   }_{-0.41 }$  &	/    &   /    &  /	& /    & /	               & -24.8  &  -24.8    &elliptical$+$mask centre\\  
NGC4649  &     5.7  $^{+0.19    }_{- 0.14  }$    & 45.7     &    3.6  $^{+0.54    }_{-0.75  }$  &	/    &   /    &  /	& /    & /	               & -25.4  &  -25.4    &elliptical$+$mask centre \\
NGC4697  &     6.6  $^{+0.01    }_{- 0.03  }$    & 39.1     &    3.8  $^{+0.03    }_{-0.01  }$  &	/ &  / &  / 	& /    &  /	                        & -20.3  &  -20.3     &elliptical       \\
NGC5576  &     7.8  $^{+0.24   }_{-  0.17 }$     & 16.9     &    5.1  $^{+1.06   }_{-1.94 }$   &	/    &   /    &  /	&  /   &  /		     & -24.2  &  -24.3     &elliptical$+$mask centre  \\
NGC5813  &     6.7  $^{+0.66    }_{- 0.65  }$    & 132.9    &    8.3$^{+2.55    }_{-2.69  }$   &	/    &   /    &  /	&  /   &  /		     & -25.9  &  -25.9     &elliptical$+$mask centre  	     \\
NGC5845  &     9.2  $^{+0.01    }_{- 0.02  }$    &  3.5     &    2.6   $^{+0.07    }_{-0.06  }$   &	/    &   /    &  /	&  /   &  /		     & -22.8  &  -22.9     &elliptical  	     \\
NGC5846  &     6.8  $^{+1.07    }_{- 0.33  }$    & 46.3     &    3.7  $^{+1.07    }_{-5.29  }$   &	/    &   /    &  /	&  /   &  /		     & -25.2  &  -25.1     &elliptical$+$mask centre    \\
NGC7052   & 10.0 $^{+0.03 }_{- 0.07  }$   &  4.3  &    1.8$^{+0.08   }_{-0.01  }$    &	 9.2 &   15.2 &  /	&  /   &  /			     & -24.2   &  -25.6    &bulge$+$disk	     \\
UGC9799  &     9.4   $^{+0.48  }_{- 0.26  }$  & 31.3     &    2.9     $^{+0.41  }_{-0.96   }$   &	/    &   /    &  /	&  /   &  /		     & -26.4  &  -26.8     &elliptical$+$mask centre  \\
\noalign{\smallskip}
\hline
\end{tabular}
\label{table:properties} 
\end{table*} 

\begin{table*} 
\caption{Second better fit chosen. The layout is as in Table~\ref{table:properties}.}
\smallskip
\centering
\begin{tabular}{lccccccccccl}
\hline
\noalign{\smallskip}
Galaxy &m$_{\rm sph}$ & r$_{\rm eff}$ &$n_{\rm sph}$ & m$_{\rm d}$ & r$_{\rm s}$ & m$_{\rm bar}$ & r$_{\rm bar}$ & $n_{\rm bar}$ & M$_{\rm sph}$ & M$_{\rm tot}$  & best fit\\
Name   &   (mag)      & (arcsec)      & Index	     &    (mag)    & (arcsec)	 & (mag)	 & (mag)	 & Index	 & (mag)	 &  (mag)	  &	    \\
 (1)   &   (2)        &   (3)	      &  (4)	     &     (5)     &  (6)	 & (7)  	 & (8)  	 & (9)  	 & (10) 	 &   (11)	  &  (12)   \\        
\noalign{\smallskip}
\hline
\hline
\noalign{\smallskip}             
NGC4258   &   8.8$^{+ 0.2   }_{-  0.09 }$   &   6.3  & 2.2$^{+0.4    }_{-0.12  }$ &    6.6 &   50.0 &   8.9  &  0.9 & 22.0     &  -20.5 & -24& bulge$+$disk$+$bar$+$psf \\
NGC4374   &   7.6$^{+0.04  }_{- 0.03 }$    &   7.8  & 1.6$^{+0.02  }_{-0.04  }$ &	 7.0 &   22.8 &  /	&  /   &   /	& -23.7  &    -24.9                &bulge$+$disk	       \\		
NGC4435   &   8.8$^{+0.02  }_{-  0.07 }$   &   4.5  & 1.5$^{+0.04  }_{- 0.03 }$ &    8.6      &   17.4 &   9.8  &  0.4 & 19.9   &  -21.9 & -22.9 &bulge$+$disk$+$bar$+$psf \\	   
NGC4486b &    11.2$^{+0.11    }_{-0.1    }$      &  0.9     &    1.8$^{+ 0.15  }_{-  0.11}$  &	10.7 &    2.2 &  /	&  /   &  /		& -20.0   &  -21.2                    &bulge$+$disk	   \\
NGC4697   &   7.9$^{+0.01  }_{- 0.03  }$   & 10.0  & 2.9$^{+0.03  }_{-0.01  }$ &	 7.2 &   23.8 &  / 	& /     &  /	&  -22.4 & -20.3            &bulge$+$disk       \\
NGC7052   &   8.6 $^{+0.04 }_{- 0.06  }$   & 18.8  & 3.5$^{+0.05  }_{-0.01  }$ &	 / &   / &  /	&  /   &  /	         & -25.6  & -25.6                     &elliptical$+$mask centre 	     \\

\noalign{\smallskip}
\hline
\end{tabular}
\label{table:secproperties} 
\end{table*}

\begin{table*} 
  \caption{Galaxy Sample; Column(2): The distances have come from \citet{tex:TD01} unless otherwise 
specified. Column(3): The black hole masses have been adjusted to the distance given in column 2.
Column(4): The method of measuring the black hole mass: s-stellar kinematics, g-gas kinematics, 
m-water masers, p-stellar proper motion and r-reverberation mapping. Column(5-6): S{\'e}rsic indices 
and their band. 
References: (1)\citealt{tex:VC02}; (2)\citealt{tex:PF04}; (3)\citealt{tex:LB03}; (4)\citealt{tex:GR03}; 
 (5)\citealt{tex:HB02}; (6)\citealt{tex:BS01}; (7)\citealt{tex:HM99};
 (9)\citealt{tex:SG09};  (10)\citealt{tex:FF96a}; (11)\citealt{tex:PM07}; (12)\citealt{tex:JB04}; 
 (13)\citealt{tex:CB99}; (14)\citealt{tex:MB01}; (15)\citealt{tex:CS06b}; (16)\citealt{tex:SR01}; 
 (17)\citealt{tex:MM97}; (18) \citealt{tex:FF96b}; (19)\citealt{tex:NS07}; (20)\citealt{tex:VM04}; 
 (21)\citealt{tex:KB97}; (22) \citealt{tex:CB08}; (23)\citealt{tex:BF09} ; (24)NED/Virgo $+$ GA $+$ Shapley
 corrected Hubble flow distances; (25)\citealt{tex:MB98} ; (26)\citealt{tex:GD07a} ; 
 (27)\citealt{tex:KF09}; (28)\citealt{tex:SG07}; (29)\citealt{tex:GR09}; (30)\citealt{tex:GT09}; 
 (31)\citealt{tex:SG10}.}	
  \smallskip
  \centering
  \begin{tabular}{lccccc}
  \hline
  \noalign{\smallskip}
Galaxy    & Distance &  $M_{\rm bh}$         & method-  &  $n_{\rm sph}$  & Band-\\
Name      &   (Mpc)  &(10$^{8}$M$_{\odot}$)  &  ref     &  (mag)      &  ref     \\
 (1)      &  (2)     &	(3)                  &    (4)   &   (5)       &  (6)     \\  
  \noalign{\smallskip}
  \hline
  \hline
  \noalign{\smallskip}
NGC221   &  0.86     &0.025$^{+0.005 }_{- 0.005 }$ & s-1     &  1.32  & R-26  \\ 
NGC863   &  7.4 (24) &0.47$^{+0.074  }_{-  0.074}$ & r-2     &   /    & /     \\
NGC1068  &  15.2 (24)& 0.084$^{+0.003}_{- 0.003 }$ & m-3     &   /    & /     \\ 
NGC2778  &  22.3     &0.15$^{+0.09   }_{- 0.10  }$ & s-4     &  1.60  & R-26  \\ 
NGC2960  &  72.8 (24)&0.12$^{+0.03   }_{- 0.03}$   & m-5     &   /    & /     \\ 
NGC3245  &  20.9     & 2.1$^{+0.5    }_{- 0.5 }$   & g-6     &   /    & /     \\ 
NGC4258  &  7.2 (7)  & 0.3$^{+0.2}_{- 0.2}$     & s-9    &  2.04  & R-26  \\ 
NGC4261  &  31.6     & 5.2$^{+1.0    }_{- 1.1 }$   & g-10    &  7.30  & R-26  \\ 
NGC4303  &  16.1 (18)& 0.006-0.16                  & g-11    &   /    & /     \\ 
NGC4342  &  17.0 (12)& 3.3$^{+  1.9  }_{- 1.1 }$   & s-13,20 &   /    & /     \\ 
NGC4374  &  18.4     & 4.64$^{+  3.46}_{- 1.83}$   & g-14    &  5.60  & V-27  \\ 
NGC4435  &  14.0(24) &$<$0.075                     & g-15    &   /    & /     \\ 
NGC4459  &  16.1     & 0.70$^{+  0.13}_{- 0.13 }$  & g-16    &  3.17  & V-27  \\ 
NGC4473  &  15.3     & 1.2$^{+  0.4  }_{- 0.9  }$  & s-4     &  2.73  & R-26  \\  
NGC4486  &  16.1     &    34$^{+  10 }_{- 10   }$  & g-17    & 11.84  & V-27  \\  
NGC4486a &  17.0 (12)&  0.13$^{+ 0.08}_{- 0.08 }$  & s-19    &  2.04  & V-27  \\  
NGC4486b &  17.0 (12)&  6.0$^{+  3.0 }_{-2.    }$  & s-21    &  2.2   & V-27  \\ 
NGC4552  &  15.3     &  4.8$^{+  0.8 }_{-0.8   }$  & s-22    &  9.22  & V-27  \\ 
NGC4564  &  14.6     &  0.60$^{+ 0.03}_{-0.09 }$   & s-4     &  3.15  & R-26  \\ 
NGC4596  &  17.0 (18)&  0.79$^{+ 0.38}_{-0.33 }$   & g-16    &   /    & /     \\ 
NGC4621  &  18.3     &  4.0$^{+ 0.6  }_{-0.6  }$   &  s-22   &  5.36  & V-27  \\ 
NGC4649  &  16.8     & 22.0$^{+  4.0 }_{-6.0  }$   & s-4     &  6.04  & R-26  \\ 
NGC4697  &  11.4     &  1.8$^{+ 0.2  }_{-0.1  }$   & s-4     &  4.00  & R-26  \\ 
NGC5576  &  24.8     &  1.8$^{+ 0.3  }_{-0.4  }$   & s-29    &   /    & /     \\
NGC5813  &  32.2     &  7.0$^{+ 1.1  }_{-1.1  }$   &  s-22   &   /    & /     \\ 
NGC5845  &  25.2     &  2.6$^{+ 0.4  }_{-1.5  }$   &  s-4    &  3.22  & R-26  \\   
NGC5846  &  24.9     & 11.0$^{+  2.0 }_{-2.0  }$   &  s-22   &   /    & /     \\  
NGC7052  &  66.4 (24)&  3.7$^{+ 2.6  }_{-1.5  }$   & g-25    &  4.55  & R-26  \\ 
UGC9799  &  141 (23) & $<$46.0                     & g-23    &  1.4   & R-28  \\ 
\noalign{\smallskip}	     
\hline
\end{tabular}		     
\label{table:masses} 
\end{table*}

\begin{table*} 
  \caption{Galaxy Sample - Additional SMBH mass - galaxy distance measurements. 
   The layout is as in Table~\ref{table:masses}.}	
  \smallskip
  \centering
  \begin{tabular}{lccc}
  \hline
  \noalign{\smallskip}
Galaxy    & Distance &  $M_{\rm bh}$         & method-  \\
Name      &   (Mpc)  &(10$^{8}$M$_{\odot}$)  &  ref     \\
 (1)      &  (2)     &	(3)                  &    (4)   \\  
  \noalign{\smallskip}
  \hline
  \hline
  \noalign{\smallskip}
NGC4486  &  17.9     &  64$^{+5}_{-5}$       & s-30   \\
NGC4649  &  15.7     &  45$^{+10}_{-10}$     & s-31   \\ 
\noalign{\smallskip}	     
\hline
\end{tabular}		     
\label{table:secmasses} 
\end{table*}

\clearpage

\clearpage

\appendix

\section{GALFIT3 sigma map}

In this paper we profiled a sample of 29 galaxies by allowing \texttt{GALFIT3} to create the sigma (weight) maps internally. In this Section we want to repeat the fit for the full sample by providing an external sigma map for each galaxy. The construction of the external maps is based on the same formulae that \texttt{GALFIT3} uses internally: 
\begin{equation}
\sigma^{[ADU]}(x,y) =\frac{\sqrt{\sigma_{d}^{[e^{-}]}(x,y)^2 + \sigma_{sky}^{[e^{-}]}(x,y)^2}}{gain}
\end{equation}
\begin{equation}
= \sqrt{\frac{f_{d}^{[ADU]}(x,y)-sky^{[ADU]}}{gain} + \sigma^{[ADU]}_{\rm sky}(x,y)^2}
\end{equation}
where
\begin{equation}
\sigma_{d}^{[e^{-}]}(x,y) =\sqrt{(f_{d}^{[ADU]}(x,y)-sky^{[ADU]})gain}
\end{equation}
where $f_{d}(x,y)$ is the image flux at pixel (x,y) in $ADU$ units, $sky$ is the sky value listed in Table~\ref{table:info}, $gain$ is equal to 4.5 e$^{-}\\ADU$, and $\sigma_{\rm sky}(x,y)$ is the full resolution noise map of the sky background created by \texttt{SExtractor}. The sigma map shows the flux uncertainty at each pixel and the construction is based on Poisson statistics. The created sigma map is compatible with \texttt{GALFIT3} and can replace the internal sigma map. 

\begin{figure}
\centering
\includegraphics[height=8.0cm,width=8.0cm]{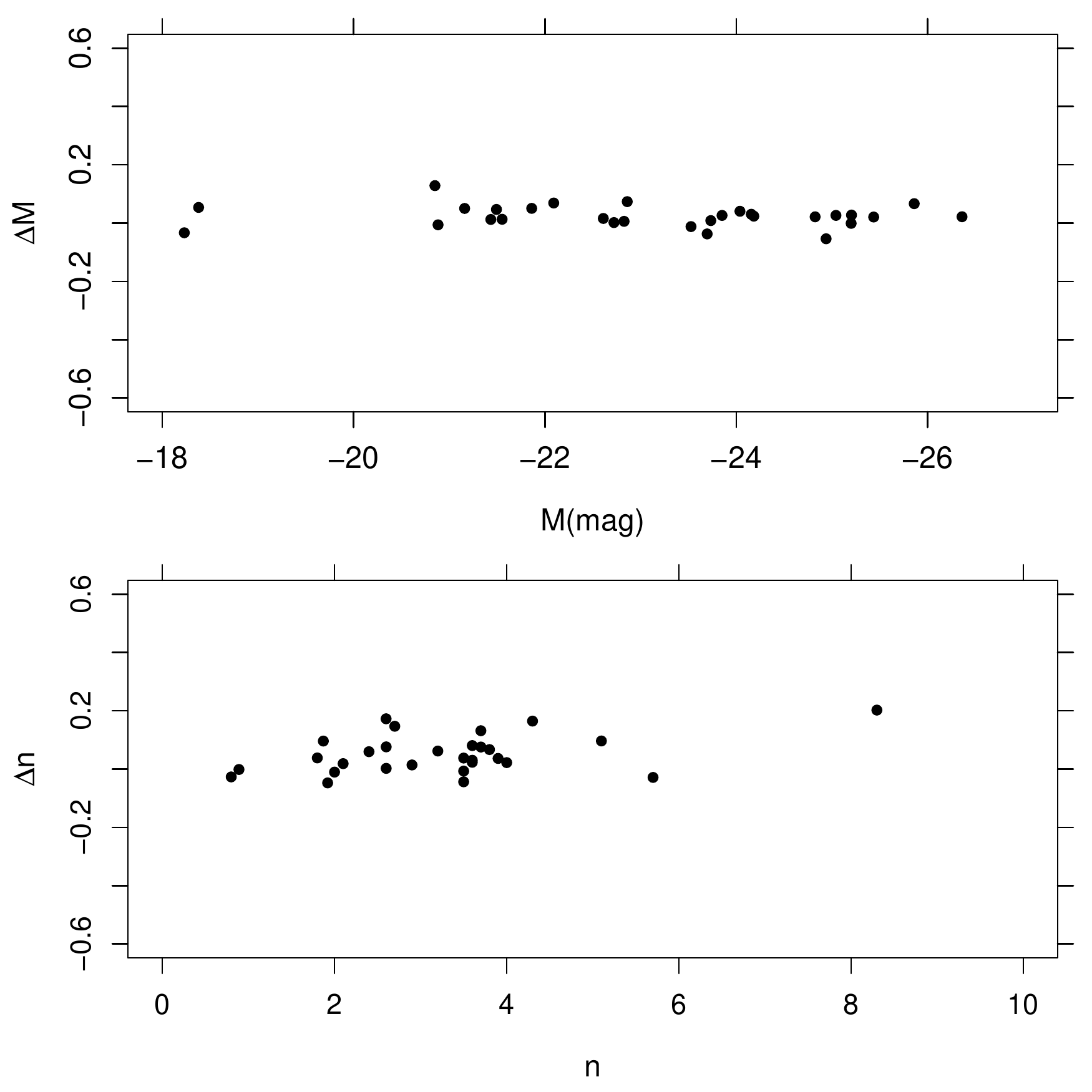}
\caption{Top panel: A plot of the \texttt{GALFIT3} magnitude differences found using either an internal sigma map or an external sigma map against the range of internal magnitudes.  Bottom panel: The same for the S{\'e}rsic index.} 
\label{fig:sigmacom}
\end{figure}

After the sigma map has been created we re-run \texttt{GALFIT3} using as  input/starting values the best fit values found in Table~\ref{table:properties} but instead of permitting an internal weight map estimation we provide the external weight map. Figure~\ref{fig:sigmacom} shows that there is no significant change in using a  \texttt{GALFIT3} internal sigma map or a \texttt{SExtractor} external sigma map.

\label{lastpage}

\end{document}